\documentclass[aps,twocolumn,showpacs,preprintnumbers,amssymb]{revtex4}

\newcommand{\be}{\begin{eqnarray}}
\newcommand{\ee}{\end{eqnarray}}
\def\bea{\begin{eqnarray}}
\def\eea{\end{eqnarray}}
\usepackage{bm} 
\usepackage{shadow,fancybox,amsmath} 
\usepackage{amscd} 

\usepackage{amsmath,amsfonts,amssymb,graphics,graphicx,epsfig,color,times,bbm}

\begin{document}



\title{Standard forms of noisy quantum operations via depolarization}

\author{W. D{\"u}r$^{1,2}$, M. Hein$^{1}$, J.I. Cirac$^{3}$ and H.-J. Briegel$^{1,2}$}

\affiliation{
$^1$ Institut f{\"u}r Theoretische Physik, Universit{\"a}t Innsbruck,
Technikerstra{\ss}e 25, A-6020 Innsbruck, Austria.\\
$^2$ Institut f\"ur Quantenoptik und Quanteninformation der \"Osterreichischen Akademie der Wissenschaften, Innsbruck, Austria.\\
$^3$ Max-Planck-Institut f\"ur Quantenoptik, Hans-Kopfermann-Str. 1, D-85748 Garching, Germany.
}

\date{\today}

\begin{abstract}
We consider noisy, non--local unitary operations or interactions, i.e. the corresponding evolutions are described by completely positive maps or master equations of Lindblad form. We show that by random local operations the completely positive maps can be depolarized to a standard form with a reduced number of parameters describing the noise process in such a way that the noiseless (unitary) part of the evolution is not altered. A further reduction of the parameters, in many cases even to a single one (i.e. global white noise), is possible by tailoring the decoherence process and increasing the amount of noise. We generalize these results to the dynamical case where the ideal unitary operation is given by some interaction Hamiltonian. The resulting standard forms may be used to compute lower bounds on channel capacities, to simplify quantum process tomography or to derive error thresholds for entanglement purification and quantum computation.
\end{abstract}


\maketitle


\section{Introduction}

Quantum systems evolve via unitary operations $U(t)$, as they are governed by the Schr\"odinger equation. This also holds for composite systems, e.g. a small quantum system $S$ which is surrounded by some environment $E$, where the evolution of the total system is described by a unitary operation $U_{SE}(t)$. The dynamics of the system $S$ alone ---which can be obtained by tracing out the (uncontrollable) degrees of freedom of the environment--- is in general no longer unitary. In fact, the system interacts with degrees of freedom of the environment, leading to entanglement between system $S$ and environment $E$ reflected in $U_{SE}(t) \not = U_S(t) \otimes U_E(t)$. The system--environment interaction leads to decoherence and the dynamics of the system can be described either by a (time dependent) completely positive map (CPM) ${\cal E}(t)$ or ---under certain assumptions on the nature of interaction--- by a master equation of Lindblad form \cite{Lidar01}. From the perspective of quantum information processing, such an interaction with environmental degrees of freedom is undesirable and leads to errors and noise in the system. As discussed below, an arbitrary noise process acting on a $d$--dimensional system $S$ is, at fixed time $t_0$, determined by $O(d^4)$ real parameters. Even for small system sizes, e.g. when $S$ consists of three qubits (i.e. $d=8$), this leads to a huge number of independent parameters (e.g. around 4000 for the three--qubit system), which makes an analytical treatment of the influence of such general noise processes on the properties of the system rather difficult. This is particularly hindering when considering either large systems or sequences of several noisy evolutions (or gates), as is e.g. required in the analysis of quantum circuits or processes such as entanglement purification.

When considering the influence of noise in quantum information processing, one hence often restricts the analysis to certain (ad hoc) noise models, such as Pauli channels or depolarizing (white) noise models. This is usually the case in the analysis of entanglement purification protocols in the presence of noisy operations as well as in the theory of fault--tolerant quantum computation. On the other hand, having a specific physical set--up in mind, one can sometimes justify these (or other) noise models by a microscopic description of the underlying system--environment interactions, where only the dominant part of noisy interactions is considered. However, when considering (abstract) quantum processes that deal with the manipulation of quantum information, one does not want to restrict oneself to a specific physical set--up, but rather would like to keep the analysis at an abstract level and as general as possible. To this aim, it would be very useful to justify the usage of simple noise models in a general context, or to provide a method to bring any noise process to a simple standard form described by a few parameters. 

In this article, we provide a method which allows one to achieve this aim. We show that indeed any noise process can be brought to a simple standard form by means of {\it depolarization}. That is, by applying appropriate (local) unitary control operations on a system before and after the noisy evolution in a correlated way, one can {\em depolarize} the noise process. This depolarization process of the corresponding CPM ${\cal E}$ or the Liouvillian ${\cal L}$ can be viewed as an analogue of the depolarization of mixed states. For bipartite states, for instance, it was shown that one can bring any mixed state $\rho$ of two $d$--level systems to a standard form specified by a single parameter using an appropriate (random) sequence of local unitary operations. Depolarization takes place in such a way that the fidelity of the state, i.e. the overlap with a maximally entangled state $|\Phi\rangle =1/\sqrt{d} \sum_{k=1}^d |k\rangle |k\rangle$, remains invariant. The resulting states (isotropic states) are equivalent to Werner states and are given by $\rho (x)= x |\Phi\rangle\langle \Phi| + (1-x) \frac{1}{d^2} \mathbf{1}$. Werner states played an important role in the investigation of the relations between entanglement and local hidden variable theories \cite{We89}, as well as in the development of entanglement purification protocols, schemes which are becoming increasingly important since it was realized that entanglement can serve as a valuable resource not only in quantum communication but also in quantum information processing. The development of these important issues was triggered by the simplified description of Werner states (still covering essential entanglement properties), and allowed at the same time to obtain necessary or sufficient conditions for separability or distillability for arbitrary bipartite states.

We are confident that also the depolarization of noisy evolutions will prove to be a fruitful tool in the analysis of noise processes. In direct analogy to the depolarization of states, the depolarization of the noise maps takes place in such a way that the fidelity of the ideal (unitary or Hamiltonian) part of the evolution is not altered. In fact, we make use of the isomorphism between completely positive maps and mixed states \cite{Jam72,Ci00} and connect the problem of depolarization of maps to the depolarization of the corresponding states, while respecting certain locality restrictions. For decoherence processes (e.g. storage errors of a system due to its interaction with the local environment, or errors resulting due to sending a system through a noisy quantum channel), where the ideal operation is the identity or, equivalently, the Hamiltonian part in the corresponding master equation is zero, we find that one can depolarize the corresponding map or master equation to a standard form which is described by correlated and uncorrelated white noise processes. In the case of two $d$--level systems, for instance, the corresponding depolarized map is described by three real parameters (the weights of ideal operation, single particle (uncorrelated) white noise processes and two--particle (correlated) white noise) as compared to $O(d^8)$ parameters of an arbitrary map. We also consider noisy interactions (i.e. the ideal evolution is given by some non--local unitary operation $U$ or some non--trivial system interaction Hamiltonian $H$), where we concentrate on two--system interactions. We find that for certain unitary operations, in particular for SWAP gates, CNOT gates, as well as phase gates with arbitrary phase $\alpha$, a depolarization is still possible. The required number of parameters to describe the (depolarized) noise process depends on the unitary operation (interaction) that has to be kept invariant, and is given by 17 in the case of arbitrary phase gate, 8 in the case of the CNOT gate and 3 for the SWAP gate.

Knowledge of the exact form of the noise process (which can e.g. be acquired by means of gate tomography) or additional control of interactions (e.g. the ability to switch a noisy interaction on and off at will) allows to further tailor the noise process. In this case, the fidelity of the ideal operation is decreased by a certain (small) amount, while the description of the noise process is simplified and the number of relevant parameters is further reduced. In many relevant cases (e.g. noisy SWAP or CNOT gate, switchable noisy phase gate), one finds that one can indeed simplify the noise process in such a way that the corresponding CPM is described by a single parameter and the noise process corresponds to correlated white noise. The total amount of noise is ---in the worst case--- increased by about an order of magnitude, as weight of the ideal operation is transferred to the noise part in an appropriate way to achieve this further simplification.

While in the case of maps a depolarization with a significant reduction of the associated parameters is only possible for certain unitary operations, one finds that, in the case of master equations, sequential application of fast intermediate local unitary control operations allows one to depolarize {\em any} master equation (of two systems) to a standard form described by at most 17 parameters. In return this depolarization protocol generally increases the noise level of the decoherence process. Under certain circumstances, one may even achieve a standard form described by a single parameter for arbitrary two--qubit interactions by accepting a further increase of the noise level.

Such standard forms for noisy evolutions may have wide spread applications in the analysis of quantum information processes under realistic conditions. For instance, our approach allows one to obtain lower bounds on the capacity of arbitrary multipartite quantum channels by considering the corresponding depolarized channels. The depolarized noise process also gives rise to a simplified process tomography. The tomography has to reveal fewer parameters (the parameters characterizing the standard form) than those necessary to describe the original decoherence process or the noisy gate. This can lead to a significant reduction of the experimental effort to sufficiently characterize the influence of noise in a given set--up. 
Also processes involving sequences of noisy gates, e.g. entanglement purification or some quantum circuits, can be analyzed by considering the standard forms for the corresponding gates. The resulting threshold values do no longer refer to specific error models but are valid in general, as any noise process can be brought to the corresponding standard form. When applying this method to derive generally valid error thresholds, e.g. in the context of fault tolerant quantum computation, some care is required. An implicit assumption in order to allow the application of such a local depolarization procedure is that the corresponding (local) control operations can in fact be (noiselessly) applied to the system. When dealing for instance with decoherence processes due to channel noise or local interaction of the system with some environment, such an assumption is perfectly reasonable. Also for two--system interaction gates (such as the CNOT), one may assume that local, single system gates are noiseless (or introduce a negligible amount of noise as compared to the two--system gate). However, when dealing also with noisy single system operations (as is e.g. required in the analysis of fault tolerant quantum computation), it is no longer straightforward to apply our results. One might argue that for sequences of gates the required (random) operations for depolarization can be incorporated in previous/subsequent noisy gates, although it is not entirely clear whether this argument justifies the assumption that any gate within a quantum circuit is already of standard form. However, whenever local, single system control operations can be assumed to be noiseless, our results are applicable and one can indeed bring an arbitrary (non--local) noise process to a simplified standard form.

This paper is organized as follows: In Sec.~\ref{Iso1party} we review basic properties of the Jamio\l kowski isomorphism between completely positive maps and states, which will be the main tool for the derivation of standard forms for CPM in the following sections. We will then apply this isomorphism in Sec.~\ref{NFoneParty} in order to provide standard forms for an arbitrary decoherence process in the case where the corresponding control operation to achieve this standard form does not have to obey any locality requirements. In Sec.~\ref{SF_Multi} we derive standard forms for maps describing arbitrary decoherence processes and some noisy unitary operations. These standard forms are achieved by control operations that are local with respect to (w.r.t) some given partitioning. In Sec.~\ref{SF_Dynamics} we suggest a protocol to bring an arbitrary noisy evolution described by a master equation into some standard form, for which the accompanying noise process is described by a reduced number of parameters. Finally we summarize our results in Sec.~\ref{summary}. Some technicalities can be found in the appendices.


\section{The Jamio\l kowski correspondence between completely positive maps and states} \label{Iso1party}

In this section we review some properties of the Jamio\l kowski isomorphism \cite{Jam72} between completely positive maps (CPM) and states. In Sec.~\ref{IsoGeneral} we state and discuss this isomorphism first on an abstract level as a correspondence between matrices and the endomorphisms of the corresponding matrix algebras. We will then restrict this general isomorphism to the physical setting of quantum states and quantum operations in Sec.~\ref{IsoPhysics}, where the isomorphism has a clear interpretation in terms of a teleportation protocol. In Sec.~\ref{distance} we review some applications of the isomorphism \cite{GLN}. Known distance measures for quantum states can be used to provide distance measures for (trace preserving) CPM, which we will use in the following.  Finally we extend the Jamio\l kowski isomorphism in Sec.~\ref{MultiPartyIso} to the multi--party setting and discuss some implications for the entanglement capabilities of CPMs. For sake of completeness the reader can find a review\cite{Ci00,Ar03}  about the relation between the spectral decomposition of states and the Kraus representation for CPM in Appendix A and about the relation between the purification for quantum states and quantum operation in Appendix B. Note that the main properties of the isomorphism are stated in the form of short propositions with a consecutive numbering (No.~{\bf 1} -- {\bf 13}) , that is continued in the Appendix.

\subsection{The Isomorphism in the general setting}\label{IsoGeneral}

Let $\mathbf{H}_A$ and $\mathbf{H}_{A'}$ be two Hilbert spaces of finite dimensions $d_A=\text{dim}_{\mathbb C} (\mathbf{H}_A)$ and $d_{A'}=\text{dim}_{\mathbb C} (\mathbf{H}_{A'})$. With $\mathcal{M}_A=\mathcal{M}(\mathbf{H}_A)$ and $\mathcal{M}_{A'}=\mathcal{M}(\mathbf{H}_{A'})$  we denote the corresponding matrix algebras over $\mathbf{H}_A$ and $\mathbf{H}_{A'}$ respectively, which contain the set of physical states (density matrices) $\mathcal{D}_A\subset \mathcal{M}_A $ and $\mathcal{D}_{A'}\subset \mathcal{M}_{A'}$ as (proper) convex subsets. Similarly, we will write $\mathcal{M}(\mathbf{H}_A,\mathbf{H}_{A'})$ for the algebra of $d_A'\times d_A$ matrices representing the linear maps from $\mathbf{H}_A$ to $\mathbf{H}_{A'}$. Moreover let $\text{\bf End}\left(\mathcal{M}_A \rightarrow \mathcal{M}_{A'}  \right)$ be the set of linear maps (endomorphisms) between the algebras $\mathcal{M}_{A}$ and $\mathcal{M}_{A'}$ , which contain the physical operations $\text{\bf CPM}\left(\mathcal{D}_A \rightarrow \mathcal{D}_{A'}  \right)$ between the two quantum systems, i.e. completely positive maps (CPM), as a proper subset. In the following we will frequently consider a copy $\bar{A}$ of system $A$ and use -- after identifying and fixing a basis in $\mathbf{H}_A$ and $\mathbf{H}_{A'}$ -- the maximally entangled state 
\be |\Phi\rangle = \frac{1}{\sqrt{d_A}} \sum_{i=1}^{d_A} \, |i\rangle^{\bar{A}}|i\rangle^A , \hspace{0.3cm} P_\Phi= |\Phi\rangle\langle\Phi|\ee
on the composite system $\mathbf{H}_{\bar{A}}\otimes\mathbf{H}_A$.

{\bf  \underline{The Jamio\l kowski Isomorphism} } \\
{\it The map $\mathbf{\mathcal{J}}: \mathcal{M}\left(\mathbf{H}_{A'} \otimes \mathbf{H}_{A}\right) \rightarrow \text{\bf End}\left(\mathcal{M}_A \rightarrow \mathcal{M}_{A'}  \right)$, that maps a matrix $E$ of the matrix algebra over the composite system $\mathbf{H}_{A'} \otimes \mathbf{H}_{A}$ to the linear map $\mathcal{E}$ given by
\be \label{Iso}  \mathcal{E}(M) := d_A^2\, \text{tr}_{A\bar{A}}\left[ E^{A'A}\, P_\Phi^{A\bar{A}}\, M^{\bar{A}} \right]\ee
(for any $ M \in \mathcal{M}_A \simeq \mathcal{M}_{\bar{A}}$)
 is an isomorphism \cite{Jam72,Ci00}, i.e. it is linear and bijective. For the inverse of $\mathbf{\mathcal{J}}$ the matrix $E \in \mathcal{M}\left(\mathbf{H}_{A'} \otimes \mathbf{H}_{A}\right)$, that corresponds to the linear map $\mathcal{E}$, is given by
\be  \label{Inv} E :=  \mathcal{E}^{\bar{A}} \otimes \text{Id}^A \left( P_\Phi^{\bar{A} A} \right) \; .  \ee
}

If the matrix $E$ and map $\mathcal{E}$ in correspondence are decomposed with respect to the chosen basis $|j\rangle$ ($j\in \mathbb{N}_{d_A}$) on $\mathbf{H}_A$ (or $\mathbf{H}_{\bar{A}}$) and $|i\rangle$ ($i\in\mathbb{N}_{d_{A'}}$) on $\mathbf{H}_{A'}$ 
\bea\label{In_basis}
E^{A' A} & = & \sum_{\genfrac{}{}{0pt}{}{i,k \in \mathbb{N}_{d_A'}}{ j,l \in\mathbb{N}_{d_{A}}}} \,E_{i j | k l} \, |i\rangle^{A'}\langle k| \otimes |j\rangle^{A}\langle l| \\ 
\mathcal{E}(M) & = & \sum_{\genfrac{}{}{0pt}{}{i,k \in \mathbb{N}_{d_A'} }{j,l \in\mathbb{N}_{d_{A}}}} \,\mathcal{E}_{i k | j l}\, \langle j | M | l \rangle \; |i\rangle^{A'} \langle k| 
\eea
the Jamio\l kowski isomorphism simply is \cite{Ar03}
\be \label{Iso_In_Basis}  \mathcal{E}_{ik|jl} = d_A \; E_{i j|k l}\; . \ee
From this relation between the coefficients the bijectivity immediately follows from the linearity of $\mathcal{J}$ together with the fact, that $\mathcal{M}\left(\mathbf{H}_{A'} \otimes \mathbf{H}_{A}\right)$ and $\text{\bf End}\left(\mathcal{M}_A \rightarrow \mathcal{M}_{A'}  \right)$ both are linear spaces of dimension $d_{A} \times d_{A'}$. The above result therefore can be shown by deriving relation (\ref{Iso_In_Basis}) separately from Eq.~(\ref{Iso}) and from Eq.~(\ref{Inv}) using the fact that 
\be d_A \, \text{tr}_{\bar{A}} \left[ P_\Phi^{A \bar{A}}\, M^{\bar{A}} \right] = (M^A)^t \ee
holds for any $ M\in \mathcal{M}_A $.

The isomorphism also turns out to be an isometry \cite{Ar03}: If one uses the inner products
\bea \langle E,F\rangle & := & \text{tr}\,E^\dagger F\\
\langle \mathcal{E}, \mathcal{F}\rangle & := & \sum_{j,l \in\mathbb{N}_{d_{A}}} \text{tr}\,\left[\left(\mathcal{E}(|j\rangle\langle l|)\right)^\dagger \mathcal{F}(|j\rangle\langle l|)\right] 
\eea
for $E,F\in \mathcal{M}\left(\mathbf{H}_{A'} \otimes \mathbf{H}_{A}\right)$ and the corresponding maps $\mathcal{E},\mathcal{F}\in \text{\bf End}\left(\mathcal{M}_A \rightarrow \mathcal{M}_{A'}  \right)$, it can readily be seen that \be \langle E, F\rangle = \langle \mathcal{E}, \mathcal{F} \rangle \ee holds.
Note that the corresponding norms $\|M \|$ on $\mathcal{M}_A$ ($\mathcal{M}_{A'}$) and $\|\mathcal{E}\|$ on $\text{\bf End}\left(\mathcal{M}_A \rightarrow \mathcal{M}_{A'}  \right)$ are Euclidean ($l_2$ norms). 

Eq.~(\ref{Iso_In_Basis}) shows, that the isomorphy does not extend to the respective compositions in $\mathcal{M}\left(\mathbf{H}_{A'} \otimes \mathbf{H}_{A}\right)$ and $\text{\bf End}\left(\mathcal{M}_A \rightarrow \mathcal{M}_{A'}  \right)$. For example, the composition $\mathcal{E} \circ \mathcal{F}$ of to maps would correspond to a matrix composition law $(E\Diamond F)_{ij|kl} = \sum_{mn} E_{im|kn} F_{mj|nl}$ which differs from the usual matrix multiplication. Nevertheless it also provides $\mathcal{M}\left(\mathbf{H}_{A'} \otimes \mathbf{H}_{A}\right)$ with a semi-group structure \cite{Ar03}.

The effect of matrix multiplication by local \cite{local} matrices is given by the following formula:
Given a map $\mathcal{E}$ and its corresponding matrix $E$ and the matrices $B_1,C_1\in \mathcal{M}_{A'}$, $B_2,C_2 \in \mathcal{M}_{A}$, then the transformed matrix 
\be 
  E' = B_1^{A'}\otimes B_2^{A} \, E^\mathcal{E} \, C_1^{A'}\otimes C_2^{A}
\ee
corresponds to a map $\mathcal{E'}$ with
\be\label{local_op_sigma}
   \mathcal{E'}(M):=\, B_1\,\mathcal{E}\left(B_2^T M C_2^T \right)\,C_1 \; . 
\ee

\subsection{The isomorphism for quantum states and quantum operations} \label{IsoPhysics}

Under which conditions on the matrix $E \in \mathcal{M}\left(\mathbf{H}_{A'} \otimes \mathbf{H}_{A}\right) $ does the linear map $\mathcal{E} \in \text{\bf End}\left(\mathcal{M}_A \rightarrow \mathcal{M}_{A'}  \right)$ correspond to a physical operation, i.e. is a (trace-preserving) CPM ? This can be answered by the following results \cite{Ar03}:
\begin{itemize}
\item[\bf{1.}] $\mathcal{E}$ is {\it Hermiticity preserving}\cite{Notation},  iff $E$ is {\it Hermitian}.
\item[\bf{2.}] $\mathcal{E}$ is {\it positivity preserving} \cite{Notation}, iff $E$ is {\it Hermitian and for all separable states $F \in \mathcal{D}\left(\mathbf{H}_{A'} \otimes \mathbf{H}_{A}\right) $  $\text{tr}\left( E F \right) \geq 0 $ holds}.
\item[\bf{3.}] $\mathcal{E}$ is {\it completely positive } \cite{Notation}, iff $E$ is {\it positive}.
\item[\bf{4.}] $\mathcal{E}$ is a {\it trace-preserving CPM} \cite{Notation}, iff $E$ is {\it positive and $\text{tr}_{A'} E^{A' A} = \frac{1}{d_A} \mathbf{1}_A$ holds}.
\end{itemize}

\begin{figure}[th]
\includegraphics[width=5cm]{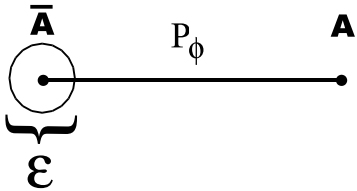}
\caption{\label{fig:Inv} In order to obtain the state $E$ the CPM $\mathcal{E}$ is applied to system $\bar{A}$ of the joint system of $A$ and $\bar{A}$, which is prepared in the maximally entangled state $P_\Phi^{\bar{A}A}$. }  
\end{figure}

\begin{figure}[th]
\includegraphics[width=8cm]{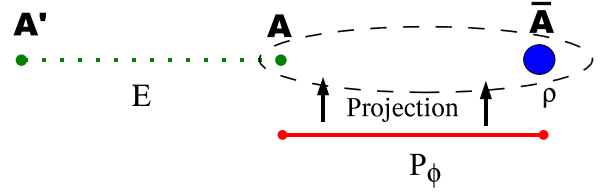}
\caption{\label{fig:Iso} Given the state $E$ on the composite system $A'$ and $A$, the CPM $\mathcal{E}$ is evaluated for an arbitrary input state $\rho$ by taking $\rho$ as an input at system $\bar{A}$. Then the joint system $A\bar{A}$ is measured in a Bell basis containing the maximally entangled state $P_\Phi^{A\bar{A}}$. With probability $\frac{1}{d_A^2}$ the desired output state $\mathcal{E}(\rho)$ is then obtained at system $A'$.} 
\end{figure}

These results imply that the Jamio\l kowski isomorphism can be restricted to $\mathbf{\mathcal{J}}: \mathcal{D}\left(\mathbf{H}_{A'} \otimes \mathbf{H}_{A}\right) \rightarrow \text{\bf End}\left(\mathcal{D}_A \rightarrow \mathcal{D}_{A'}  \right)$ yielding a correspondence between {\it trace-preserving CPM} and {\it states} on the composite system of $A$ and $A'$. In this case the isomorphism can be given a natural interpretation in terms of a {\it teleportation protocol} (without classical communication): In order to obtain the state $E$ corresponding to a CPM $\mathcal{E}$ according to Eq.~(\ref{Inv}) the CPM $\mathcal{E}$ simply has to be applied at the system $\bar{A}$ of the composite system in the maximally entangled state $|\Phi\rangle$ (see Fig.~\ref{fig:Inv}).
 Conversely, given the state $E$, the CPM $\mathcal{E}$ can be evaluated for an arbitrary input state $\rho$ according to Eq.~(\ref{Iso}) as follows (see Fig~\ref{fig:Iso}). Considering the composite system consisting of parties $A'$ and $A$ in the state $E$ together with the input state $\rho$ at system $\bar{A}$, i.e. the total state $E^{A'A}\otimes\rho^{\bar{A}}$, the joint system $A\bar{A}$ is measured in a Bell basis containing the maximally entangled state $P_\Phi^{A\bar{A}}$. With probability $\frac{1}{d_A^2}$ the desired output state $\mathcal{E}(\rho)$ is then obtained at system $A'$. 

According to Eq.~(\ref{local_op_sigma}) any operation $\mathcal{N}$ on $E$, which is separable w.r.t. the partitioning $(A',A)$, i.e. 
\be E' = \sum_{j} B_j^{A'} \otimes C_j^A E (B_j^{A'} \otimes C_j^A)^\dagger \; ,\ee 
translates to a probabilistic application of combined operations before and after the CPM $\mathcal{E}$:
\be \label{local_manipulation}
\mathcal{E'}(M):= \sum_{j}\, B_j\,\mathcal{E}\left( C_j^T M C^*_j \right)\, B_j^\dagger \; .
\ee
In particular, the application of local unitaries or measurements to $E$ on party $A$ [$A'$] corresponds to the application of local unitaries or measurements before [after] the CPM $\mathcal{E}$. On the other hand, not all separable operations can be implemented by local quantum operations and classical communication (LOCC) \cite{Be98}. Since only the measurement results before the CPM $\mathcal{E}$ can influence operations performed afterward, we have to restrict the separable operations on side $A$ and $A'$ even  to be local quantum operations and one-way classical communication (1-LOCC) from party $A$ to party $A'$. The separable operations in question thus correspond to the state
\be 
E' = \sum_{i j} B_{i j}^{A'} \otimes C_j^A E (B_{i j}^{A'} \otimes C_j^A)^\dagger \; ,\nonumber 
\ee
where
\begin{itemize} 
\item[(i)] $\sum_{j} C_j^\dagger C_j = \mathbf{1}$, i.e. the quantum operation $\mathcal{C}(\rho) = \sum_{j} C_j \rho C_j^\dagger$ on party $A$ is a trace preserving CPM;
\item[(ii)] $\mathcal{C}$ is bi-stochastic $\mathcal{C}(\mathbf{1})=\mathbf{1}$ and hence the corresponding CPM $\tilde{\mathcal{C}}(\rho) = \sum_{j} C_j^T \rho C^*_j$ before the application of $\mathcal{E}$ is also a trace preserving CPM, i.e. $\sum_{j} C^*_j C_j^T = \mathbf{1}$;
\item[(iii)] for each measurement outcome $j$ on party $A$ [before the application of $\mathcal{E}$] the corresponding operation $\mathcal{B}_j (\rho) = \sum_{i} B_{i j} \rho B_{ij}^\dagger$, that is performed on party $A'$ according to the classical information sent by $A$, is a trace preserving CPM, i.e. $\sum_{i} B_{ij}^\dagger B_{ij} = \mathbf{1}$.
\end{itemize}
Condition (i) and (iii) specify the notion of a general 1-LOCC protocol, that we consider in the following.
In many cases such as for local projective measurements or for probabilistic applications of local unitaries, property (ii) follows from (i), but in general (ii) provides an separate condition, which reflects the fact that before $\mathcal{E}$ not  $\mathcal{C}$ but $\tilde{\mathcal{C}}$ with transposed Kraus operators is applied. To simplify notations we will therefore consider those 1-LOCC protocols, that satisfy all three conditions.
The above discussion indicates the two directions, in which one can try to manipulate $\mathcal{E}$ with the help of the corresponding state $E$:
\begin{itemize}
\item[\bf{(A)}] If one really has the above teleportation protocol available {\it in practice}, any (non-local) operation on $E$ can be considered in order to manipulate the corresponding CPM.
\item[\bf{(B)}] If the isomorphism is only a helpful {\it theoretical} tool, then one should only consider {\it 1-LOCC} operations on $E$ in order to manipulate a given CPM, since these operations can be implemented by a coordinated application of operations before and after the evaluation of the CPM.  
\end{itemize}
We emphasize that only one direction of the isomorphism protocol can be implemented with unit probability of success. This implies that the case {\bf{(A)}} corresponds to a {\it probabilistic} modification of the CPM $\mathcal{E}$ whereas case {\bf{(B)}} gives rise to a {\it deterministic} manipulation protocol. 
Case {\bf{(A)}} is also equivalent to all protocols, in which one does not only allow arbitrary local operations before and after the application of $\mathcal{E}$ (and therefore the use of independent ancillary systems to perform these operations) but also to make an (arbitrary) ancillary system available to store {\it quantum} information. This information is obtained during the operations before the CPM $\mathcal{E}$ and later used in the operations performed after $\mathcal{E}$ \cite{ancEquiv}. 

\subsection{Distance measures for quantum states and quantum operations}\label{distance}

In the remainder of this paper we derive standard forms $\mathcal{E}'$ for some noisy CPM $\mathcal{E}$, that approximates some ideal operation $\mathcal{I}$. A reasonable requirement for such a standard form is that it is also a considerably good approximation to the ideal operation. In order to assess this requirement some kind of distance measure between quantum operation is needed. As a first application of the Jamio\l kowski correspondence we thus review the derivation \cite{GLN} of distance measures for quantum operations from those for quantum states. 

Concerning the isometry properties discussed in Sec.~\ref{IsoGeneral} note that the Euclidean norm does not provide a proper distance measure $d$ for quantum states, since it does not obey the contractivity property, that is \be d(\mathcal{E}(\rho),\mathcal{E}(\sigma))\leq d (\rho,\sigma) \ee for all states $\rho, \sigma$ and trace-preserving quantum operations $\mathcal{E}$ \cite{Oz00}. This property expresses the physical condition that no quantum process should allow to increase the distinguishability of two quantum states. In the literature (see e.g. \cite{GLN,Ni00,Ki02}) there are mainly two metrics \cite{metric} considered that also obey the contractivity property, namely:
\begin{itemize}
\item {\it trace distance}: $d_1(\rho,\sigma):=\frac{1}{2} | \rho -\sigma|_\text{tr}$, where $|M|_\text{tr}:=\text{tr}\left(\sqrt{M^\dagger M}\right)$ is the {\it trace norm};
\item {\it fidelity-based distances}, that are monotonically decreasing functions of the fidelity  $F(\rho,\sigma):=\text{tr}\left(\sqrt{\sqrt{\rho}\sigma\sqrt{\rho}} \right)^2$ such as $d_2(\rho,\sigma):= \sqrt{1-F(\rho,\sigma)}$; note that $F(|\psi\rangle,\sigma)= \langle \psi | \sigma | \psi \rangle$ if $\rho=|\psi\rangle\langle\psi|$ is pure.
\end{itemize}
By again using the Jamio\l kowski isomorphism both distance measures for quantum states have a natural counterpart as a distance measure for quantum operation. Given two trace-preserving quantum operations $\mathcal{E}$ and $\mathcal{F}$ one can define the distance $\Delta(\mathcal{E},\mathcal{F})=d (E,F)$ as the distance $d$ between the corresponding states $E$ and $F$, which is easily shown to yield a metric $\Delta$ on the set of trace-preserving quantum operations as long as $d$ is a metric on the corresponding set of quantum states. Choosing $d=d_1$ or $d=d_2$ the corresponding distance measures $\Delta_1$ and $\Delta_2$ also have the following two properties, which seem to be reasonable requirements for any distance measure for quantum operations \cite{GLN}:
\begin{itemize}
\item {\it Stability} \cite{Ki02}: $\Delta(\text{Id}\otimes\mathcal{E},\text{Id}\otimes\mathcal{F})=\Delta(\mathcal{E},\mathcal{F})$, i.e. the distance measure of two quantum processes should not depend on whether they are considered to occur in an environment together with some unrelated ancillary quantum system;
\item {\it Chaining} \cite{chainingCond}: $\Delta (\mathcal{E}_1\circ\mathcal{E}_2,\mathcal{F}_1\circ\mathcal{F}_2) \leq \Delta(\mathcal{E}_1,\mathcal{F}_1) + \Delta(\mathcal{E}_2,\mathcal{F}_2)$, i.e. for a composed process, the total error will be less than the sum of the errors in each individual step.
\end{itemize} 
Note that the evaluation of the above distance measures in practice requires some quantum process tomography. Moreover both measures can be shown to have some physical interpretation in the sense of a bound to the average--case--error in function computation and sampling computation \cite{GLN}. But in the following we will consider a slightly different application. Unfortunately the natural approach for defining error measures for quantum operations by averaging the distances between the output states, i.e.
$\Delta (\mathcal{E},\mathcal{F}) := \int d\psi\; d(\mathcal{E}(\psi),\mathcal{F}(\psi)) $ so far could not be modified in such a way that it would also fulfill the stability property. Nevertheless, for the case that one operation $\mathcal{F}=U$ is a unitary operation, the average fidelity 
\be \bar{F}(\mathcal{E},U):=\int d\psi\; \langle \psi| U^\dagger \, \mathcal{E} (|\psi\rangle \langle \psi |)\, U |\psi \rangle \ee
 has at least a plausible interpretation in terms of the average 'overlap' between the two outputs $U|\psi\rangle $ and $\mathcal{E} (|\psi\rangle \langle \psi |)$, although it does not even define a metric. It was shown in \cite{HoNi} that this average fidelity $\bar{F}(\mathcal{E},U)$ is linearly related to the 'Jamio\l kowski' fidelity $F(\mathcal{E},U)= F(E,|\psi_U\rangle)$ by:
\be \bar{F}(\mathcal{E},U) = \frac{F(\mathcal{E},U) d +1}{d+1} \; ,\ee
where $|\psi_U\rangle$ denotes the pure state corresponding to the unitary operation $U$. 

In the following we will be interested in standard forms for noisy operations $\mathcal{E}$, which approximate some ideal operation, that will be either the identity $\text{Id}$ or some unitary $U$. These standard forms $\mathcal{E'}$ are obtained by different protocols, which might introduce additional noise to the operation $\mathcal{E}$. Apart from simplicity of the obtained standard form $\mathcal{E}'$, it should only differ in the same order of magnitude from the ideal operation $U$ or $\text{Id}$ as the original imperfect operation $\mathcal{E}$. The above argument shows that the 'Jamio\l kowski' fidelity $F(\mathcal{E},\text{Id})$ or $F(\mathcal{E},U)$ can in both cases be used to measure this distance: On the one hand the fidelity is related to a decent distance measure for quantum operation by a monotonic decreasing function. On the other hand for our applications the fidelity has a physical interpretation in terms of the average error in approximating an ideal (unitary) quantum operation $U$. In the following we therefore try to provide standard forms $\mathcal{E}'$ of noisy operations $\mathcal{E}$, that have either the same or a slightly decreased fidelity $F(\mathcal{E}',U)$ with the ideal operation as the original one ($F(\mathcal{E},U)$).

\subsection{The Isomorphism in the multi-party setting}\label{MultiPartyIso}
The Jamio\l kowski isomorphism has a natural extension to multi-party scenarios, which are of special interest in quantum information theory.
For this let the system $A=(A_1,\ldots,A_N)$ consist of $N$ parties, each representing Hilbert spaces $\mathbf{H}^{A_i}$ of different dimensions $d_{A_i}$, such that
$\mathbf{H}^A=\mathbf{H}^{A_1} \otimes \ldots \otimes\mathbf{H}^{A_N} $ and $d_A=\prod_{i=1}^N d_{A_i}$. In order to keep the argumentation simple we consider only CPM $\mathcal{E}$, whose input and output Hilbert spaces are of the same type, i.e. $\mathbf{H}^{A'}=\mathbf{H}^{A'_1} \otimes \ldots \otimes\mathbf{H}^{A'_N}$ with $\mathbf{H}^{A'_i}\simeq \mathbf{H}^{A_i}$. 
\begin{figure}[th]
\includegraphics[width=6cm]{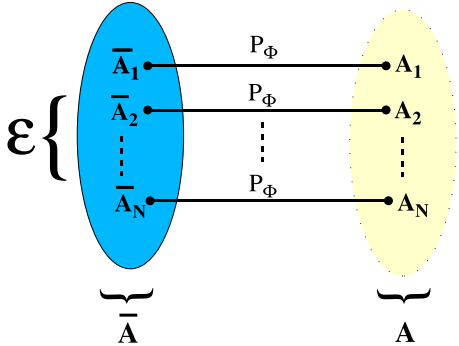}
\caption{\label{fig:MultiInv} In order to obtain the state $E$ the CPM $\mathcal{E}$ is applied to the systems $\bar{A}_i$ of the joint system $\bar{A}=(\bar{A}_1,\ldots,\bar{A}_N)$, which are (locally) prepared in the maximally entangled states $P_\Phi^{\bar{A}_iA_i}$. }  
\end{figure} 

The main point in extending the isomorphism to the multi-party setting is to choose the maximally entangled state $|\Omega\rangle_{\bar{A}A}$ to be the tensor product of the respective maximally entangled states 
\be |\Phi\rangle^{\bar{A}_iA_i}=\frac{1}{\sqrt{d_{A_i}}} \sum_{k=1}^{d_{A_i}} |k\rangle^{\bar{A}_i}|k\rangle^{A_i}\ee
 between the subsystem $A_i$ and its copy $\bar{A}_i$ at each individual party $i=1,\ldots,N$. The maximally entangled state $\omega$ is therefore
\be\label{Omega_Multi} P_\Phi^{\bar{A}A} = P_\Phi^{\bar{A}_1 A_1} \otimes \ldots \otimes P_\Phi^{\bar{A}_N A_N} \ee
with $P_\Phi^{\bar{A}_iA_i}=|\Phi\rangle^{\bar{A}_iA_i}\langle \Phi |$.
In this notation the isomorphism will have exactly the same form as stated above with the only difference that the maximally entangled state $\Phi$ used in both directions now also respects the partitioning $A=(A_1,\ldots,A_N)$:
\bea  \mathcal{E} (M) & := & d_A^2\, \text{tr}_{A\bar{A}}\left[ E^{A'A}\, P_\Phi^{A\bar{A}}\, M^{\bar{A}} \right] \\
 E & := &  \mathcal{E}^{\bar{A}} \otimes \text{Id}_A \left( P_\Phi^{\bar{A} A} \right) \; .\eea
\begin{figure}[th]
\includegraphics[width=8.5cm]{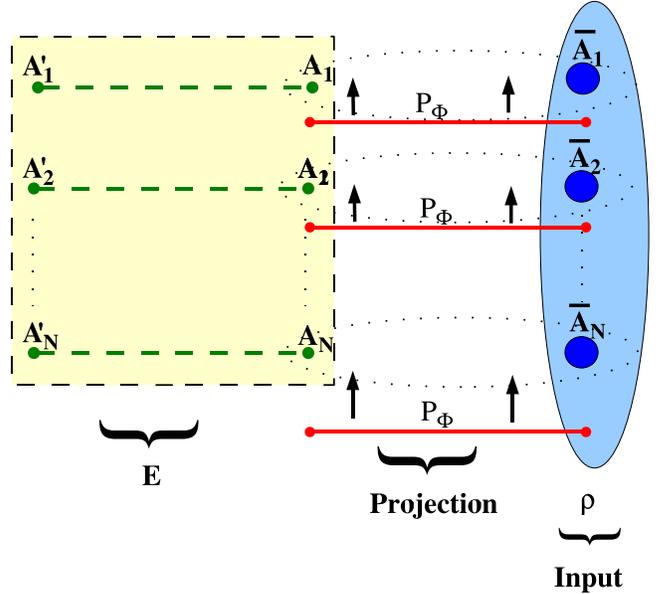}
\caption{\label{fig:MultiIso} Given the state $E$ on the composite system $A'=(A'_1,\ldots,A'_N)$ and $A=(A_1,\ldots,A_N)$, the CPM $\mathcal{E}$ is evaluated for an arbitrary multipartite input state $\rho$ by taking $\rho$ as an input at system $\bar{A}=(\bar{A}_1,\ldots,\bar{A}_N)$. Then the joint systems $A_i\bar{A}_i$ are (locally) measured in a Bell basis containing the maximally entangled state $P_\Phi^{A_i\bar{A}_i}$. With probability $\frac{1}{d_A^2}$ the desired output state $\mathcal{E}(\rho)$ is then obtained at system $A'$.} 
\end{figure} 
 
For the interpretation in terms of a teleportation protocol the preparation of the maximally entangled state $P_\Phi^{\bar{A}A}$ and the corresponding Bell measurement can be performed locally at each party separately, since the entanglement present in $P_\Phi$ is only with respect to the systems $A_i$ and their copies $\bar{A}_i$ but not with respect to the partitioning itself (see Fig.~\ref{fig:MultiInv} and Fig.~\ref{fig:MultiIso}). For the index notation it is convenient to take the same formula as in Eq.~(\ref{Iso_In_Basis})
\be \mathcal{E}_{\mathbf{ik}|\mathbf{jl}} = d_A \; E_{\mathbf{i j}|\mathbf{k l}}\; , \ee
but to consider the indices $\mathbf{i}, \mathbf{j}, \mathbf{k}$ and $\mathbf{l}$ as multi-indices, e.g. $\mathbf{i}=(i_1,\ldots,i_N) \in \mathbb{N}_{d_{A_1}} \times \ldots \times \mathbb{N}_{d_{A_N}} $. It turns out, that many of the {\it entangling capabilities} of the CPM $\mathcal{E}$ are directly related to the {\it entanglement properties} of the corresponding state $E$ (see \cite{Ci00,Du0102}):
\begin{itemize}
\item[\bf{5.}] $\mathcal{E}$ is {\it separable} w.r.t. parties $A_k$ and $A_l$ (and therefore not capable to create entanglement between them), iff $E$ is {\it separable} w.r.t. parties $A_k$ and $A_l$. In particular, we find that the CPM  corresponding to the tensor product of states $E\otimes F$ simply is the tensor product of the corresponding CPMs $\mathcal{E}\otimes\mathcal{F}$.
\item[\bf{6.}] For the partial transposition $^{T_{A_k}}$ w.r.t. party $A_k$ we have:
\be \left[ \mathcal{E} \left(M \right)\right]^{T_{A'_k}} =  \mathcal{E}' \left(M^{T_{A_k}} \right)\ee with 
$ E' = E^{T_{A'_k A_k}} $.
\\ In particular, $\mathcal{E}$ is {\it PPT preserving} w.r.t. party $A_k$ \cite{PPT_preserving}, iff $E$ is {\it PPT} w.r.t. the joint transposition of $A'_k$ and $A_k$.
\item[\bf{7.}] The CPM $\mathcal{E}$ can {\it simulate} another CPM $\mathcal{F}$ {\it under SLOCC} \cite{slocc}, iff the corresponding positive operator $E$ can be {\it converted} into $F$ {\it by means of SLOCC}.
\item[\bf{8.}] Two CPMs $\mathcal{E}$ and $\mathcal{F}$ are  {\it equivalent under local unitaries (LU)}, iff the corresponding positive operators $E$ and $F$ are {\it LU-equivalent w.r.t. the finest partitioning $(A_1, A_1', \ldots, A_N, A_N')$} .
\item[\bf{9.}] The CPM $\mathcal{E}$ can {\it generate} a state $\rho$ of the composite system $\mathbf{H}^{A'} \otimes \mathbf{H}^{A}$ {\it with non-zero probability of success}, iff the corresponding positive operator $E$ can be {\it converted} into $\rho$ {\it by means of SLOCC}. 
\end{itemize}
Since the classification of pure states in bipartite and three-qubit systems under SLOCC is known in detail, the results can be transferred to the corresponding maps via No.~{\bf 7} and {\bf 9} (see \cite{Du0102}).  For further applications to {\it purification, storage, compression, tomography and probabilistic implementation} of non-local operations and its use in quantum computation we refer the reader to Ref.~\cite{Du00} and \cite{Du03}. 

In the following we will mainly consider two-qubit gates, which are of special interest in quantum computation and quantum information. Note that in Ref.~\cite{Ci00} and \cite{Du00} (see Sec.III) it was shown for the case of two-qubit unitary operations, how to modify the teleportation protocol to implement an arbitrary two-qubit unitary with unit probability of success. We will illustrate the  results No.~{\bf 7}, No.~{\bf 8} and  No.~{\bf 9}  for these gates, namely for the
\begin{itemize}
\item CNOT-gate: $U^{AB} = |0\rangle^A\langle 0 | +   |1\rangle^A\langle 1| \sigma_x^B$\\
{\small \be\label{CNOT_E} |\psi_\text{CNOT}\rangle = \frac{1}{\sqrt{2}} \left(|00\rangle^{AA'}|\psi_0\rangle^{BB'} + |11\rangle^{AA'}|\psi_1\rangle^{BB'}\right) \ee }
\item Phase-gate: $U^{AB}(\alpha) = e^{-i\alpha \sigma_y^A\otimes \sigma^B_y}$\\
{\small \be\label{PhaseGate_E} |\psi(\alpha) \rangle = \cos(\alpha) |\psi_0\rangle^{AA'}|\psi_0\rangle^{BB'}  - i \sin(\alpha) |\psi_2\rangle^{AA'}|\psi_2\rangle^{BB'}\ee }
\item SWAP-gate: $U^{AB} = |00\rangle^{AB}\langle 00 | +  |01\rangle^{AB}\langle 10 | + |10\rangle^{AB}\langle 01 | +  |11\rangle^{AB}\langle 11|  $\\
\be\label{SWAP_E} |\psi_\text{SWAP} \rangle =  |\psi_0\rangle^{AB'}|\psi_0\rangle^{BA'} \ee 
\end{itemize}
Note that  $|\psi_\text{SWAP} \rangle$ is a product state w.r.t. the partitioning $AB'$ versus $A'B$ but not w.r.t. the partitioning $AA'$ versus $BB'$. In fact $|\psi_\text{SWAP} \rangle$ has a Schmidt decomposition into  $4$  $(AA',BB')$-product terms, whereas $|\psi_\text{CNOT} \rangle$ and $|\psi(\alpha) \rangle$ can be decomposed into $2$ $(AA',BB')$-product states. From basic facts \cite{Conversion} about bi-partite entanglement for pure states it follows from No.~{\bf 7} that the SWAP-gate can simulate the phase gate and the CNOT-gate by means of SLOCC operation to be performed before and after the SWAP operation (but not vice versa). Moreover the CNOT-gate and the phase gate can simulate each other under SLOCC for arbitrary $\alpha \in ]0,2\pi[ $. For the case of  $\alpha=\frac{\pi}{4}$ they actually coincide up to some local unitaries \cite{Kr00,Du0102}:
\be \label{LU_CNOT_PhaseGate} 
U_\text{CNOT}^{AB} = U_1^A \otimes U_2^B \, U^{AB}(\frac{\pi}{4})\,  V_1^A \otimes V_2^B \hspace{0.5cm} \text{with}
\ee
 \bea 
 U_1 =\frac{1}{\sqrt{2}}\begin{pmatrix}1 & -i \\ -1 & -i \end{pmatrix}  & &  U_2 =\begin{pmatrix}1 & 0 \\ 0 & -i \end{pmatrix} \nonumber \\
 V_1 =\frac{1}{\sqrt{2}}\begin{pmatrix}1 & i \\ i & 1 \end{pmatrix}  & &  V_2  =\frac{1}{\sqrt{2}}\begin{pmatrix}1 & i \\ -1 & i \end{pmatrix}  \nonumber
\eea
This corresponds to the fact that the corresponding states $|\psi_\text{CNOT}\rangle$ and $|\psi(\frac{\pi}{4}) \rangle$ are LU-equivalent w.r.t. to the partitioning $(A,A',B,B')$ (see No.~{\bf 8}), i.e. 
\be|\psi_\text{CNOT}\rangle = U_1^{A'} \otimes U_2^{B'} \otimes  \left(V_1^A\right)^T \otimes \left(V_2^B\right)^T  \, |\psi(\frac{\pi}{4})\rangle \; .\ee
 According to No.~{\bf 9} the SWAP-gate will moreover be capable to create more entanglement than the CNOT-gate and the phase-gate, which can -- up to SLOCC -- create the same type of entanglement.


\section{Standard form for decoherence in the single-party-setting}\label{NFoneParty}

In this section we apply the Jamio\l kowski isomorphism in order to derive a standard form for an arbitrary decoherence process, that is described by some CPM $\mathcal{E}$. We show that this standard form can be achieved by randomly choosing appropriate unitaries to be performed before and after the actual CPM occurs. We first consider the case of a qubit system and then discuss a generalization to $d$-level systems. 

Let denote $\sigma_0=\mathbf{1}$, $\sigma_1=\sigma_x$, $\sigma_2=\sigma_y$ and $\sigma_3=\sigma_z$ the Pauli matrices.
Note that starting with the maximally entangled state $|\Phi\rangle^{A\bar{A}} = \frac{1}{\sqrt{2}} \left(|00\rangle^{A\bar{A}} + |11\rangle^{A\bar{A}} \right)$ we obtain a complete Bell basis $(|\psi_0\rangle,|\psi_1\rangle,|\psi_2\rangle,|\psi_3\rangle)$ simply by applying $\sigma_i$ locally on system $A$, i.e. $ |\psi_{i}\rangle^{A\bar{A}}=\sigma_{i}^A \otimes \text{Id}^{\bar{A}} \left(|\psi_{0}\rangle^{A \bar{A}}\right) $.
Thus any decomposition of a state $E= \sum_{ij} E_{ij} |\psi_i\rangle \langle \psi_j|$ in terms of the Bell basis corresponds to a canonical representation of the CPM $\mathcal{E}$ in terms of Pauli matrices
\begin{equation}\label{GenCPM}
 {\cal E} (\rho) =  \sum_{i,j=0}^3 E_{ij} \sigma_i \rho \sigma_j \; ,
\end{equation}
where the conditions on the matrix $\mathbf{E}=(E_{ij})$ can directly be read off from the isomorphism, i.e. $\mathbf{E}$ must be density matrix.
A case of particular interest in quantum information theory, especially in the study of fault-tolerance of quantum computation, is when $E$ is a diagonal matrix.
In many applications the corresponding CPM, the so called  Pauli channel
\begin{equation}\label{PauliChannel}
 {\cal E} (\rho) =  \sum_{i=0}^3 E_i \sigma_i \rho \sigma_i \; \text{with} \; (\sum_{i=0}^3 E_i =1)\, ,
\end{equation}
describes some underlying noise model or decoherence process. This class contains for $E_0=\frac{1+3p}{4}$ and $E_1=E_2=E_3=\frac{1-p}{4}$ the {\it depolarizing channel} (white noise) $ {\cal E}\rho =   p \rho + (1-p)\frac{1}{2} \mathbf{1} $, for $E_0=\frac{1+p}{2}$, $E_1=E_2=0$ and $E_3= \frac{1-p}{2}$ the {\it dephasing channel} $ {\cal E}\rho =   p \rho + \frac{1-p}{2}\left( \rho + \sigma_z \rho \sigma_z \right)$ and for $E_0=\frac{1+p}{2}$, $E_2=E_3=0$ and $E_1=\frac{1-p}{2} $ the {\it bit-flip channel} $ {\cal E}\rho =   p \rho + \frac{1-p}{2}\left( \rho + \sigma_x \rho \sigma_x \right)$.

We show now that the decoherence process specified by an arbitrary CPM $\mathcal{E}$ as in Eq.~(\ref{GenCPM}) can be transformed into a Pauli channel $\mathcal{E'}$ (see Eq.~(\ref{PauliChannel})) with the same diagonal elements $E_i=E_{ii}$. This can be achieved by a probabilistic but correlated application of one of the four Pauli matrices $\sigma_i$ before and after the actual noise occurs:
\be \label{DepolCPM2} \mathcal{E'}(\rho) = \frac{1}{4} \; \sum_{i=0}^3 \;  \sigma_i \, \mathcal{E}\left( \sigma_i \rho \sigma_i\right)\,  \sigma_i \; . \ee 
In other words, by randomly choosing one of the four Pauli matrices with probability $\frac{1}{4}$ to apply to a system before and after the noise process affects the system (e.g. some memory device) and ignoring the information about which Pauli matrix has been applied, an experimenter will actually (only) have to deal with noise of the form of a Pauli channel.
The fact that the CPM $\mathcal{E}$ can be brought to this form $\mathcal{E}'$ follows from the Jamio\l kowski isomorphism used as in case {\bf (B)} and the fact, that the corresponding state $E$ can be diagonalized to $E'$ by a mixing procedure, in which each of the local Pauli operators $\sigma^A_i\otimes\sigma^{A'}_i$ is applied with probability $\frac{1}{4}$:
\be E'^{AA'} = \frac{1}{4} \;\sum_{i=0}^3 \; \sigma^A_i\otimes\sigma^{A'}_i\,E^{AA'} \, \sigma^A_i\otimes\sigma^{A'}_i  \; .\ee 

The achieved standard form in Eq.~(\ref{PauliChannel}) can be further depolarized, by considering the following three Clifford operations $Q_k=e^{i\frac{\pi}{4}\sigma_k}=\sqrt{i\sigma_k}$ with $k=1,2,3$. 
Starting with a state of standard form Eq.~(\ref{PauliChannel}) 
 one can in fact compute that 
\bea \label{finiteDepolqubit}
& & \frac{1}{3}\sum_{k=1}^3 Q_k\otimes Q^*_k E Q_k^\dagger \otimes Q_k^T \\
& = & E_0 |\psi_0\rangle\langle\psi_0 | + \left(E_1+E_2+E_3\right) \; \times \nonumber \\
& \times & \; \left(  |\psi_1\rangle\langle\psi_1 | +  |\psi_2\rangle\langle\psi_2 | +  |\psi_3\rangle\langle\psi_3 |\right)\; . \nonumber
\eea
This means that by uniformly choosing one of the $12$ unitaries $U_{ki}=Q_k\sigma_i$ for $k=1,2,3$ and $i=0,1,2,3$ and applying $U_{ki}^\dagger$ before and $U_{ki}$ after the application of an arbitrary CPM $\mathcal{E}$ (see Eq.~(\ref{GenCPM}))  the resulting CPM $\mathcal{E}'$:
\be \mathcal{E'}(\rho) \,=\, \frac{1}{12}\,\sum_{ki} \,U_{ki} \mathcal{E}\left( U_{ki}^\dagger \rho \ U_{ki} \right) U_{ki}^\dagger  \ee
 will be of the form of depolarizing channel 
\be\label{DepolSFqubit}\mathcal{E}'(\rho) = \alpha(f)\, \rho + (1-\alpha(f))\,\text{tr}\rho\,\frac{1}{d}\mathbf{1}  
\ee
 with $f=E_{00}$ and $\alpha(f)=\frac{4 f-1}{3}$.
Note that a similar twirling procedure is also used in the recurrence protocol \cite{IBM} for entanglement purification. Both depolarization procedures Eq.~(\ref{DepolCPM2}) and Eq.~(\ref{finiteDepolqubit}) leave the state $|\Phi\rangle$ and thus the identity operation Id invariant. Hence the Jamio\l kowski fidelity remains the same, i.e. $F(\mathcal{E},\text{Id})=E_{00}=F(\mathcal{E}',\text{Id})$. Since the Jamio\l kowski fidelity represents the noise level of the respective operations $\mathcal{E}$ and $\mathcal{E}'$, both standard forms can be achieved without introducing additional noise to the system.

Let us now turn to the case of general qudit systems with $d=\text{dim}_\mathbb{C} (\mathbf{H}^A) = \text{dim}_\mathbb{C} (\mathbf{H}^{ A'})$. Here the following complete basis of maximally entangled states can be chosen: 
\be \label{GenBellBasis} |\psi_{kl}\rangle^{AA'} := \frac{1}{\sqrt{d}} \sum_{m =0}^{d-1} e^{i \frac{2\pi }{d} k \cdot m}|m + l\rangle^A |m\rangle^{A'}\, , \ee  where addition $m + l$ and multiplication $k \cdot m$ is meant modulo $d$. Note that the Bell basis can be generated by acting on only one of the systems by means of unitaries $U_{kl}$ (generalized Pauli group) 
\be\label{GenPauli} U_{kl}|m\rangle:= e^{i \frac{2\pi}{d} k \cdot m}|m + l\rangle \ee
 out of the maximally entangled state $|\psi_{00}\rangle=|\Phi\rangle$, e.g.
 \be |\psi_{kl}\rangle^{AA'}=U_{kl}^A \otimes \text{Id}_{A'} \left(|\psi_{00}\rangle^{A A'}\right) \; .\ee 
Similar to Eq.~(\ref{GenCPM}) the canonical form for an arbitrary CPM in terms of the generalized Pauli operators is 
\be \label{GendCPM} \mathcal{E}(\rho)= \sum_{kl,k'l'} E_{kl,k'l'}\, U_{kl}\rho U_{k'l'}^\dagger  \; .\ee
With respect to this Bell basis the corresponding state has the decomposition 
\be E = \sum_{kl,k'l'} E_{kl,k'l'}\, |\psi_{kl}\rangle \langle\psi_{k'l'} | \; .\ee  

By generalizing the depolarization procedure in Eq.~(\ref{DepolCPM2}) we can again diagonalize the state $E$ and thus bring the corresponding CPM $\mathcal{E}$ to the form of a (generalized) Pauli channel.

{\bf Standard form: \underline{Pauli Channel} } \\
{\it By uniformly choosing one of the $d^2$ Pauli operators $U_{kl}$ and applying $U_{kl} ^\dagger$ before and $U_{kl}$ after the application of an arbitrary CPM $\mathcal{E}$ (see Eq.~(\ref{GendCPM}))  the resulting CPM $\mathcal{E}'$:}
\be \label{TwirlPauliGen}\mathcal{E'}(\rho) = \frac{1}{d^2} \; \sum_{k,l=0}^{d-1} \; U_{kl} \, \mathcal{E}\left(  U_{kl}^\dagger \rho U_{kl} \right)\,  U_{kl}^\dagger \ee
{\it will be of the form}
\be \label{PauliChan} \mathcal{E}'(\rho)= \sum_{kl} E'_{kl}\, U_{kl}\rho U_{kl}\dagger   \ee
{\it with $E'_{kl} = E_{kl,kl}$.}

A proof of this statement can be found in the Appendix C. Whereas an arbitrary hermitian matrix $E$ is described by $\frac{1}{2}d^2(d^2-1)$ real parameters, which in addition have to fulfill the constraints No.~4 (Sec.~\ref{IsoPhysics}) in order to correspond to a trace preserving CPM $\mathcal{E}$, an arbitrary Pauli channels $\mathcal{E}'$ can be described by only $d^2-1$ positive parameters $E_{kl}$ \cite{parametercount}.

The number of parameters can even be decreased by considering for the symmetrization procedure not only the Pauli group $S:=\{\pm 1,\pm i\}\times\{\sigma_0,\sigma_1,\sigma_2,\sigma_3 \}$ but the larger group 
\be S'\,=\,\{\, U^A\otimes {U^*}^{ A'}\; |\;  U \in \mathbf{U}(d)\, \}\ee 
of  all local unitaries of the form $U^A\otimes {U^*}^{ A'}$. Since the group $S'$ contains $S$, it has at most a smaller commutant. Whereas the commutant of $S$ is the set of all Bell diagonal states, the commutant of $S'$ is indeed \cite{Hor99} only generated by the (orthogonal) states $P_{\Phi}$ and  
\be \gamma :=\frac{1}{d^2-1} {\sum_{\genfrac{}{}{0pt}{}{k,l}{(k,l)\neq (0,0)}}} |\psi_{kl}\rangle \langle \psi_{kl}| =\frac{1}{d^2-1} \left( \mathbf{1} - P_{\Phi} \right)\,. \ee 
 In other words the set of states, that is invariant under $S'$, is determined by a fewer set of parameters. In fact, the states $E'=\mathcal{D}(E)$ obtained by this 'twirling' operation 
\be \mathcal{D}(E) \,=\, \int \,(U\otimes{U^*})\, E \, (U^\dagger\otimes U^T)\, dU \;  ,\ee
where $dU$ denotes the uniform probability distribution on the unitary group $\mathbf{U}(d)$ proportional to the Haar measure, is determined by a single real parameter :
\bea E' & = & f\, P_{\Phi} + (1-f)\, \gamma \\ 
 & = & \alpha(f)\, P_{\Phi} + (1-\alpha(f))\,\frac{1}{d^2}\mathbf{1} \; , \label{isotropic}
\eea
with $\alpha(f)=\frac{d^2 f-1}{d^2-1}$. Note that the fidelity $f=\langle \Phi | E' |\Phi\rangle = \langle \Phi | E |\Phi\rangle$ ($0\leq f\leq 1$) is left unchanged under the twirling procedure $\mathcal{D}$, since $\mathcal{D}$ simply is a projection onto the subset of states invariant under this 'isotropic symmetry':
\be\mathcal{D}(E)= \text{tr}(P_\Phi E) \,P_\Phi + (d^2-1)\, \text{tr}(\gamma E)\, \gamma\; . \ee
We remark that by partial transposition these isotropic states $E'$ are in one-to-one correspondence \cite{Wo03} to the set of Werner states \cite{We89}. 
Since $E'$ is a mixture of the maximally entangled state $P_\Phi$ and a maximally mixed state, we find that the normal form of the corresponding CPM $\mathcal{E}'$ is the (generalized) depolarizing channel \cite{IsoOf1}:

{\bf Standard form: \underline{Depolarizing Channel} } \\
{\it By uniformly choosing a unitary $U\in\mathbf{U}(d)$ and applying $U^\dagger$ before and $U$ after the application of an arbitrary CPM $\mathcal{E}$ (see Eq.~(\ref{GendCPM}))  the resulting CPM $\mathcal{E}'$:}
\be \mathcal{E'}(\rho) \,=\, \int \,U \mathcal{E}\left( U^\dagger \rho \ U \right) U^\dagger dU \ee
{\it will be of the form}
\bea \mathcal{E}'(\rho) & = & f \, \rho + \frac{1-f}{d^2-1}\, \sum_{\genfrac{}{}{0pt}{}{k,l}{(k,l)\neq (0,0)}} U_{kl} \rho U_{kl}^\dagger   \nonumber \\
& = & \alpha(f)\, \rho + (1-\alpha(f))\,\text{tr}\rho\,\frac{1}{d}\mathbf{1}  \label{DepolChan}
\eea
{\it with $f=E_{00}$ and $\alpha(f)=\frac{d^2 f-1}{d^2-1}$.}

Since an isotropic state $E'$ as in Eq.~(\ref{isotropic}) is separable iff  $f\leq \frac{1}{d}$ \cite{Hor99}, we note that, according to No.~{\bf 13.} in Appendix A, the corresponding depolarizing channel becomes entanglement breaking at this point.

Let us briefly address the question of possible practical implementations of the twirling protocol described above. As it is shown in the Appendix C it is actually sufficient for the depolarization protocol to uniformly choose some unitaries from a finite set of Clifford unitaries. 

{\it To summarize we have shown that both standard forms $\mathcal{E}'$, the Pauli channel and the depolarizing channel, can be obtained by a random application of quantum operations applied before and after the actual CPM $\mathcal{E}$. These operations are chosen uniformly at random from a finite set of unitaries. Moreover we have seen that these depolarization protocols do not introduce additional noise to the system.}


\section{Standard forms for CPM in the multi-party setting}\label{SF_Multi}

In this section we will continue the discussion of standard forms for noisy quantum operations. We will consider a partitioning of the system $A=(A_1,\ldots,A_N)$ into $N$ parties, which might be located at distant places. Any depolarization protocol that brings a given (non-local) CPM into its standard form therefore should be local w.r.t. this partitioning.  In the following we will consider an {\it ideal operation} $\mathcal{I}$, that can only be realized imperfectly as a CPM $\mathcal{E}$. We are now interested in the possible normal forms $\mathcal{E}'$, into which we can transform $\mathcal{E}$ by means of LOCC operation (w.r.t. to the given partitioning), that are carried out before and after $\mathcal{E}$ \cite{dir/indir} is actually applied. If one is interested in the standard form for a map describing a given decoherence process itself, the ideal operation is the identity $\mathcal{I}=\text{Id}$. Apart from the identity we will in the multi-party setting also consider  the case, where the ideal operation is some 2-qubit unitary operation $\mathcal{I}=\mathcal{U}$ ($\mathcal{U}(\rho)=U\rho U^\dagger$), which can only be realized in form of some noisy quantum operation $\mathcal{E}$. In contrast to the case discussed in the previous section, the locality requirements now impose rather severe constraints on the allowed operations to manipulate a given CPM. Note that the state $I$ corresponding to the ideal operation (identity or unitary) is pure. That is for $\mathcal{I}=\mathcal{U}$ [$\mathcal{I}=\text{Id}$] we have $I=|\psi_U\rangle \langle\psi_U|$ [$I=|\Phi\rangle \langle \Phi|$] respectively. Before we give an outline of this section we mention several aspects of the problem of finding such a standard form. 
\begin{itemize}
\item One can distinguish the two cases where only {\it deterministic} or also {\it probabilistic} transformations are considered, i.e. whether it is possible to transform $\mathcal{E}$ into the respective normal form $\mathcal{E}'$ in all of the possible measurement branches of the LOCC protocol or in at least one.
\item Closely related to this distinction is the question whether one uses the teleportation protocol {\it directly}  as in {\bf (A)} or {\it indirectly} as in {\bf (B)}, since a direct use of the isomorphism protocol in general has only a certain probability of success.
\item Firstly, one would like the transformation protocol $\mathcal{D}$ (on state level) to {\it leave the ideal operation invariant}, i.e. $\mathcal{D}(I)=I$. In this case the fidelity $F(\mathcal{E}',\mathcal{I})= \text{tr} I E'$ of the ideal operation with the transformed noisy operation $\mathcal{E}'$ will be the same as the fidelity $F(\mathcal{E}, \mathcal{I})= \text{tr} I E $ of the ideal operation with the noisy operation $\mathcal{E}$. Since the Jamio\l kowski fidelity with the ideal operation can be regarded as some kind of distance measure, the transformation will keep $\mathcal{E}$ as close to the ideal operation as before. For the case of the ideal operation being the identity, the protocol $\mathcal{D}$ simply should be unital. On the other hand one might as well be allowed to {\it sacrifice} some fidelity with the ideal operation in order derive simpler standard forms.
\item The transformation protocol might bring any CPM onto its respective standard form ({\it universal protocol}) or it might be designed to transform a specific CPM into standard form .
\end{itemize}
Note that most of the differences in these versions of the problem only become important in the multi-party setting. This is mainly due to the fact that the depolarizing channel already provides a standard form for an arbitrary noise process, which can be achieved deterministically by a unital transformation and which is already specified by a single noise parameter $f$. 

 We generalize the results of Sec.~\ref{NFoneParty} in Sec.~\ref{MultiDecSF} and derive standard form for decoherence processes (i.e. for the case $\mathcal{I}=\text{Id}$) under the constraint that the underlying control operation have to be local w.r.t. the given partitioning. In Sec.~\ref{Unitary_SF} we discuss the case where the ideal operation is one of the unitary gates SWAP, CNOT or a phase gate with some arbitrary angle. In this section we restrict first to those depolarization procedures that are universal, deterministic and leave these unitary gates invariant.
In Sec.~\ref{Sacrificing_SF}  we also discuss the case where the fidelity $f$ of the operation is decreased by a certain amount (i.e. additional noise is introduced) in order to obtain a simpler standard form describing the noise process, that is to reduce the number of required parameters. For gates locally equivalent to SWAP and CNOT, this leads to noise processes described by global white noise. A similar result is obtained for all phase gates, provided that one has control over switching the noisy operation on and off at will. The noise is -- in the worst case -- increased by an order of magnitude. Finally we briefly mention the problems in Sec.~\ref{arbitraryUnitary} that occur when trying to transfer the techniques developed in Sec.~\ref{Unitary_SF} and Sec.~\ref{Sacrificing_SF} to the more general case of an arbitrary unitary operation as the ideal operation.

\subsection{Standard forms for decoherence in the multi--party setting}\label{MultiDecSF}

\subsubsection{Depolarization without sacrificing}
Let us now consider possible standard forms for noise operations (i.e. ideal operation is the identity) in the multi-partite setting. Note that the twirling operation $\mathcal{D}$  used in Sec.~\ref{NFoneParty} corresponds to a projection into the space of states $E$ over $\mathbf{H}^A\otimes\mathbf{H}^{A'}$ (recall that we chose $d_A=\text{dim}_\mathbb{C} (\mathbf{H}^A) = \text{dim}_\mathbb{C} (\mathbf{H}^{ A'})=\prod_{i=1}^N d_{A_i}$ ). Thus  it is straightforward to derive the corresponding standard form for the multi-party case, which is obtained after sequential application of the twirling operation $\mathcal{D}$ locally at each party. Since the invariant group in question are $\bigotimes_{i=1}^N S_{A_i}$ and $\bigotimes_{i=1}^N S'_{A_i}$ respectively, the commutants of these groups (i.e. the subspace of invariant matrices in $\mathcal{M}\left(\mathbf{H}^{A'} \otimes \mathbf{H}^{A}\right)$, onto which the projection $\bigotimes_{i=1}^N \mathcal{D}_{A_i}$ projects) are just given by the tensor products of the commutants (subspaces) for each party. Hence, a probabilistic application of local unitaries $g_{\mathbf{kl}}=g_{k_1l_1}^{A_1A'_1}\otimes \ldots \otimes g_{k_Nl_N}^{A_NA'_N}$\cite{MultiIndex} with probability $\frac{1}{d_A}=\frac{1}{d_{A_1}}\cdots\frac{1}{d_{A_N}}$ diagonalizes $E$ and gives the generalized multi-partite Pauli channel 
\be \label{MultiPauli}\mathcal{E}'(\rho)= \sum_{\mathbf{k},\mathbf{l}\, \in\, \mathbb{N}_{d_{A_1}} \times\cdots \times \mathbb{N}_{d_{A_N}}} E_{\mathbf{kl}}\; U_{\mathbf{kl}}\,\rho\, U_{\mathbf{kl}} \; ,\ee
which is again specified by $d_A^2-1$ positive parameters ($\mathbb{N}_{d_i}:=\{0,\ldots,d_i-1\}$). In the case of equal dimensions $d$ the channel is determined by $d_A=d^{2N}-1$ parameters. For a noise operation on two qubits, for example, the corresponding standard form is given by
\be  \mathcal{E}'(\rho)= \sum_{i,j=0}^3 E_{ij}\, \sigma^{A_1}_i\otimes\sigma^{A_2}_{j}\,\rho\, \sigma^{A_1}_i\otimes\sigma^{A_2}_{j}  \; .\ee

As in the case of a single system, further depolarization is possible and hence a simpler standard form can be achieved. To this aims one performs a complete twirl over the larger group $S'$. The result of this twirl is that one projects $E$ into the set of states of the form 
\be\label{isotropic1} E = \sum_{\mathbf{k}\, \in\, \mathbb{N}_{2} \times\cdots \times \mathbb{N}_{2}} E_{\mathbf{k}}\, \gamma_{\mathbf{k}}\; ,\ee
where $\gamma_{\mathbf{k}}=\gamma_{k_1}^{A_1}\otimes \ldots\otimes\gamma_{k_N}^{A_N}$ and at each party $A_i$ $\gamma_{k_i}^{A_i}$ denotes one of the two orthogonal states $\gamma_0=P_\Phi$ or $\gamma_1=\gamma$ spanning the respective isotropic subspace.
If $E'$ is decomposed w.r.t. the (non-orthogonal) basis $(\gamma_0,\gamma_1)=(P_\Phi,\frac{1}{d^2_{A_i}} \mathbf{1})$, the corresponding CPM $\mathcal{E}'$ has a natural interpretation in terms of different white noise factors. To be more precise let us consider the example of two qudits with $d=d_{A_1}=d_{A_2}$. In this case the state $E'$ is of the form 
\bea E' & = & \alpha_{00}\, P_\Phi^{A_1}\otimes P_\Phi^{A_2} + \alpha_{01}\, P_\Phi^{A_1}\otimes \frac{1}{d^2}\mathbf{1}_{A_2}\nonumber\\
& & + \alpha_{10}\, \frac{1}{d^2}\mathbf{1}_{A_1}\otimes P_\Phi^{A_2} +\alpha_{11}\, \frac{1}{d^2}\mathbf{1}_{A_1}\otimes\frac{1}{d^2}\mathbf{1}_{A_2} \; .\eea
Since $(\delta_0,\delta_1)=(P_\Phi-\gamma,d^2 \gamma)$ is a dual basis for $(P_\Phi,\frac{1}{d^2}\mathbf{1})$,
 we have that 
\bea \label{Alpha} \alpha_{00} & = & \text{tr} \left[\delta_0^{A_1} \otimes \delta_0^{A_2}\, E^{A_1A_2}\right]\nonumber\\ 
& = & E_{00} -\frac{E_{01}}{d^2-1}-\frac{E_{10}}{d^2-1}+\frac{E_{11}}{(d^2-1)^2}\nonumber\\
\alpha_{01} & = & \text{tr} \left[\delta_0^{A_1} \otimes \delta_1^{A_2}\, E^{A_1A_2}\right]\nonumber\\ 
& = & \frac{d^2 E_{01}}{d^2-1}-\frac{d^2 E_{11}}{(d^2-1)^2}\nonumber\\
\alpha_{10} & = & \text{tr} \left[\delta_1^{A_1}\otimes\delta_0^{A_2}\, E^{A_1A_2}\right] \nonumber\\ 
& = & \frac{d^2 E_{10}}{d^2-1}-\frac{d^2 E_{11}}{(d^2-1)^2}\nonumber\\
\alpha_{11} & = & \text{tr} \left[\delta_1^{A_1} \otimes \delta_1^{A_2}\, E^{A_1A_2}\right] =\frac{d^4}{(d^2-1)^2} E_{11} \; .
\eea
For the corresponding normal form $\mathcal{E}'$ of the CPM $\mathcal{E}$ we obtain
\bea \mathcal{E}'(\rho) & = & \alpha_{00}\, \rho  + \alpha_{01}\, \rho_{A_1}\otimes \frac{1}{d}\mathbf{1}_{A_2} + \alpha_{10}\, \frac{1}{d}\mathbf{1}_{A_1}\otimes \rho_{A_2} \nonumber \\
& &  +\alpha_{11}\, \frac{1}{d^2}\text{tr}(\rho)\mathbf{1}_{A_1A_2}\; ,\eea
where $\rho_{A_1}=\text{tr}_{A_2}(\rho)$ and  $\rho_{A_2}=\text{tr}_{A_1}(\rho)$. The second and third term correspond to white noise introduced locally at each party, whereas the last summand introduces global white noise. Note that some of the coefficients $\alpha_{00}$, $\alpha_{01}$ or $\alpha_{10}$ can be negative. In the case of equal dimensions at each party ($d_{A_i}=d$) we have reduced the number of parameters from initially $O(d^{4N})$ for a general CPM $\mathcal{E}$ over $d^{2N}-1$ for a general multi-party Pauli channel to $2^N-1$ parameters (independent of the dimension $d$) to describe different types of multi-party white noise.

So far we have only considered twirling protocols, that were deterministic, made direct use of the isomorphism and left the ideal operation invariant, namely the identity operation. But can we further reduce the number of parameters for a different standard form, which is achieved by a LOCC protocol, that has only a certain probability of success or that makes indirect use of the isomorphism or allows to sacrifice some of the initial fidelity $E_{00}$ with the ideal operation ? To be more precise, is it possible by weakening one of these conditions to achieve e.g. just a global white noise channel 
\be \label{GlobalWhiteNoise}\mathcal{E}'(\rho) = p\, \rho + (1-p)\, \frac{1}{d^{2}} \mathbf{1} \; , \ee
i.e. the case, for which also all the diagonal elements vanish except $\alpha_{00}$ and $\alpha_{11}$?
 
If one allows for a direct use of the isomorphism protocol and thus to perform local Bell measurements in the basis $|\psi_i\rangle$, this is certainly possible, since the ideal operation $\text{Id}$ is local w.r.t. any partitioning and the coefficients $E_{\mathbf{kl}}$ in Eq.~(\ref{MultiPauli}) can deliberately be adapted without increasing the noise level by simply 'twirling' all components $E_{\mathbf{kl}}$ but $E_{\mathbf{00}}$ into $\frac{1}{d}\mathbf{1}$. Note that this procedure gives rise to a probabilistic LOCC protocol, since a direct use of the Jamio\l kowski isomorphism can in general not be achieved with unit probability of success (see Sec.~\ref{IsoPhysics}).  As discussed in Sec.~\ref{IsoPhysics} this use of the isomorphism will therefore in general not be of great interest for all applications, in which one would like to work with a standard form of a given CPM rather than with the CPM itself.

Using the isomorphism only indirectly, we have to restrict the transformations to the class of SLOCC operation that are also local  w.r.t. to $(A_i,\bar{A}_i)$. Note that one cannot increase the fidelity $f$ with the ideal operation $\mathcal{I}=\text{Id}$ by means of any physical protocol, whenever $f$ corresponds to the largest eigenvalue in $E$ \cite{noIncrease}, e.g. Eq.~(\ref{MultiPauli}) with $E_{\mathbf{00}}> E_{\mathbf{kl}}$ ($\neg \, \mathbf{k}=\mathbf{l}=0$). In all these cases any SLOCC operation $C_1^A\otimes C_2^{\bar{A}}$, that is contained in the transformation protocol and that does not leave $|\Phi\rangle$ invariant, will cause a decrease in the fidelity $f'<f$ of the respective standard form $\mathcal{E}'$. Thus any {\it universal} transformation protocol, that yields a respective standard form for all CPM without decreasing the fidelity, consists in a probabilistic application of operation $C_1^A\otimes C_2^{\bar{A}}$ with $C_1^A\otimes C_2^{\bar{A}}|\Phi \rangle =|\Phi\rangle$, i.e. $C_1\otimes C_2=C\otimes(C^{-1})^T$ for some invertible matrix $C$ \cite{AllInvTrafo}. Since all such transformation will also leave all $\gamma_{\mathbf{k}}$ in Eq.~(\ref{isotropic1}) invariant, the respective standard form cannot contain fewer terms than the multi-party white noise channel. In this sense the above twirling procedure already yields a standard form, that is {\it optimal} among all forms achieved by some {\it universal protocol}, that only use the isomorphism indirectly and does not sacrifice any fidelity with the ideal operation.

\subsubsection{Depolarization by means of sacrificing}

In the remainder of this subsection we will show how to design twirling protocols that bring a specific CPM into the standard form of global white noise by introducing additional noise. The procedure described below does therefore satisfy neither the universality property nor the no--sacrificing condition, but it will - especially for the many party case - significantly reduce the number of parameters of the standard form to a single one. By applying the above universal depolarization procedures, we can start with considering only states $E$, that are already in isotropic form (see Eq.~(\ref{isotropic1})). 

We will illustrate the procedure for the case of two qubits. Generalization to the multipartite case and higher dimensions are straightforward. 
For such two--qubit maps, the corresponding state $E$ (after applying the universal depolarization protocol described in the previous section) is given by  
\bea
\label{SacrifDecomp} E & = & E_{00} P^{A_1}_{\psi_0}P^{A_2}_{\psi_0} +  E_{01} \sum_{j=1}^3 P^{A_1}_{\psi_0}P^{A_2}_{\psi_j} + \nonumber \\ 
& & E_{10} \sum_{i=1}^3 P^{A_1}_{\psi_i} P^{A_2}_{\psi_0}  +  E_{11}\sum_{i,j=1}^3 P^{A_1}_{\psi_i} P^{A_2}_{\psi_j}\; , \eea
where $P_{\psi_i}=|\psi_i \rangle\langle\psi_i|$ is the projector onto one of the Bell states $|\psi_i\rangle$. In the following we will collect the parameters in a vector $\mathbf{E}=\left(E_{00},E_{01},E_{10},E_{11}\right)^T$. The fact, that $E$ corresponds to a trace preserving CPM then simply reads (i) $\mathbf{E}\geq 0$ component-wise (i.e. $E_{ij}\geq 0$ for $i,j=0,1$)  and (ii) $N(\mathbf{E}):= E_{00} + 3\left( E_{01} + E_{10} + 3 E_{11} \right) = 1$.
We now consider the following type of depolarization 
\bea \mathcal{D}(E) & = & p_{00} E + p_{01} \sum_{j=1}^3 \sigma_j^{A_2} E \sigma_j^{A_2} + p_{10} \sum_{i=1}^3 \sigma_i^{A_1} E \sigma_i^{A_1} \nonumber\\
 & &  +\, p_{11}\, \sum_{i,j=1}^3 \sigma_i^{A_1}\otimes\sigma_j^{A_2} E \sigma_i^{A_1}\otimes\sigma_j^{A_2} \; .
\eea
In a similar notation as before $\mathcal{D}$ corresponds to a trace preserving CPM iff $N(\mathbf{p}) = 1$  and $\mathbf{p}\geq 0$. It is straight forward to calculate, that the resulting state $E'=\mathcal{D}(E)$ is again in isotropic form Eq.~(\ref{SacrifDecomp}) with new parameters $E'_{ij}$, that are given by the linear transformation 
\be \label{LinEq} \mathbf{E}' = \mathbf{D}[\mathbf{E}]\cdot \mathbf{p} \; ,\ee
 where the matrix $\mathbf{D}[\mathbf{E}]$ depends on the initial state $E$:
{\scriptsize
\be 
\begin{pmatrix}
E_{00} & 3 E_{01} & 3 E_{10}  & 9 E_{11} \\
E_{01} &  E_{00} + 2 E_{01} & 3 E_{11}  & 3 (E_{10} + 2 E_{11}) \\
E_{10} & 3 E_{11} & E_{00} + 2 E_{10}   & 3 (E_{01} + 2 E_{11}) \\
E_{11} & E_{10} + 2 E_{11} & E_{01} + 2 E_{11}  & E_{00} + 2 (E_{01} + E_{10} + 2 E_{11})  
 \end{pmatrix}
\ee}
Our goal is to bring $E$ into a form, that corresponds only to global white noise (see Eq.~(\ref{GlobalWhiteNoise})), i.e. $E'_{01}=E'_{10}= \frac{1}{3} E'_{11}$. Since we require $\mathbf{E}'\geq 0$ and $N(\mathbf{E}')=1$, $E'$ should be of the form $\mathbf{E}'(f')=\left(f',\frac{1-f'}{9},\frac{1-f'}{9},\frac{1-f'}{27}\right)^T$ with the new fidelity $f'\in[0,1]$. Thus we look for solutions $\mathbf{p}$ to the linear system Eq.~(\ref{LinEq}) for the specific choice of $\mathbf{E}'(f')$, that additionally fulfills the constraints $N(\mathbf{p})=1$ and $\mathbf{p}\geq 0$. One computes that $N(\mathbf{E}')=N(\mathbf{p})N(\mathbf{E})$ and we can therefore omit the trace preservation condition $N(\mathbf{p})=1$, since we already require (chose) $E$ and $E'$ to be trace preserving. Using $N(\mathbf{E})=1$ one can compute for the determinant
$\text{det}(\mathbf{D}[\mathbf{E}])= r s t$, where $r= 1 - 4 (E_{01} + 3 E_{11}) $ , $s = 1 - 4 (E_{10} + 3 E_{11})$ and $t = 1 - 4(E_{01} + E_{10} + 2 E_{11})$.
Thus the linear system Eq.~(\ref{LinEq}) will definitely have a unique solution, whenever $\text{max}\left(E_{01},E_{10},E_{11}\right)<\frac{1}{16}$ (or alternatively the initial fidelity $f=E_{00}> 4 \,\text{max}\left(E_{01},E_{10},E_{11}\right)$).
Let us consider a fixed vector $\mathbf{E}$ with $\text{det}(\mathbf{D}[\mathbf{E}])\neq 0$ first. For the corresponding CPM $\mathcal{E}$ we want to design a standard form $\mathcal{E}'$ with maximal fidelity $f'$. In contrast to the standard forms discussed so far the depolarization process $\mathcal{D}$, which translates into applying Pauli operators before and after the actual CPM $\mathcal{E}$ occurs, will be specifically designed for the given initial CPM $\mathcal{E}$, since the corresponding probabilities are given by the unique solution $\mathbf{p}(f')= \mathbf{D}[\mathbf{E}]^{-1} \cdot \mathbf{E}'(f') $. Note that each of the inequalities $p_{ij}\geq 0$ is linear in $f'$, i.e. of the form $a_{ij} f' + b_{ij} \geq 0$, where the coefficients are
\bea
 a_{00} =  \frac{1}{6}\left(\frac{1}{r} + \frac{1}{s} + \frac{4}{t} \right) &  & b_{00} = \frac{1}{48}\left(3 + \frac{1}{r} + \frac{1}{s} - \frac{5}{t}\right) \nonumber \\
 a_{01} =  \frac{1}{18}\left(\frac{3}{s} - \frac{1}{r} - \frac{4}{t} \right) &  & b_{01} = \frac{1}{144}\left(9 + \frac{3}{s} + \frac{5}{t} - \frac{1}{r}\right) \nonumber \\ 
 a_{10} =  \frac{1}{18}\left(\frac{3}{r} - \frac{1}{s} - \frac{4}{t} \right) &  & b_{10} = \frac{1}{144}\left(9 + \frac{3}{r} + \frac{5}{t} - \frac{1}{s}\right) \nonumber \\ 
 a_{11} =  \frac{1}{54}\left(\frac{4}{t} - \frac{3}{r} - \frac{3}{s} \right) &  & b_{11} = \frac{1}{432}\left(27 - \frac{3}{r} - \frac{3}{s} - \frac{5}{t}\right) \nonumber \; . \eea
Depending on the signs of $a_{ij}\neq 0$ the constraints are thus represented by intervals starting or ending at $f'_{ij}=-\frac{b_{ij}}{a_{ij}}$ (if $a_{ij}=0$ the corresponding condition is either always or never satisfied). In the following we will discuss the (complete) positivity condition $\mathbf{p}\geq 0$ in terms of the parameter $(r,s,t)$ instead of  $(E_{01},E_{10},E_{11})$ since they are linearly related.  Because of  $0\leq E_{ij}\leq 1$ we generally have $-15\leq r,s,t \leq 1$. The back transformation is given by $E_{01}=\frac{1}{16}\left(1 + 3s - r - 3 t \right) $, $E_{10}=\frac{1}{16}\left(1 +3r-s-3t \right) $ and  $E_{11}=\frac{1}{16}\left(1+ t- r-s\right)$, which implies $E_{00}=\frac{1}{16}\left( 1 + 3 (r + s + 3 t)\right)$. 

Let us consider the situation, in which the initial CPM $\mathcal{E}$ is close enough to the ideal operation, such that $\text{max}\left(E_{01},E_{10},E_{11}\right)<\frac{1}{16}$. Then there exists a unique solution with $1 \geq r, s, t >0$. Moreover a further restriction of $r,s,t$ to the interval $]\frac{2}{3},1]$ will ensure $a_{01},a_{10},a_{11}<0$ and $f'_{00}\leq f'_{01},f'_{10},f'_{11}$. Thus for a fixed initial vector $\mathbf{E}$ we are left with the three constraints $f'\leq  f'_{01},f'_{10},f'_{11} $ (and $f'\leq 1$) and the maximal achievable fidelity is $f'_\text{max}=\text{min}\left(1,f'_{01},f'_{10},f'_{11}\right)$. Although restricting $E_{10},E_{01},E_{11}$ to the interval  $[0,\frac{1}{48}]$ will be sufficient to guarantee  that $r,s,t \in ]\frac{2}{3},1]$, not all $r,s,t$ in the interval $]\frac{2}{3},1]$ correspond to $E_{01}, E_{01}, E_{01} \in [0,1]$. A minimization of $f'_\text{max}$ for $r,s,t \in ]\frac{2}{3},1]$ therefore yields only an upper bound to the fidelity decrease $f'_\text{max}$ (versus $f$) or the increase  of $1-f'_\text{max}$ (versus $1-f$), which represents the noise level of the CPM. A numerical minimization in this region shows that the relative fidelity  $\frac{f'_\text{max}}{f}$ 
is at least $37.14 \%$ and the relative noise level $\frac{1-f'_\text{max}}{1-f}$ is at most increased by a factor $5.5$. 

Note that for an initial fidelity $f> \frac{15}{16}$ we have $0\leq E_{10},E_{01},E_{11}<\frac{1}{48}$. In other words any decoherence process on two qubits, that introduces only little noise ( i.e. $f>\frac{15}{16}$), can be brought into the form of global white noise by increasing the noise level by a factor less than $5.5$. This standard form is achieved by application of local Pauli operations, that are chosen randomly according to some probability distribution specified by the parameters $p_{ij}(f'_\text{max})$. Thus the protocol is specifically designed for the initial form of the decoherence process (more precisely it depends on the vector $\mathbf{E}$, that is obtained after the decoherence is brought into the form Eq.~(\ref{SacrifDecomp}) by the methods described above).

If the initial CPM does not belong to the region with $0\leq E_{10},E_{01},E_{11}<\frac{1}{48}$, a similar derivation can be applied. The constraints $a_{ij} f' + b_{ij} \geq 0$ again determine, whether a standard form can be obtained in this way and how much fidelity has to be sacrificed in order to achieve the normal form with $\mathbf{E}'(f')$. Moreover a generalization to the case of $d$-level systems with $d>2$ and to the multi-party setting with $N>2$ can be developed along the lines of the previous discussion. Note that for increasing $N$, although the achieved standard forms will also be global white noise and thus be specified by one parameter only, the derivation and the transformation protocol itself will become more involved, since the number of parameters $p_{i_1\ldots i_N}$ will be $2^N$ and thus increase exponentially.

\subsection{Standard forms for noisy unitary operations} \label{Unitary_SF}

We now turn to standard forms of noisy operations where the ideal operation is given by some unitary operation $U$. We concentrate on two--particle operations and will illustrate our approach with help of several examples, including gates which are up to local unitary operations equivalent to the SWAP gate, the CNOT gate and a general phase gate with arbitrary phase $\alpha$. We will show that one can depolarize these noisy gates to standard forms with a reduced number of parameters, without changing the fidelity of the ideal operation. To this aim, we decomposes a CPM ${\cal E}$ into a unitary part $U$ and some remaining (orthogonal) part ${\cal E'}$ (where ${\cal \tilde{E}}$ is in general no longer a CPM), i.e.  
\be
{\cal E}\rho = f U\rho U^{\dagger} + (1-f) {\cal \tilde{E}}\rho,\label{decomposemap}
\ee
and both, $f$ and ${\cal \tilde{E}}$, are determined by the Isomorphism. We have that 
\bea
f&=&\langle\Psi_U|E|\psi_U\rangle \\
\tilde{E}&=&\frac{E-f |\Psi_U\rangle\langle\Psi_U|}{1-f},\label{Ep}
\eea
where $\tilde{E}$ is the operator corresponding to the map ${\cal \tilde{E}}$ and $f$ specifies the initial fidelity of the operation $U$. 
It is not necessary to make such a decomposition, however in this notation it is immediately evident that only the noise part, namely 
${\cal \tilde{E}}$ is altered by the depolarization procedure. We remark that $\text{tr}(\tilde{E})=1$, however $\tilde{E}$ may have negative eigenvalues 
and is hence not a density operator. Nevertheless, we can formally decompose $E$ (and thus ${\cal E}$) into these two parts. 
We will show that one can depolarize the map ${\cal E}$ to
\be
{\cal E}'\rho = f U\rho U^{\dagger} + (1-f) {\tilde {\cal E}}'\rho,
\ee
where ${\tilde{\cal E}}'$ is a (generally non--positive) map of certain standard form, specified by a few parameters. Clearly, the total map ${\cal E}'$ remains completely positive. Note that the depolarization of ${\cal E}$ takes place in such a way that the weight of the ideal operation is not altered. In particular, if the operation is initially noiseless (i.e. $f=1$), it will remain noiseless after the depolarization. This is achieved by considering depolarization processes that leave the unitary operation $U$ (or equivalently the state $|\Psi_U\rangle$ when considering the operator $E$ corresponding to the operation ${\cal E}$) invariant. The number of required parameters and the explicit form of $\tilde{\cal E}$ depends on the ideal operation $U$, as the group of local operations that leave $U$ invariant is determined by the structure of the state $|\Psi_U\rangle$. 

\subsubsection{The noisy SWAP gate}\label{SWAP_SF}

In this section we determine a standard form for noisy SWAP operations. The ideal $d$--level SWAP operation is defined via its action on product basis states, namely $U_{\rm SWAP} |i\rangle^A|j\rangle^B = |j\rangle^A|i\rangle^B$, where $\{|k\rangle\}_{k=0,1,\ldots,d-1}$ is a basis of $\mathbf{H}=\mathbb{C}^{d}$. The state $E_{\rm SWAP}=|\Psi_{\rm SWAP}\rangle\langle\Psi_{\rm SWAP}|$ corresponding to $U_{\rm SWAP}$ is specified by (see Eq.~(\ref{SWAP_E}))
\be
|\Psi_{\rm SWAP}\rangle=|\Phi\rangle^{AB'}|\Phi\rangle^{BA'}.
\ee
Consider the mixed state $E$ describing ---via the isomorphism--- a noisy SWAP gate. We have that all operations of the form $U^{A}\otimes {U^*}^{B'}\otimes V^{B}\otimes {V^*}^{A'}$ leave $|\Psi_{\rm SWAP}\rangle$ invariant and hence can be used to depolarize $E$. This implies that we can essentially use the same depolarization procedure as in the case where the ideal operation is given by the identity (see Sec.~\ref{MultiDecSF}), only the role of particles $A'$ and $B'$ is exchanged. This implies that the resulting standard form can again be interpreted as a local and global white noise processes with three independent parameters,that occur before the application of an ideal SWAP operation, i.e. $\mathcal{E}'(\rho)= U_{\rm SWAP}\mathcal{D}( \rho) U_{\rm SWAP}^\dagger$ with
\bea \mathcal{D}(\rho) & = & \alpha_{00}\, \rho  + \alpha_{01}\, \rho_A\otimes \frac{1}{d}\mathbf{1}_{B} + \alpha_{10}\, \frac{1}{d}\mathbf{1}_{A}\otimes \rho_{B} \nonumber \\
& &  +\alpha_{11}\, \frac{1}{d^2}\text{tr}(\rho)\mathbf{1}_{AB}\; .\eea
Note that the parameters $\alpha_{kl}$ are again given by Eq.~(\ref{Alpha}), where $E_{kl}$ are the coefficients in a decomposition Eq.~(\ref{isotropic1}) of $E$ according to the basis 
{\footnotesize \be \gamma_\mathbf{k}\in \left\{ P_\Phi^{AB'}\otimes P_\Phi^{BA'}, P_\Phi^{AB'}\otimes \gamma^{BA'},\gamma^{AB'}\otimes P_\Phi^{BA'}, \gamma^{AB'}\otimes \gamma^{BA'} \right\}\nonumber \ee}
with $\gamma=\frac{1}{d^2-1}\left(\mathbf{1}- P_\Phi\right)$. In particular by the twirling procedure the Jamio\l kowski fidelity remains the same, i.e. $F(\mathcal{E},U_\text{SWAP}) =F(\mathcal{E}',U_\text{SWAP})$.

\subsubsection{The noisy phase gate and CNOT gate}\label{PhaseGateSF}

In this section we consider the unitary operation
\be
U_{AB}(\alpha)=e^{-i \alpha \sigma_y^{A}\otimes\sigma_y^{B}},
\ee
for arbitrary angles $\alpha$. Up to the local unitary operations, $U(\alpha)$ is equivalent to a controlled phase gate, while for 
$\alpha=\pi/4$, $U(\alpha)$ is equivalent to the CNOT gate (see Eq.~(\ref{LU_CNOT_PhaseGate})), i.e. $U_{\rm CNOT}=U^A_1\otimes U^B_2 U(\pi/4) V^A_1 \otimes V^B_2$. 

We will obtain a standard form for noisy operations, given in the ideal case by $U(\alpha)$, by depolarizing the corresponding CPM ${\cal E}$. The depolarization takes place by applying appropriate random local unitary operations that leave the state 
\be
|\Psi_{\alpha}\rangle=\cos(\alpha)|\psi_0\rangle_{\bm A}|\psi_0\rangle_{\bm B} -i\sin(\alpha) |\psi_2\rangle_{\bm A}|\psi_2\rangle_{\bm B},
\ee
invariant (up to an irrelevant phase), where $|\Psi_\alpha\rangle\langle\Psi_\alpha|$ is the state corresponding to $U(\alpha)$ via the Isomorphism, and 
\be |\psi_j\rangle = \mathbf{1}\otimes \sigma_j |\Phi\rangle \ee
 are Bell states.
Note that such a depolarization procedure for $U(\alpha)$ automatically provides a depolarization procedure for all operations that are local unitary equivalent to $U(\alpha)$, leading to a standard form with the same number of parameters for these noisy gates. The depolarization procedure simply has to be adopted according to the local unitary operations. To be specific, consider for instance the noisy $U(\pi/4)$ gate and the noisy CNOT gate. If $W^{{\bm A}{\bm B}}$ is a local unitary operation that keeps $|\Psi_{\pi/4}\rangle$ invariant , then the operation 
{ 
\bea 
W'^{{\bm A}{\bm B}}  & =  & U_1^{A'} \otimes U_2^{B'}\otimes (V_1^{A})^T\otimes (V_2^{B})^T \, W^{{\bm A}{\bm B}} \nonumber \\ 
& & \; (U_1^{A'})^\dagger \otimes (U_2^{B'})^\dagger\otimes {V^*}_1^{A}\otimes {V^*}_2^{B} 
\nonumber \eea } 
keeps $|\Psi_{\rm CNOT}\rangle$ invariant. That is, one obtains a depolarization procedure for the noisy CNOT gate from the depolarization procedure for the $U(\pi/4)$ gate by replacing each unitary operation $W$ by $W'$.

We now present an explicit depolarization procedure for the noisy $U(\alpha)$ gate, described by the CPM ${\cal E}$, with arbitrary $\alpha$. We will consider the depolarization of the corresponding state by means of 4--local operations. We remark that any sequence of depolarization steps can be translated into a single step with multiple possibilities by considering all possible combinations. Such a single step procedure can then be translated to appropriate random operations applied to the system before and after the application of ${\cal E}$ and hence to depolarize the corresponding map. For notational convenience, we define the four--qubit states 
\be
|\Psi_{ij}\rangle^{AA'BB'}\equiv |\psi_i\rangle^{AA'}\otimes |\psi_j\rangle^{BB'}.
\ee
Given an arbitrary CPM $\cal E$ specified by
\be
{\cal E}\rho = \sum_{i,j,k,l=0}^{3} \lambda_{ij,kl} \sigma_i\sigma_j \rho \sigma_k\sigma_l,
\ee
the corresponding state $E$ is given by
\be
E=\sum_{i,j,k,l=0}^{3} E_{ij,kl} |\Psi_{ij}\rangle \langle\Psi_{kl}|,
\ee
where $\lambda_{ij,kl}=\lambda^*_{kl,ij}$. We define two--qubit unitary operations ${\cal U}, \tilde {\cal U}, {\cal V}$ by
\bea
{\cal U} &\equiv& (i\sigma_y)\otimes(i\sigma_y), \nonumber\\ 
\tilde{\cal U} &\equiv& \sigma_x\otimes \sigma_x, \\
{\cal V} &\equiv& e^{-i\pi/4\sigma_y} \otimes e^{-i\pi/4\sigma_y}. \nonumber 
\eea
The action of these operations on Bell-basis states $\{|\psi_j\rangle\}$ can be readily obtained and one finds that ${\cal U}, \tilde{\cal U}$ introduce relative phases between the Bell states, while ${\cal V}$ exchanges two of them. To be specific, we have
,
{\small 
\bea
{\cal U}\{|\psi_0\rangle,|\psi_1\rangle,|\psi_2\rangle,|\psi_3\rangle\}&=& \{|\psi_0\rangle,-|\psi_1\rangle,|\psi_2\rangle,-|\psi_3\rangle\}, \nonumber\\
\tilde{\cal U}\{|\psi_0\rangle,|\psi_1\rangle,|\psi_2\rangle,|\psi_3\rangle\}&=& \{|\psi_0\rangle,|\psi_1\rangle,-|\psi_2\rangle,-|\psi_3\rangle\}, \nonumber\\
{\cal V}\{|\psi_0\rangle,|\psi_1\rangle,|\psi_2\rangle,|\psi_3\rangle\}&=& \{|\psi_0\rangle,-|\psi_3\rangle,|\psi_2\rangle,|\psi_1\rangle\}. \label{calU}
\eea }
All local operations of the form $\mathbf{1}_{\bm A}\mathbf{1}_{\bm B}$, $\mathbf{1}_{\bm A}{\cal U}_{\bm B}$, ${\cal U}_{\bm A}\mathbf{1}_{\bm B}$, ${\cal U}_{\bm A}{\cal U}_{\bm B}$, $\tilde{\cal U}_{\bm A}\tilde{\cal U}_{\bm B}$, $\mathbf{1}_{\bm A}{\cal V}_{\bm B}$, ${\cal V}_{\bm A}\mathbf{1}_{\bm B}$, ${\cal V}_{\bm A}{\cal V}_{\bm B}$ keep the state $|\Psi_\alpha\rangle$ (and the fidelity $f=\langle\Psi_\alpha |E|\Psi_\alpha\rangle$ of the ideal operation) invariant and can thus be used for depolarization.

We decompose ${\cal E}$ into the unitary part $U(\alpha)$ and the remaining noise part ${\cal \tilde E}$ (see Eq.~(\ref{decomposemap})) and consider the corresponding (non--positive) operator $\tilde{E}$ (see Eq.~(\ref{Ep})) in the following,
\be
\tilde{E}=\sum_{i,j,k,l=0}^{3} \lambda_{ij,kl} |\Psi_{ij}\rangle \langle\Psi_{kl}|.
\ee
We randomly apply $\mathbf{1}_{\bm A}\mathbf{1}_{\bm B}$, $\mathbf{1}_{\bm A}{\cal U}_{\bm B}$, ${\cal U}_{\bm A}\mathbf{1}_{\bm B}$ or ${\cal U}_{\bm A}{\cal U}_{\bm B}$, each with probability $1/4$, which leads to an operator 
{\small \be
\tilde E'=\frac{1}{4}\left(\tilde{E} + U^{\bm B} \tilde{E} (U^{\bm B})^\dagger + U^{\bm A} \tilde{E} (U^{\bm A})^\dagger + U^{\bm A}U^{\bm B} \,\tilde{E}\, (U^{\bm A} U^{\bm B})^\dagger\right). \nonumber
\ee }
One finds that $\tilde E'$ is of block--diagonal form with coefficients $\lambda'_{ij,kl}$, that fulfill $\lambda'_{ij,kl}=0$ whenever $(i \mod 2) \not = (k \mod 2)$ or $(j \mod 2) \not = (l \mod 2)$ and remain invariant otherwise. This follows from Eq.~(\ref{calU}), as ${\cal U}$ introduces a phase $(-1)$ for Bell states $|\psi_i\rangle$ with $(i \mod 2 =1)$ while states with even parity ($i \mod 2 =0$) remain invariant, which results in the cancellation of the corresponding off--diagonal elements. Thus only elements $\lambda'_{ij,kl}$ with $(i \mod 2) = (k \mod 2)$ and $(j \mod 2) = (l \mod 2)$ remain, which can be grouped into four $4\times 4$ blocks $\Gamma_{ab}$ with $a=(i \mod 2) = (k \mod 2), b=(j \mod 2) = (l \mod 2)$. For instance, $\Gamma_{01}= \sum_{i,k\in\{0,2\};j,l \in\{1,3\}} \lambda_{ij,kl} |\Psi_{ij}\rangle \langle\Psi_{kl}|$.

In the following, we consider the depolarization of the subspaces $\Gamma_{ab}$ separately. We start with $\Gamma_{00}$, which is spanned by the states $\{|\Psi_{00}\rangle,|\Psi_{02}\rangle,|\Psi_{20}\rangle,|\Psi_{22}\rangle\}$. By randomly applying $\mathbf{1}_{\bm A}\mathbf{1}_{\bm B}$ or $\tilde{\cal U}_{\bm A}\tilde{\cal U}_{\bm B}$ with probability $1/2$, we find that the resulting operator $\Gamma'_{00}$ has coefficients $\lambda'_{00,02}= \lambda'_{00,20} = \lambda'_{02,22} =0$, while the other coefficients remain invariant, i.e. 
\be 
\label{lam1}
\lambda'_{00,00} = \lambda_{00,00}; \lambda'_{02,02} = \lambda_{02,02}; \lambda'_{20,20} = \lambda_{20,20}; \nonumber \\
\lambda'_{22,22} = \lambda_{22,22}; \lambda'_{00,22} = \lambda_{00,22}; \lambda'_{02,20} = \lambda_{02,20}. 
\ee
We thus find that $\Gamma_{00}$ is of the form
\be
\Gamma'_{00} = \left(%
\begin{array}{cccc}
  \lambda'_{00,00} & 0 & 0 & \lambda'_{00,22} \\
  0 & \lambda'_{02,02} & \lambda'_{02,20} & 0 \\
  0 & \lambda'^*_{02,20} & \lambda'_{20,20} & 0 \\
  {\lambda'}^*_{00,22} & 0 & 0 & \lambda'_{22,22} \\
\end{array}%
\right),
\ee
which are 8 independent real parameters as $\lambda_{ij,kl}=\lambda^*_{kl,ij}$.

The effect of these (random) operations on the other subspaces $\Gamma_{01},\Gamma_{10},\Gamma_{11}$ is similar, i.e. the corresponding off--diagonal term vanish. However, in these subspaces further depolarization is possible. Consider $\Gamma_{01}$ which is spanned by the states $\{|\Psi_{01}\rangle,|\Psi_{03}\rangle,|\Psi_{21}\rangle,|\Psi_{23}\rangle\}$. Applying randomly $\mathbf{1}_{\bm A}\mathbf{1}_{\bm B}$ or $\mathbf{1}_{\bm A}{\cal V}_{\bm B}$ with probability $1/2$ leads to coefficients 
\bea
\label{lam2}
\lambda'_{01,01}&=&\lambda'_{03,03} = \frac{1}{2}(\lambda_{01,01}+\lambda_{03,03}) \nonumber\\
\lambda'_{21,21}&=&\lambda'_{23,23} = \frac{1}{2}(\lambda_{21,21}+\lambda_{23,23}) \\
\lambda'_{01,23}&=& -\lambda'_{03,21}= \frac{1}{2}(\lambda_{01,23}-\lambda_{03,21})\nonumber
\eea
This can readily be seen by using that 
\be
\mathbf{1}_{\bm A}{\cal V}_{\bm B} \{|\Psi_{01}\rangle,|\Psi_{03}\rangle,|\Psi_{21}\rangle,|\Psi_{23}\rangle\} \nonumber \\= \{-|\Psi_{03}\rangle,|\Psi_{01}\rangle,-|\Psi_{23}\rangle,|\Psi_{21}\rangle\}.
\ee
Thus we find that $\Gamma_{01}$ is of the form
\be
\Gamma'_{01} = \left(%
\begin{array}{cccc}
  \lambda'_{01,01} & 0 & 0 & \lambda'_{01,23} \\
  0 & \lambda'_{01,01} & -\lambda'_{01,23} & 0 \\
  0 & -{\lambda'}^*_{01,23} & \lambda'_{21,21} & 0 \\
  {\lambda'}^*_{01,23} & 0 & 0 & \lambda'_{21,21} \\
\end{array}%
\right),
\ee
and is thus described by 4 independent, real parameters ($\lambda'_{01,01}$ and $\lambda'_{21,21}$ are real, $\lambda'_{01,23}$ is complex). 

Similarly, by randomly applying $\mathbf{1}_{\bm A}\mathbf{1}_{\bm B}$ or ${\cal V}_{\bm A}\mathbf{1}_{\bm B}$ with probability $1/2$, one depolarizes the subspace $\Gamma_{10}$ ---spanned by the states $\{|\Psi_{10}\rangle,|\Psi_{30}\rangle,|\Psi_{12}\rangle,|\Psi_{32}\rangle\}$--- to the form
\be
\Gamma'_{10} = \left(%
\begin{array}{cccc}
  \lambda'_{10,10} & 0 & 0 & \lambda'_{10,32} \\
  0 & \lambda'_{10,10} & -\lambda'_{10,32} & 0 \\
  0 & -{\lambda'}^*_{10,32} & \lambda'_{12,12} & 0 \\
  {\lambda'}^*_{10,32} & 0 & 0 & \lambda'_{12,12} \\
\end{array}%
\right),
\ee
where
\bea
\label{lam3}
\lambda'_{10,10}&=&\lambda'_{30,30} = \frac{1}{2}(\lambda_{10,10}+\lambda_{30,30}) \nonumber \\
\lambda'_{12,12}&=&\lambda'_{32,32} = \frac{1}{2}(\lambda_{12,12}+\lambda_{32,32}) \\
\lambda'_{10,32}&=& -\lambda'_{30,12}= \frac{1}{2}(\lambda_{10,32}-\lambda_{30,12}), \nonumber
\eea
which is again described by 4 independent, real parameters.

Finally, the subspace $\Gamma_{11}$ ---spanned by the states $\{|\Psi_{11}\rangle,|\Psi_{13}\rangle,|\Psi_{31}\rangle,|\Psi_{33}\rangle\}$--- can be further depolarized by randomly applying one of the operations $\mathbf{1}_{\bm A}\mathbf{1}_{\bm B}$, $\mathbf{1}_{\bm A}{\cal V}_{\bm B}$, ${\cal V}_{\bm A}\mathbf{1}_{\bm B}$ or ${\cal V}_{\bm A}{\cal V}_{\bm B}$ with probability $1/4$. One finds that
\bea
\lambda'_{11,11}&=& \lambda'_{11,11}=\lambda'_{13,13}=\lambda'_{31,31}=\lambda'_{33,33}=\nonumber \\
&=&\frac{1}{4}(\lambda_{11,11}+\lambda_{13,13}+\lambda_{31,31}+\lambda_{33,33}) \nonumber\\
\lambda'_{11,33}&=&-\lambda'_{13,31}=\frac{1}{2}\left( \Re(\lambda_{11,33}) - \Re(\lambda_{13,31})\right )
\label{lam4}
\eea
where $\Re(x)$ denotes the real part of $x$.
Thus $\Gamma_{11}$ is described by 2 independent, real parameters and is of the form
\be
\Gamma'_{11} = \left(%
\begin{array}{cccc}
  \lambda'_{11,11} & 0 & 0 & \lambda'_{11,33} \\
  0 & \lambda'_{11,11} & -\lambda'_{11,33} & 0 \\
  0 & -\lambda'_{11,33} & \lambda'_{11,11} & 0 \\
  \lambda'_{11,33} & 0 & 0 & \lambda'_{11,11} \\
\end{array}%
\right),
\ee
We remark that the depolarization process described in this final step leaves the subspaces $\Gamma_{00},\Gamma_{01},\Gamma_{10}$ ---which were already depolarized earlier--- invariant. The final depolarized CPM $\tilde{\cal E}_S$ is specified by $(8+4+4+2-1)=17$ real parameters (where the (-1) results from the normalization condition $\text{tr}(\tilde E)=1$) and is of Block--diagonal form. The coefficients $\lambda'_{ij,kl}$ are given by Eqs. \ref{lam1},\ref{lam2},\ref{lam3},\ref{lam4} and are zero otherwise. This leads to the standard form,
\be
\label{ES}
{\cal E}\rho = f U(\alpha)\rho U(\alpha)^{\dagger} + (1-f) \sum_{ij,kl} \lambda'_{ij,kl} \sigma_i\sigma_j \rho \sigma_k\sigma_l,
\ee
where $f'=\langle \Psi_\alpha|E|\Psi_\alpha\rangle = \langle \Psi_\alpha|E|\Psi_\alpha\rangle$, i.e. the fidelity of the ideal operation remains invariant.
To summarize, we can achieve the following standard form:

{\bf  \underline{Standard form for the Phase Gate} } \\
{\it By uniformly choosing one of the unitaries $U_k$ from $\mathcal{U}_1 \cdot \mathcal{U}_2 \cdot\mathcal{U}_3 $, where
\bea 
\mathcal{U}_1 & = & \{\mathbf{1}_A\mathbf{1}_B,\,  e^{-i\frac{\pi}{4}\sigma_y^A}\mathbf{1}_B,\, \mathbf{1}_A e^{-i\frac{\pi}{4}\sigma_y^B},\, e^{-i\frac{\pi}{4}\sigma_y^A}e^{-i\frac{\pi}{4}\sigma_y^B}\} \nonumber \\
\mathcal{U}_2 & = & \{\mathbf{1}_A\mathbf{1}_B,\, \sigma_x^A \sigma_x^B \} \\
\mathcal{U}_3 & = & \{\mathbf{1}_A\mathbf{1}_B,\, \sigma_y^A\mathbf{1}_B,\, \mathbf{1}_A \sigma_y^B, \,\sigma_y^A \sigma_y^B\}\; , \nonumber
\eea
and applying $U_k^\dagger$ before and $U_k$ after the application of the noisy phase gate $\mathcal{E}$ the resulting CPM $\mathcal{E}'$ is of the standard form $\mathcal{E}'(\rho)=\sum_{i,j,k,l=0}^{3} E'_{ij,kl} \sigma_i\sigma_j \rho \sigma_k\sigma_l$ with
\be \label{NoisyPhaseGateSF}
 E'=
\left(%
\begin{array}{cccc}
  \Gamma'_{00} & 0 & 0 & 0 \\
  0 & \Gamma'_{01} & 0 & 0 \\
  0 & 0 & \Gamma'_{10} & 0 \\
  0 & 0 & 0 & \Gamma'_{11} \\
\end{array}%
\right) \hspace{1cm} {\rm where}
\ee
\bea
\Gamma'_{00}=
\left(
\begin{array}{cccc}
  a & 0 & 0 &  u \\
  0 & b & v & 0 \\
  0 & v^* & \tilde b & 0 \\
  u^* & 0 & 0 & \tilde a \\
\end{array}
\right)&,&
\Gamma'_{01}=
\left(
\begin{array}{cccc}
  c & 0 & 0 & w \\
  0 & c & -w & 0 \\
  0 & -w^* & \tilde c & 0 \\
  w^* & 0 & 0 & \tilde c \\
\end{array}
\right),\nonumber\\
\Gamma'_{10}=
\left(
\begin{array}{cccc}
  d & 0 & 0 & x \\
  0 & d & -x & 0 \\
  0 & -x^* & \tilde d & 0 \\
  x^* & 0 & 0 & \tilde d \\
\end{array}
\right)&,&
\Gamma'_{11}=
\left(
\begin{array}{cccc}
  e & 0 & 0 & \tilde e \\
  0 & e & - \tilde e  & 0 \\
  0 & -\tilde e  & e & 0 \\
  \tilde e  & 0 & 0 & e \\
\end{array}
\right) \nonumber
\eea
with the following choice of basis
\bea \mathcal{B} & = &
\{ \, |\Psi_{00}\rangle, \, |\Psi_{02}\rangle, \, |\Psi_{20}\rangle, \,|\Psi_{22}\rangle, \nonumber \\
& & \, |\Psi_{01}\rangle, \, |\Psi_{03}\rangle, \, |\Psi_{21}\rangle, \, |\Psi_{23}\rangle, \nonumber \\
& & \, |\Psi_{10}\rangle, \, |\Psi_{30}\rangle, \, |\Psi_{12}\rangle, \, |\Psi_{32}\rangle, \nonumber \\
& & \, |\Psi_{11}\rangle, \, |\Psi_{13}\rangle, \, |\Psi_{31}\rangle, \, |\Psi_{33}\rangle \, \} 
\label{BasisOrdering} \eea
and the parameters $a,\tilde a, b, \tilde b, c,\tilde c,d, \tilde d, e, \tilde e \in\mathbb{R}$ and $u,v,w,x \in \mathbb{C}$. This depolarization does not increase the noise level, i.e. $f'=f$.
}

\subsubsection{The CNOT--type gate}\label{CNOT_SF}

For certain values of $\alpha$, further depolarization is possible. In particular, we consider $\alpha=\pi/4$, i.e. the operations $U(\pi/4)$ which is local unitary equivalent to the CNOT gate. In this case, the state $|\Psi_{\pi/4}\rangle$ is a maximally entangled state (with respect to systems ${\bm A},{\bm B}$), which remains invariant under a larger set of local unitary operations than a non--maximally entangled state $|\Psi_{\alpha}\rangle$. In particular, we consider the unitary operations 
\bea
\label{calW}
{\cal W}&\equiv&\mathbf{1}\otimes (i\sigma_y),\\
\tilde{\cal W}&\equiv&\sigma_z\otimes \sigma_x.
\eea
which act on Bell states as follows
{\small \bea
{\cal W}\{|\psi_0\rangle,|\psi_1\rangle,|\psi_2\rangle,|\psi_3\rangle\}&=& \{i|\psi_2\rangle,|\psi_3\rangle,i|\psi_0\rangle,-|\psi_1\rangle\}, \nonumber\\
\tilde{\cal W}\{|\psi_0\rangle,|\psi_1\rangle,|\psi_2\rangle,|\psi_3\rangle\}&=& \{-i|\psi_2\rangle,|\psi_3\rangle,i|\psi_0\rangle,|\psi_1\rangle\}.\nonumber
\eea }
The operation ${\cal W}_{\bm A}\tilde{\cal W}_{\bm B}$ 
leaves the state $|\Psi_{\pi/4}\rangle$ ---up to an irrelevant global phase factor $(-i)$--- invariant. Note that this is not true for $|\Psi_{\alpha}\rangle$ with $\alpha \not=\pi/4$. We take the standard form Eq.~(\ref{ES}) as initial map, and apply randomly either $\mathbf{1}_{\bm A}\mathbf{1}_{\bm B}$ or ${\cal W}_{\bm A}\tilde{\cal W}_{\bm B}$. One finds that the resulting operator $\tilde E''$ is significantly simplified and is described by 8 independent, real parameters. We denote the coefficients of $\tilde E''$ by $\mu_{ij,kl}$.

To be specific, for $\Gamma''_{00}$ we find 
\be
\label{mu1}
\mu_{00,00}= \mu_{22,22} = \frac{1}{2}(\lambda'_{00,00} + \lambda'_{22,22}); \nonumber \\
\mu_{02,02}= \mu_{20,20} = \frac{1}{2}(\lambda'_{02,02}+\lambda'_{20,20});\\
\mu_{00,22}= i\Im(\lambda'_{00,22}); \mu_{02,20}= i\Im( \lambda'_{02,20}), \nonumber
\ee 
where $\Im(x)$ denotes the imaginary part of $x$, i.e. $i \Im(x)= (x-x^*)/2$ and we thus have 4 real parameters. This follows from ${\cal W}_{\bm A}\tilde{\cal W}_{\bm B}\{|\Psi_{00}\rangle,|\Psi_{02}\rangle,|\Psi_{20}\rangle,|\Psi_{22}\rangle\} = \{|\Psi_{22}\rangle,-|\Psi_{20}\rangle,|\Psi_{02}\rangle,-|\Psi_{00}\rangle\}$.

Similarly, for $\Gamma''_{01}$ we find
\bea
\label{mu2}
\mu_{01,01}&=&\mu_{03,03} = \mu_{21,21}=\mu_{23,23} = \frac{1}{2}( \lambda'_{01,01}+ \lambda'_{21,21}); \nonumber \\
\mu_{01,23}&=& -\mu_{03,21} = \Re(\lambda'_{01,23}),
\eea
while $\Gamma''_{10}$ simplifies to
\bea
\label{mu3}
\mu_{10,10}&=&\mu_{30,30} = \mu_{12,12}=\mu_{32,32} = \frac{1}{2}(\lambda'_{10,10}+ \lambda'_{12,12}); \nonumber \\
\mu_{10,32}&=& -\mu_{30,12} = \Re(\lambda'_{10,32}),
\eea
where we have 2 real parameters in each case.

Finally, for $\Gamma''_{11}$ we have
\bea
\label{mu4}
\mu_{11,11}&=&\mu_{13,13} = \mu_{31,31}=\mu_{33,33} = \lambda'_{11,11}; \nonumber \\
\mu_{11,33}&=& -\mu_{13,31} = 0,
\eea
which is a single, real parameter. It follows that the standard form for the depolarized gate $U(\pi/4)$ is given by
\be
\label{ECNOT}
{\cal E''}\rho = f U(\frac{\pi}{4})\rho U(\frac{\pi}{4})^{\dagger} + (1-f) \sum_{ij,kl} \tilde\mu_{ij,kl} \sigma_i\sigma_j \rho \sigma_k\sigma_l,
\ee
where the coefficients $\tilde \mu_{ij,kl}$ are defined in Eqs.~(\ref{mu1}), (\ref{mu2}), (\ref{mu3}), (\ref{mu4}) and are zero otherwise. Note that the fidelity of the ideal operation $U(\pi/4)$ remains invariant.

{\bf  \underline{Standard form for the CNOT--type Gate} } \\
{\it The total state $\tilde E$ is thus of the form

{\it \scriptsize \be 
\label{bigGamma}
\left(%
\begin{array}{cccccccccccccccc}
  a & 0 & 0 & i u & 0 & 0 & 0 & 0 & 0 & 0 & 0 & 0 & 0 & 0 & 0 & 0 \\
  0 & b & i v & 0 & 0 & 0 & 0 & 0 & 0 & 0 & 0 & 0 & 0 & 0 & 0 & 0 \\
  0 & -i v & b & 0 & 0 & 0 & 0 & 0 & 0 & 0 & 0 & 0 & 0 & 0 & 0 & 0 \\
  -i u & 0 & 0 & a & 0 & 0 & 0 & 0 & 0 & 0 & 0 & 0 & 0 & 0 & 0 & 0 \\
  0 & 0 & 0 & 0 & c & 0 & 0 & w & 0 & 0 & 0 & 0 & 0 & 0 & 0 & 0 \\
  0 & 0 & 0 & 0 & 0 & c & -w & 0 & 0 & 0 & 0 & 0 & 0 & 0 & 0 & 0 \\
  0 & 0 & 0 & 0 & 0 & -w & c & 0 & 0 & 0 & 0 & 0 & 0 & 0 & 0 & 0 \\
  0 & 0 & 0 & 0 & w & 0 & 0 & c & 0 & 0 & 0 & 0 & 0 & 0 & 0 & 0 \\
  0 & 0 & 0 & 0 & 0 & 0 & 0 & 0 & d & 0 & 0 & x & 0 & 0 & 0 & 0 \\
  0 & 0 & 0 & 0 & 0 & 0 & 0 & 0 & 0 & d & -x & 0 & 0 & 0 & 0 & 0 \\
  0 & 0 & 0 & 0 & 0 & 0 & 0 & 0 & 0 & -x & d & 0 & 0 & 0 & 0 & 0 \\
  0 & 0 & 0 & 0 & 0 & 0 & 0 & 0 & x & 0 & 0 & d & 0 & 0 & 0 & 0 \\
  0 & 0 & 0 & 0 & 0 & 0 & 0 & 0 & 0 & 0 & 0 & 0 & e & 0 & 0 & 0 \\
  0 & 0 & 0 & 0 & 0 & 0 & 0 & 0 & 0 & 0 & 0 & 0 & 0 & e & 0 & 0 \\
  0 & 0 & 0 & 0 & 0 & 0 & 0 & 0 & 0 & 0 & 0 & 0 & 0 & 0 & e & 0 \\
  0 & 0 & 0 & 0 & 0 & 0 & 0 & 0 & 0 & 0 & 0 & 0 & 0 & 0 & 0 & e \\
\end{array}%
\right)\nonumber
\ee}

 or equivalently

\be
\tilde E=
\left(%
\begin{array}{cccc}
  \Gamma_{00} & 0 & 0 & 0 \\
  0 & \Gamma_{01} & 0 & 0 \\
  0 & 0 & \Gamma_{10} & 0 \\
  0 & 0 & 0 & \Gamma_{11} \\
\end{array}%
\right)\hspace{0.5cm} {\rm with}
\ee
\bea
\Gamma_{00}=
\left(
\begin{array}{cccc}
  a & 0 & 0 & i u \\
  0 & b & i v & 0 \\
  0 & -i v & b & 0 \\
  -i u & 0 & 0 & a \\
\end{array}
\right)&,&
\Gamma_{01}=
\left(
\begin{array}{cccc}
  c & 0 & 0 & w \\
  0 & c & -w & 0 \\
  0 & -w & c & 0 \\
  w & 0 & 0 & c \\
\end{array}
\right),\nonumber\\
\Gamma_{10}=
\left(
\begin{array}{cccc}
  d & 0 & 0 & x \\
  0 & d & -x & 0 \\
  0 & -x & d & 0 \\
  x & 0 & 0 & d \\
\end{array}
\right)&,&
\Gamma_{11}=
\left(
\begin{array}{cccc}
  e & 0 & 0 & 0 \\
  0 & e & 0 & 0 \\
  0 & 0 & e & 0 \\
  0 & 0 & 0 & e \\
\end{array}
\right),\nonumber
\eea
 where we use the basis $\mathcal{B}$ in Eq.~(\ref{BasisOrdering}) and $a= \mu_{00,00}$, $b=\mu_{02,02}$ etc. are all real parameters.}


\subsection{Standard forms by means of sacrificing}\label{Sacrificing_SF}

While the standard forms for the general $U(\alpha)$ gate or the CNOT--type gate $U(\pi/4)$ are already relatively simple (as the number of relevant parameters is significantly reduced, namely from 255 to 17 or 8 respectively), for many practical applications a further simplification might still be desirable. If, for instance, one would like to analyze error thresholds for processes involving several particles and/or operations (as is e.g. the case in fault tolerant quantum computation or entanglement purification with imperfect means) where noisy operations are described by these standard forms, the corresponding CPMs are still rather complex.

In this section we will provide such a further simplification of the corresponding noise process, where we find that in many relevant cases a single parameter is sufficient and the noise process can be described by (global) white noise. In contrast to the previous depolarization procedure, here the exact form of the noise process (equivalently the corresponding state $E$) has to be known. This may e.g. achieved by performing a process tomography of the CPM resulting after the universal depolarization protocol describe in the previous section (note that only the knowledge of the depolarized map is required). In addition, the fidelity of the ideal operation is no longer conserved but decreased by a certain amount. That is, by ''sacrificing`` a (small) amount of the fidelity of the operation, one can modify the resulting noise process in such a way that one obtains a very simple standard form. This is done by transferring weight from the ideal operation to the noisy part in an appropriate way and hence tailor the noise process. 

\subsubsection{The noisy SWAP gate}

Let us consider a noisy SWAP gate in the two qubit case.
For a specific noisy SWAP operation with sufficiently large Jamio\l kowski fidelity $f=F(\mathcal{E},U_\text{SWAP})=\langle\psi_\text{SWAP}|E|\psi_\text{SWAP}\rangle > \frac{15}{16}$ a depolarization procedure can be designed that brings the noisy operation $\mathcal{E}$ to the standard form 
\be \mathcal{E}'(\rho) = f'\, U_\text{SWAP} \rho U_\text{SWAP}^\dagger + (1-f')\,\frac{1}{16} \mathbf{1}_{AB} \ee
with $f'>f/3$. Thus the noise in the standard form corresponds to white noise, where the noise level $(1-f')$ is at most increased by a factor of $5.5$. As in Sec.~\ref{SWAP_SF} this immediately follows from the results for the case, where the ideal operation is the identity, by simply applying the designed depolarization procedure from the end of the Sec.~\ref{MultiDecSF} with the role of particles $A'$ and $B'$ exchanged. Note that the corresponding twirling procedure remains local w.r.t. the physical partitioning $(AA',BB')$.

\subsubsection{The noisy CNOT--type gate}

We consider now the noisy CNOT--type gate $U(\pi/4)$, described by the standard form given in Eq.~(\ref{ECNOT}). We will further depolarize the corresponding noise process in such a way that the fidelity of the ideal operation is decreased (as few as possible) and the noise is global white noise, i.e. the simplified standard form is given by
\bea
\label{whitenoise}
{\cal E}'\rho &=& \tilde q U(\pi/4)\rho U(\pi/4)^{\dagger} + (1- \tilde q) \frac{1}{16}\sum_{ij} \sigma_i\sigma_j \rho \sigma_i\sigma_j \nonumber\\
&=& \tilde q U(\pi/4)\rho U(\pi/4)^{\dagger} + \frac{1- \tilde q}{16}\mathbf{1},
\eea
where the fidelity of the ideal operation $\tilde f = \tilde q + (1-\tilde q)/16$. We find that the amount of noise is increased at most by (approximately) an order of magnitude, i.e. $(1- \tilde f)/(1- f) \approx 20$. Clearly, such a further depolarization is only useful if the fidelity of the ideal operation is initially sufficiently large, i.e. $f \gtrsim 0.96$, as otherwise the completely depolarizing operation would be obtained.

We will first demonstrate that a depolarization to global white noise is possible , and will then discuss the resulting decrease of fidelity. The state $E_S$ can be written as 
\be
E_S= f E_{\pi/4} + (1-f) \tilde E,
\ee
where $\tilde E$ is the operator corresponding to the noise part of the CPM ${\cal E}_S$ (see Eqs.~(\ref{ECNOT}), (\ref{bigGamma})) and 
\bea
E_{\pi/4} & = &|\Psi_{\pi/4}\rangle\langle \Psi_{\pi/4}| = \frac{1}{2}( |\Psi_{00}\rangle\langle \Psi_{00}| + |\Psi_{22}\rangle\langle \Psi_{22}| \nonumber\\
&& + i|\Psi_{00}\rangle\langle \Psi_{22}| - i|\Psi_{22}\rangle\langle \Psi_{00}|), 
\eea
is the operator corresponding to $U(\pi/4)$. 

We will first show that by means of local unitaries, one can change the off--diagonal elements of $E_{\pi/4}$ in such a way that each off--diagonal element in $\tilde E$ (or equivalently $E$) can be erased by probabilistically applying either the corresponding unitary operation or the identity with appropriate probability. Here, we are no longer restricted to operations that keep $|\Psi_{\pi/4}\rangle$ invariant, but can use arbitrary local unitaries. We first note that by applying $\mathbf{1}_{AA'}\otimes\sigma^B_z\otimes\sigma_z^{B'}$, one can change the sign of the off--diagonal elements $|\Psi_{00}\rangle\langle \Psi_{22}|$ and $|\Psi_{02}\rangle\langle \Psi_{20}|$, which implies that in the following discussion the sign of the off--diagonal elements do not play a role. 

By the depolarization procedure in Sec.~\ref{CNOT_SF} we can assume the sub--block $\Gamma_{00}$ ---spanned by the states $\{|\Psi_{00}\rangle,|\Psi_{02}\rangle,|\Psi_{20}\rangle,|\Psi_{22}\rangle\}$--- of the (total) state $E$ to be in the form
\be
\Gamma_{00}= \left(%
\begin{array}{cccc}
  A & 0 & 0 & i Y \\
  0 & B & i X & 0 \\
  0 & - i X & B & 0 \\
  -i Y & 0 & 0 & A \\
\end{array}%
\right),
\ee
with 
\bea
A = f/2 + (1-f) a,  && B = (1-f) b, \\
Y  = f/2 + (1-f) u, && X  = (1-f) v,
\eea
where $a=\mu_{00,00}, b=\mu_{02,02}, i u= \mu_{00,22}, i v = \mu_{02,20}$ (see Eqs.~(\ref{ECNOT}), (\ref{bigGamma})).  

We consider the operations ${\cal U}_x=(\mathbf{1}\otimes\sigma_x)$ and ${\cal U}_z=(\mathbf{1}\otimes\sigma_z)$. We have that the action of these operations on Bell states is given by 
{\small \bea
{\cal U}_x\{|\psi_0\rangle,|\psi_1\rangle,|\psi_2\rangle,|\psi_3\rangle\}&=& \{|\psi_1\rangle,|\psi_0\rangle,i|\psi_3\rangle,-i|\psi_2\rangle\}, \nonumber \\
{\cal U}_z \{|\psi_0\rangle,|\psi_1\rangle,|\psi_2\rangle,|\psi_3\rangle\}&=& \{|\psi_3\rangle,i|\psi_2\rangle,-i|\psi_1\rangle,|\psi_0\rangle\},\nonumber 
\eea}
It follows that applying with probability $1/2$ the operation ${\mathbf{1}}_{\bm A}{\cal U}_x^{\bm B}$ or ${\mathbf{1}}_{\bm A}{\cal U}_z^{\bm B}$ transforms the state $E$ to the state $E^1$. In particular, the subspace $\Gamma_{00}$ is transformed to the subspace $\Gamma^1_{01}$ ---spanned by $\{|\Psi_{01}\rangle,|\Psi_{03}\rangle,|\Psi_{21}\rangle,|\Psi_{23}\rangle\}$--- and some coefficients are aligned. One finds that the resulting elements are given by
\be
\label{gamm01s}
\Gamma^1_{01} = \left(%
\begin{array}{cccc}
  C & 0 & 0 & Z \\
  0 & C & - Z & 0 \\
  0 & - Z & C & 0 \\
  Z & 0 & 0 & C \\
\end{array}%
\right),
\ee
where 
\bea
C=(A+B)/2, && Z=(X+Y)/2,
\eea
Note that the element $Z$ is real. At the same time, the subspace $\Gamma_{01}$ is transformed to $\Gamma^1_{00}$ (where the off diagonal elements are given by $i (1-f) w$ after the transformation, while the diagonal elements are still given by $(1-f)c$). Also the subspaces $\Gamma_{10}$ and $\Gamma_{11}$ are transformed into each other, where one finds that $\Gamma^1_{10}$ is diagonal with elements $(1-f) e$, and also $\Gamma^1_{11}$ is diagonal with elements $(1-f) d$.

Similarly, if one applies the operations ${\cal U}_x$, ${\cal U}_z$ in ${\bm A}$ instead of ${\bm B}$, the state $E$ is transformed to a state $E^2$. In particular, the subspace $\Gamma_{00}$ is transformed to the subspace $ \Gamma^2_{10}$ ---spanned by $\{|\Psi_{10}\rangle,|\Psi_{30}\rangle,|\Psi_{12}\rangle,|\Psi_{32}\rangle\}$---, where the elements of $\Gamma^2_{10}$ are the same as of $\Gamma_{01}$ (see Eq.~(\ref{gamm01s})). The transformation of the other subspaces follows accordingly, only the role of systems ${\bm A}$ and ${\bm B}$ is exchanged (e.g. $\Gamma_{01} \rightarrow \Gamma^2_{11}$ etc.). 
Note that one can simultaneously change the sign of all off--diagonal elements of the resulting state $E^2$ by applying $\mathbf{1}_{AA'}\otimes\sigma^B_z\otimes\sigma_z^{B'}$.

If one thus mixes the resulting states $E$ (with probability $p_0$), $\tilde E^1$ (with probability $p_1$) and $E^2$ (with probability $p_2$), ---and by appropriately choosing the phases of the corresponding off diagonal elements--- one can achieve that the final state $E''$ has no off diagonal elements in the subspaces $\Gamma''_{01}$ and $\Gamma''_{10}$. This is guaranteed by choosing 
\be
p_0=\frac{Z}{Z+(|w|+|x|)(1-f)},\nonumber\\
p_1=\frac{|w|(1-f)}{Z+(|w|+|x|)(1-f)},\\
p_2=\frac{|x|(1-f)}{Z+(|w|+|x|)(1-f)},\nonumber
\ee
The other coefficients of the resulting state can be readily calculated, taking into account whether a change of sign was necessary for $E^1$ or $E^2$, where we denote $\sigma_1={\rm sign}(w)+1, \sigma_2={\rm sign}(x)+1$ with $(-1)^{\sigma_1}={\rm sign}(w)$. One finds that each of the subspaces $\Gamma''_{01}, \Gamma''_{10}, \Gamma''_{11}$ is diagonal (with all coefficients equal) and described by a coefficient, $\gamma_{01}, \gamma_{10}, \gamma_{11}$ respectively, where 
{\small \bea
\label{gammaij}
\gamma_{01}&=&p_0 (1-f) c +(-1)^{\sigma_1} p_1 C +(-1)^{\sigma_2} p_2 (1-f) e, \nonumber \\
\gamma_{10}&=&p_0 (1-f) d + (-1)^{\sigma_1} p_1 (1-f) e + (-1)^{\sigma_2} p_2 C, \\
\gamma_{11}&=&p_0 (1-f) e + (-1)^{\sigma_1} p_1 (1-f) d + (-1)^{\sigma_2} p_2 (1-f) c \,. \nonumber 
\eea }
The subspace $\Gamma''_{00}$ is given by
\be
p_0\Gamma_{00} + (-1)^{\sigma_1} p_1 \Gamma^1_{00} + (-1)^{\sigma_2} p_2 \Gamma^2_{00},
\ee
where we find that resulting diagonal elements are 
{\small \be
A'' = p_0 A+ (-1)^{\sigma_1} p_1 (1-f) c + (-1)^{\sigma_2} p_2 (1-f) d, \nonumber \\
B'' = p_0B+ (-1)^{\sigma_1} p_1 (1-f) c + (-1)^{\sigma_2} p_2 (1-f) d,
\ee}
while the off diagonal elements are given by
{\small \be
i Y'' = p_0 i Y+ (-1)^{\sigma_1} i p_1 (1-f) w + (-1)^{\sigma_2} i p_2 (1-f) x, \nonumber \\
i X'' = p_0 i X+ (-1)^{\sigma_1} i p_1 (1-f) w + (-1)^{\sigma_2} i p_2 (1-f) x.
\ee }
It remains to show that one can erase the off--diagonal element $|\Psi_{02}\rangle\langle \Psi_{20}|$, $i\tilde X$. This can be accomplished by randomly applying the operation $\mathbf{1}_{\bm A}\mathbf{1}_{\bm B}$ or ${\cal W}^{\bm A}\mathbf{1}_{\bm B}$ (see Eq.~(\ref{calW})) with probabilities $p$ and $(1-p)$. If ${\rm sign}(\tilde X) = {\rm sign}(\tilde Y)$, one applies in a addition $\sigma_z^B\otimes \sigma_z^{B'}$ in the second case in order to change the sign of the corresponding off--diagonal elements. Choosing 
\be
p=(| X''|+ | Y''|)^{-1},
\ee 
ensures that the off--diagonal element $|\Psi_{02}\rangle\langle \Psi_{20}|$ vanishes, while $|\Psi_{00}\rangle\langle \Psi_{22}|$ becomes 
\be
Y' = i p  Y'' + (-1)^{{\rm sign}  X''+1} i (1-p)  X''
\ee
and the diagonal elements become 
\be
 A'= p A'' + (1-p) B'', \\
 B'= p B'' + (1-p) A''. 
\ee
Note that all other elements of $\tilde E'$ remain invariant. Finally, the remaining off--diagonal element $|\Psi_{00}\rangle\langle \Psi_{22}|$, $i\tilde Y'$, which corresponds in part to the ideal operation and in part to the noise part, can be formally incorporated into the ideal part of the evolution, i.e. the resulting state can be formally rewritten as
\be
E_f=f'|\Psi_{\pi/4}\rangle\langle\Psi_{\pi/4}| + (1-f')D,
\ee
where $D$ is diagonal in the basis $\{|\Psi_{ij}\rangle\}$ with elements $d_{ij}$ and $f'=2Y'$. The elements $d_{ij}$ of $D$ in the blocks $\Gamma_{01},\Gamma_{10},\Gamma_{11}$ are given by Eq.~(\ref{gammaij}) (i.e. $d_{01}=d_{03}=d_{21}=d_{23}= \gamma_{01}$ etc.), while in the block $\Gamma_{00}$ we have 
\bea
d_{02}&=&d_{20}= B',\\
d_{00}&=&d_{22}= A'- Y'.
\eea

The coefficients $d_{ij}$ can even be made equal by further reducing the fidelity of the ideal operation. In this case, $D$ corresponds to the completely mixed state, and the corresponding map is given by global white noise. This is done as follows: Using that the probabilistic operation $\tilde {\cal U}^{\bm A}\mathbf{1}_{\bm B}$ or $\mathbf{1}_{\bm A}\mathbf{1}_{\bm B}$, one produces a state diagonal in the basis $\{|\Psi_{ij}\rangle\}$ with the same diagonal coefficients as $E_f$, where we consider the situation where $d_{00} = d_{22} \gg d_{ij}$, i.e. the fidelity of the ideal operation is sufficiently large. In this sense, weight from the ideal operation can be transferred to the other states. In particular, one uses
$\mathbf{1}_A{\cal W}^B$ to increase weight in $|\Psi_{02}\rangle \langle\Psi_{02}|$ and $|\Psi_{20}\rangle \langle\Psi_{20}|$;
${\mathbf{1}}_{\bm A}{\cal U}_x^{\bm B}$ and ${\mathbf{1}}_{\bm A}{\cal U}_z^{\bm B}$ to increase weight of $\Gamma_{01}$; 
${\cal U}_x^{\bm A}{\mathbf{1}}_{\bm B}$ and ${\cal U}_z^{\bm A}{\mathbf{1}}_{\bm B}$ to increase weight of $\Gamma_{10}$; 
${\cal U}_x^{\bm A}{\cal U}_x^{\bm B}$, ${\cal U}_z^{\bm A}{\cal U}_z^{\bm A}$, 
${\cal U}_x^{\bm A}{\cal U}_z^{\bm B}$ and ${\cal U}_z^{\bm A}{\cal U}_x^{\bm B}$
to increase weight in $\Gamma_{11}$.

One thus ends up with a standard form described by global white noise as announced (see Eq.~(\ref{whitenoise})). We will evaluate the loss factor for the fidelity for an alternative protocol discussed in the next section. This protocol is capable of depolarizing also noisy phase gates to a one--parameter standard form.

\subsubsection{The noisy phase gate}\label{tnpg}

In principle, it may be possible to obtain a simplified standard form for the gate $U(\alpha)$ with arbitrary $\alpha$ by similar means as in the case of the noisy CNOT--type gate $U(\pi/4)$, i.e. by manipulating the noisy evolution in such a way that weight is transferred from the ideal evolution to the appropriate noise parts. However, for small $\alpha$ one encounters a difficulty which may be hard to circumvent. When using parts of the operator corresponding to the ideal evolution to eliminate off diagonal elements in other parts of the density matrix corresponding to the noisy evolution, we have that automatically also the diagonal elements are transferred and hence the noise part is further increased. While in the case of $U(\pi/4)$, the increase of diagonal elements of the noise part is of the same order of magnitude as the off--diagonal elements, for small $\alpha$ this is no longer the case. We have that the off diagonal element of the ideal operations, $E_{\alpha}$, is given by $\lambda_{00,22}=i \cos(\alpha)\sin(\alpha)$, while the larger diagonal element, $\lambda_{00,00}$, is given by $\cos^2(\alpha)$. Imagine we have elements in the noise part of order $\epsilon\ll 1$, where both diagonal and off diagonal terms of order $\epsilon$ appear. If one wants to eliminate an off--diagonal element in the noise part which is of order $\epsilon$, we need $(1-p) \cos(\alpha)\sin(\alpha) = |\epsilon|$, i.e. with probability $(1-p)$ the off--diagonal element of the ideal evolution is transferred to the off--diagonal element of the noise part with the sign chosen in such a way that the total off--diagonal element in the noise part vanishes. However, by doing such a transformation, one of the diagonal elements of the noise part is automatically increased by $(1-p)\cos^2(\alpha)$ which is of the order $|\epsilon|/\tan(\alpha)$. Note that for small $\alpha$, $1/\tan(\alpha) \gg 1$, i.e. the amount of noise may be increased by orders of magnitude. This is clearly not acceptable, as our goal was to obtain a simplified standard form by sacrificing a relatively small amount of the fidelity and not to decrease the fidelity by (several) orders of magnitude.

However, under certain conditions one may use an alternative method which still allows one to obtain a standard form corresponding to global white noise, specified by a single parameter. In particular, if one can switch the noisy operation on and off at will, i.e. one can decide whether one wants to apply the noisy operation or does not want to apply it (and instead apply something else), then such a depolarization is possible. If one considers the case where $U(\alpha)$ is some non--local gate, then it is natural to assume such a controllability. 
In this case, one can either apply the noisy evolution with certain probability $p$ or apply some other operation with probability $(1-p)$. In particular, one can apply any separable operation. This may, however, involve the application of arbitrary local operations (including measurements), rather than the application of local unitaries as we have assumed so far. Considering the corresponding states, this amounts to mixing of the state $E$ with some separable (in the sense ${\bm A} - {\bm B}$) state $D$. The separable operation $\mathcal{D}$ associated to $D$ can be implemented by some random application of local operations, $\mathcal{D}=\sum_i\, p_i\, A_iB_i^T \rho (A_iB_i^T)^\dagger$. The corresponding Kraus operators can be obtained from the spectral decomposition of $D$ as indicated in Appendix A. Let us now consider a density matrix $A$ of a separable two--qubit state written in the standard basis. Then all states of the block diagonal form (see e.g. Eq.~(\ref{NoisyPhaseGateSF}) with separable block matrices $\Gamma_{ij}=A$ for $i,j \in \{0,1\}$ are again separable (recall that $\Gamma_{ij}$ denotes subspaces spanned by $|\Psi_{kl}\rangle$ with $k \mod 2 = i, l\mod 2=j)$. In particular, any matrix $A$ of the form
\be
\label{Asep}
A =\frac{1}{4}\left(%
\begin{array}{cccc}
  1& 0 & 0 & \beta \\
  0 & 1 & \alpha & 0 \\
  0 & \alpha^* & 1 & 0 \\
  \beta^* & 0 & 0 & 1 \\
\end{array}%
\right),
\ee
with $\alpha,\beta \in \{0,1,-1,i,-i\}$ as well as any diagonal matrix $A$ (with positive coefficients summing up to one) is separable. This can be checked by calculating the partial transposition of these states, where one finds that the partial transposition is positive in all cases which is (for two--qubit states) sufficient to ensure separability \cite{Pe96,Ho96}. The corresponding separable maps can be implemented by a simple sequence of random local unitary operations (in the case of diagonal $A$), or measurements (in the case of matrices of the form \ref{Asep}). The Kraus representation of the state can be obtained as shown in Appendix A. 

It is now straightforward to see that mixing $E$ with separable operators of the form $\Gamma_{ij}=A$ with $A$ given by Eq.~(\ref{Asep}) or an appropriate diagonal matrix, allows one to eliminate all (unwanted) off--diagonal elements as well as to align all diagonal coefficients of the noise part. Thus the resulting simplified normal form of the noisy operation is given by
\bea
{\cal E}_S\rho &=& \tilde q U(\alpha)\rho U(\alpha)^{\dagger} + (1- \tilde q) \frac{1}{16}\sum_{ij} \sigma_i\sigma_j \rho \sigma_i\sigma_j \nonumber\\
&=& \tilde q U(\alpha)\rho U(\alpha)^{\dagger} + \frac{1- \tilde q}{16}\mathbf{1},
\label{simplestandardform}
\eea
The fidelity of the ideal operation $\tilde f = \tilde q + (1-\tilde q)/16$ is reduced, where we find e.g. for $\alpha=\pi/4$ that if $f=1-\epsilon$, then 
\be
\tilde f \geq 1-17\epsilon.
\ee 
That is, the fidelity of the operation is reduced by about an order of magnitude. This can be seen as follows. Consider for example the CNOT like gate $U(\pi/4)$ with corresponding standard form of noise $\tilde E$ given by Eq.~(\ref{bigGamma}). We have $\text{tr}(\tilde E)= 2a+2b+4c+4d+4e = (1-f)$ and denote $y = \max{(a,b,c/2,d/2,e/2)}$, i.e. $f \leq 1-2y$. One can make $\tilde E$ diagonal by mixing with matrices $\Gamma'_{ij}=A$.  Consider for instance the case where all noise is concentrated in $b,v$ (this in fact turns out to correspond to a (non unique) worst case scenario), i.e. $y=b$. One mixes $\tilde E$ (with probability $p$) with a matrix of the form $\Gamma_{00}'=A$ with $\beta=0, \alpha=-i$ (with probability $(1-p)$). The resulting matrix is diagonal for $p=(4v+1)^{-1}$ and has diagonal elements $b'=b p + (1-p)/4 \leq 2y (4y+1)^{-1}$, $a'=(1-p)/4$ and $c'=d'=e'=0$. Note that the worst case corresponds to $v=b$. By mixing the resulting matrix (with probability $q$) with a diagonal matrix (with probability $1-q$), one can make all diagonal elements equal (which corresponds to white noise). We have $q \geq (4y+1)/(32y +1)$. This leads to a total final fidelity $\tilde f = p q f \geq  f/(32y+1) \geq f/(17-16f)$. For $f=1-\epsilon$, we thus have
\be
\tilde f \geq f/(17-16f) \geq 1- 17 \epsilon,
\ee
i.e. the fidelity is decreased by about an order of magnitude. Note that this formula also holds in the general scenario with arbitrary coefficients $a,b,c,d,e,u,v,w,x$. In the first step, the worst case is given when all off diagonal elements are maximal, $u=a, v=b, w=c, x=d$. In the second step (making all diagonal elements equal), it is clearly the worst case if all weight is concentrated in one element (e.g. $b$) and all others are zero, as one has to fill up the weights of the other (14) diagonal elements. Non--zero diagonal elements would require less mixing, leading to a larger final fidelity. Thus $\tilde f \geq 1- 17(1-f)$ is a conservative bound on the final fidelity, where in many situations one will end up with a much larger final fidelity of the depolarized noisy operation.

\subsection{Standard form for arbitrary two--qubit unitary operation}\label{arbitraryUnitary}

Let us briefly discuss the possibility to generalize the results in Sec.~\ref{Unitary_SF} and Sec.~\ref{Sacrificing_SF} to the more general case of an arbitrary unitary operation as the ideal operation.
Consider an arbitrary unitary $U$ (or even a class unitaries $U_\alpha$). Note that similar to Eq.~(\ref{LU_CNOT_PhaseGate}) any unitary two-qubit gate can be represented uniquely as  \cite{Kr00}
\be U^{AB} = U_1^A\otimes U_2^B \, e^{-i \sum_{i=1}^3 \mu_i \sigma^A_i\sigma^B_i}\,  V_1^A\otimes V_2^B  \; \ee
with $\frac{\pi}{4} \geq \mu_1 \geq \mu_2 \geq |\mu_3| \geq 0$ and some single-qubit unitaries $U_i, V_i$ ($i=1,2$). The main block of the depolarization procedures in Sec.~\ref{Unitary_SF}  was  to find a set of $(A,A')$--local unitaries, that leave the corresponding pure state $|\Psi_U\rangle$ (or a class pure states $|\Psi_{U_\alpha}\rangle$ ) invariant. However, in general, ---apart from some special cases where e.g. all $\mu_i$ equal or $\mu_2=\mu_3=0$--- the only local unitary operation that keeps the state invariant is given by the identity operation. Hence, in generic cases, a depolarization of the operation ---under the condition that the fidelity of the operation remains invariant--- is impossible following this approach. Thus no universal standard form for arbitrary two--qubit unitary operations can be obtained along these lines.

It seems more appropriate to follow the ideas of Sec.~\ref{Sacrificing_SF} and to {\it specifically} design such a standard form for a noisy unitary taking the given form of decoherence into account, and allow for an increase of noise. We do not deepen such a discussion at this point but rather refer to an alternative approach. Instead of regarding the noisy unitary gate as a CPM and allowing for manipulation before and after the application of this operation one might as well consider the dynamical evolution realizing this gate and allow for a manipulation of this evolution by several short intermediate pulses of local unitary control operation. Inspired by the results in \cite{HamSim} it is shown in the following section that by this procedure it is possible to depolarize an {\em arbitrary} master equation (of two systems) to a standard form described by at most 17 parameters.


\section{Standard form for noisy evolutions described by a master equation}\label{SF_Dynamics}

In this section, we will consider the evolution of two qubits described by a master equation of Lindblad form, where the ideal evolution is given by an arbitrary two--qubit Hamiltonian. The results can be readily generalized to multi--qubit systems whose (noiseless) interaction is described by Hamiltonians that are sums of two--body Hamiltonians. 

We will consider a continuous evolution $\rho(t)=\mathcal{E}_t\rho(0)$ of the system due to Markovian quantum dynamics starting at $t=0$ in the state $\rho(0)$. Thus the family of quantum operations $\mathcal{E}_t$ forms a Markovian semi-group \cite{Markovian}, determined by some generator $\mathcal{Z}$, that in the case of two qubits $A$ and $B$ and the convention $\hbar\equiv 1$ satisfies the following differential equation ({\it master equation}):
\bea   &\frac{\partial}{\partial t} \rho & =  \mathcal{Z} \rho := -i \mathcal{H}\, \rho + -i \mathcal{H}_\text{l}\, \rho + \mathcal{L}\, \rho \;\; \text{with}\\
 \mathcal{H}\,\rho & := &  [H^{AB}, \rho] \; , \nonumber \\
 \mathcal{H}_\text{l}\, \rho & :=&  [H^{AB}_\text{l}, \rho] \;\; \text{and} \nonumber  \\
 \mathcal{L}\, \rho & := & \sum_{\genfrac{}{}{0pt}{}{\mathbf{k},\mathbf{l} }{ \mathbf{k} \neq \mathbf{0}\, \vee\, \mathbf{l} \neq \mathbf{0}}} L_\mathbf{kl}
\left([ \sigma_\mathbf{k}^{AB} \,\rho,  \sigma_\mathbf{l}^{AB}] + 
 [ \sigma_\mathbf{k}^{AB} , \rho\,  \sigma_\mathbf{l}^{AB}]\right) \, .\nonumber
\eea
We have separated the unitary evolution of the dynamics into one part $H$, which corresponds to the {\it ideal unitary process} in question, and into a 'Lamb shift' $H_\text{l}$, which is some unitary dynamics, that is induced by the coupling between the system and its environment and therefore corresponds to noise. The more relevant influence of the decoherence process $\mathcal{L}$ is however incorporated in the 'Liouvillian', which is determined by the {\it positive} 'GKS'-matrix $\mathbf{L}=(L_\mathbf{kl})$ \cite{GKS}. Note that the corresponding sum is over all multi--indices $\mathbf{k}=(k_1,k_2)$ and $\mathbf{l}=(l_1,l_2)$ with $k_i,l_j=0,1,2,3$ except $\mathbf{k}=\mathbf{0}:=(0,0)$ or $\mathbf{l}=\mathbf{0}$. Thus the totally mixed state $\sigma_{\mathbf{0}}^{AB}=\frac{1}{4}\mathbf{1}_{AB}$ does not occur in the sum and $\mathbf{L}$ is a positive $15\times 15$--matrix.  

Our goal is now to bring some dynamical evolution $\mathcal{E}_t = e^{\mathcal{Z} t}$ into an appropriate {\it standard form} $\mathcal{E}'_t$. Here, $\mathcal{E}_t$ approximates the ideal unitary evolution $\mathcal{I}$ given by $\mathcal{H}$, where 
\be
\mathcal{Z}= -i\mathcal{H} -i\mathcal{H}_\text{l}+\mathcal{L}, 
\ee 
while the standard form $\mathcal{E}'_t$ is specified by 
\be
\mathcal{Z}'= -i\mathcal{H} -i\mathcal{H}'_\text{l}+\mathcal{L}'. 
\ee
That is, the decoherence process corresponding to the standard form is described by $H'_\text{l}$ and $\mathbf{L}'$, rather than $H_\text{l}$ and $\mathbf{L}$ in the original evolution. As in the case of CPMs, our goal is to obtain a simplified standards form in the sense that the number of relevant parameters describing the decoherence process are decreased, while the desired Hamiltonian evolution is not altered. 
To achieve this we will make use of the following facts \cite{MarkovSim}:
\begin{itemize}
\item[{\bf (i)}] Let $U$ be some unitary matrix and $\mathcal{U}\rho:=U\rho U^\dagger$ be the corresponding operation. By {\it unitary conjugation} the Markovian evolution $\mathcal{E}_t=e^{\mathcal{Z}t}$ can be transformed into the Markovian evolution $\mathcal{E}'_t= \mathcal{U}\circ \mathcal{E}_t\circ\mathcal{U}^\dagger=e^{\mathcal{Z}'t}$ described by 
\bea H'& =& UHU^\dagger \; ,\\
H'_\text{l} & = & UH_\text{l}U^\dagger \; ,\\
\mathbf{L}'& = & O \mathbf{L} O^T \; ,
\eea
 where $O$ is the orthogonal matrix corresponding to $U$, that describes the basis change $\sigma_\mathbf{k}\mapsto U\sigma_\mathbf{k}U^\dagger$ for the linear basis of hermitian traceless operators $\sigma_\mathbf{k}$.

\item[{\bf (ii)}] A Markovian evolution $e^{\mathcal{Z}' t}$ according to a linear combination $\mathcal{Z}'=\sum^R_{i=1} p_i \mathcal{Z}_i$ ($\sum_i p_i =1$) can be simulated by repeatedly applying a sequence $e^{\mathcal{Z}_1 \Delta t} \cdots e^{\mathcal{Z}_R \Delta t}$ for some small time intervals $\Delta t=\frac{t}{M}$ ($M$ is the number of repetitions), i.e. \be \label{Lio}\begin{CD}\left( \prod_{i=1}^R e^{p_i \mathcal{Z}_i  \frac{t}{M}}\right)^M @>{M \rightarrow \infty}>>e^{\sum^R_{i=1} p_i \mathcal{Z}_i t } \; . \end{CD} \ee  
Note that the approximated GKS matrix is $\mathbf{L}'= \sum_{i} p_i \mathbf{L}_i$ and the error  in approximation is of order $O(\Delta t^2)$ \cite{error}.

An alternative method consists in the {\em random} application of the time evolutions $e^{\mathcal{Z}_i\Delta t}$ with probability $p_i$ in each of the time intervals $\Delta t =t/M$. That is, the evolution in the time interval $\Delta t$ is given by $\sum_i p_i e^{\mathcal{Z}_i\Delta t}$, which accurately approximates the desired operation in first order $\Delta t$. A sequential application of these random operations $M$ times reproduces the desired operation $e^{\sum^R_{i=1} p_i \mathcal{Z}_i t }$ in the limit $M \to \infty$, i.e. \cite{note_randomops} 
\be
\begin{CD} \left( \sum_{i=1}^R p_i e^{\mathcal{Z}_i  \frac{t}{M}}\right)^M @>{M \rightarrow \infty}>>e^{\sum^R_{i=1} p_i \mathcal{Z}_i t } \; \end{CD}.
\ee
\end{itemize}

Following the structure of the previous section we first consider the case of decoherence itself, i.e. we set $H=0$. We propose a depolarization protocol that, after integration of the corresponding master equation, yields the same standard forms as obtained in Sec.~\ref{NFoneParty} and Sec.~\ref{MultiDecSF}. Second, we consider the case where $H$ corresponds to some Ising-type interaction, e.g. $H=\sigma_y^A\otimes \sigma_y^B$. We make use of the results obtained in Sec.~\ref{PhaseGateSF} for the corresponding unitary $e^{-iHt}$. We show that a depolarization procedure exists for which ---in the limit of infinite intermediate local control operations--- the system evolves according to some standard form which has the standard form Eq.~(\ref{NoisyPhaseGateSF}), when regarded as a CPM $\mathcal{E}'_t$. The depolarization procedures we describe in the following (Sec.~\ref{SF_Dynamics_Dec} and Sec.~\ref{SF_Dynamics_Ising}) are only relevant in cases where one is interested in the standard form for the complete dynamics and not only after some given time (the latter corresponding to the case of CPMs discussed in the previous Sections). Otherwise, one may use the conceptually simpler depolarization procedure for CPMs. Nevertheless subsections \ref{SF_Dynamics_Dec} and \ref{SF_Dynamics_Ising} provide the necessary tools for the procedure proposed in the subsequent Sec.~\ref{SF_Dynamics_Gen}, where we show how to achieve a standard form for some {\it arbitrary} unitary dynamics. Although this procedure overcomes the restrictions to the area of applications in Sec.~\ref{Unitary_SF}, this depolarization protocol generally increases the noise level of the decoherence process.
 In the following we assume that local unitary control operations can be performed on {\it time scales negligible compared to the interaction time for the dynamics}. We will thus refer to these operations as being {\it instantaneous}.

\subsection{Standard form for decoherence processes}\label{SF_Dynamics_Dec}

We first consider maps describing pure decoherence processes ($H=0$) for a single qubit. For the depolarization we consider the same twirling procedures as in Sec.~\ref{NFoneParty}, but now we intend to bring the Markovian {\it generator} $\mathcal{Z}$ of the initial dynamics into the standard form 
\be\label{Dyn_SF_Z_Dec} \mathcal{Z}'= \,\sum_k \,u_k\, U_k \mathcal{Z} U_k^\dagger\; ,\ee 
where $U_k$ denote the unitaries which were applied  in Sec.~\ref{NFoneParty} with equal probability $u_k$ to achieve the standard form of a {\it Pauli channel}, i.e. $U_k=\sigma_k$ is one of the Pauli matrices and $u_k=\frac{1}{4}$, or the simpler standard form of a {\it depolarizing channel}, i.e. $U_k$ is of the form $Q_l\sigma_i$ ($Q_l=e^{i\frac{\pi}{4}\sigma_l}$, $l=1,2,3$) and $u_k=\frac{1}{12}$. In Sec.~\ref{NFoneParty} the number $u_k$ represented the probability to apply the different twirling unitaries. According to {\bf (ii)}, a similar procedure still works, where intermediate random applications of the corresponding unitaries lead to a standard form of the Markovian generator $\mathcal{Z}$. Alternatively, one can consider a further splitting of the time interval $Delta t$ (see Eq. \ref{Lio}), where all possible unitary operations corresponding to the depolarization process are applied sequentially. We will consider the second approach in the following.    

More precisely, we will consider the following depolarization protocol:
Let the actual dynamics of the system (i.e. the decoherence process) evolve for some time $t$ and choose a split of the total time $t$ into $M$ sufficiently small time intervals $\Delta t$.  During each of these small time intervals the system dynamics is accompanied by the sequence of instantaneous local control operations $U_{k+1}U_k$ ($U_0=\mathbf{1}$) applied in arbitrary order but with equal distance $u_k\,{\Delta t}$.  From fact {\bf (i)} it follows that instead of the original dynamics $e^{i\mathcal{Z} \,u_k\,\Delta t}$ during each of the time intervals $u_k \, {\Delta t}$  the system evolves according to the Markovian generator $U_k \mathcal{Z} U_k^\dagger$. In the time interval $\Delta t$ the evolution is thus given by 
\be \prod_{k} \,e^{u_k \Delta t \,\mathcal{U}_k \mathcal{Z} \mathcal{U}_k^\dagger }  \; .\ee 
If these intervals are chosen sufficiently small (i.e. $M\rightarrow \infty$) fact {\bf (ii)} implies that the overall system dynamics can effectively be approximated by the Markovian generator in Eq.~(\ref{Dyn_SF_Z_Dec}).

Let us consider the effect of this depolarization on $\mathbf{L}$ and $H_\text{l}$ more closely. According to {\bf (i)} the GKS matrix is brought into the standard form Eq.~(\ref{PauliChannel}) or Eq.~(\ref{DepolSFqubit}), except that the first row and column of Eq.~(\ref{GenCPM}) is disregarded in both equations. Thus we obtain 
\be\label{Dyn_SF_Dec_Pauli} \mathcal{L}' \rho = - 2 \,\sum_{k=1}^3 \,L_k\,\left(\rho -\sigma_k \rho \sigma_k \right)  \ee 
with $L_k=L_{kk}$ in the case of twirling with Pauli operators $U_k=\sigma_k$ and 
\be\label{Dyn_SF_Dec_Depol}  \mathcal{L}' \rho = - 2 L \,\left(3 \rho - \sum_{k=1}^3 \,\sigma_k \rho \sigma_k \right) = - 4 L\, \left(2 \rho - \mathbf{1}\right) \ee with $L=\frac{1}{3}\left(L_{11}+L_{22}+L_{33}\right)$ in the case of complete depolarization. Similarly for the Hamiltonian $H_\text{l}$ the twirling by means of the Pauli matrices $U_k=\sigma_k$ yields (see {\bf (i)}):
\be H'_\text{l}  = \frac{1}{4}\,\sum_{k=0}^3 \, \sigma_k H_\text{l} \sigma_k = \frac{1}{2}\, \text{tr} (H_\text{l})\, \mathbf{1} \;.\ee
Since this twirling is also performed before each of the Clifford unitaries $Q_l$ applied, we obtain the same result in the case of complete depolarization. Thus in both cases the Lamb shift Hamiltonian in the standard form gives only rise to some overall phase factor $e^{- \frac{i}{2} \text{tr} (H_\text{l})\, t}$, which can be neglected.  

We briefly examine the dynamics $\mathcal{E}_t'$ due to the Standard forms, i.e. the solutions of the master equation 
\be \dot \rho = \mathcal{L}' \rho\; ,\ee
where the Liouvillian $\mathcal{L}'$ is given by Eq.~(\ref{Dyn_SF_Dec_Pauli}) or Eq.~(\ref{Dyn_SF_Dec_Depol}). It is straightforward to see that in case of Eq.~(\ref{Dyn_SF_Dec_Pauli}) the Pauli matrices diagonalize the Liouvillian $\mathcal{L}'$ \cite{BriegelEnglert}, i.e. 
\bea 
\mathcal{L}' \sigma_0 = & 0   & \mathcal{L}' \sigma_1  =  - 4(L_2+L_3) \sigma_1  \nonumber\\
\mathcal{L}'\sigma_2  = & - 4(L_1+L_3) \sigma_2  & \mathcal{L}' \sigma_3  =  - 4(L_1+L_2) \sigma_3  \; .\nonumber
\eea
For an arbitrary initial state $\rho(0)= \frac{1}{2} \left[\mathbf{1} + \mathbf{n} \cdot {\pmb \sigma}\right]$ ($|\mathbf{n}|=1$, ${\pmb \sigma}\equiv (\sigma_1,\sigma_2,\sigma_3)^T$) we thus obtain
\be \rho(t) = e^{\mathcal{L}'t}\rho(0) = \frac{1}{2} \left[\mathbf{1} + \mathbf{n}(t) \cdot {\pmb \sigma}\right] \; ,\ee
where $\mathbf{n}(t) = \left(n_1 e^{-4(L_2+L_3)t},\, n_2 e^{-4(L_1+L_3)t},\, n_3 e^{-4(L_1+L_2)t}\right)^T$.
The action of $\mathcal{E}_t'$ in terms of Pauli matrices as in Eq.~(\ref{GenCPM}) is actually that of a Pauli channel 
\be \mathcal{E}_t\, \rho  =  \sum_{k=0}^3 \, E_k(t) \, \sigma_k \rho \sigma_k \; ,\ee
where 
{\small
\bea
E_0(t) & = & \frac{1}{4}\, \left(1 + e^{-4(L_2+L_3)t} + e^{-4(L_1+L_3)t} + e^{-4(L_1+L_2)t} \right) \nonumber \\
E_1(t) & = & \frac{1}{4}\, \left(1 + e^{-4(L_2+L_3)t} - e^{-4(L_1+L_3)t} - e^{-4(L_1+L_2)t} \right) \nonumber \\
E_2(t) & = & \frac{1}{4}\, \left(1 - e^{-4(L_2+L_3)t} + e^{-4(L_1+L_3)t} - e^{-4(L_1+L_2)t} \right) \nonumber \\
E_3(t) & = & \frac{1}{4}\, \left(1 - e^{-4(L_2+L_3)t} - e^{-4(L_1+L_3)t} + e^{-4(L_1+L_2)t} \right) \nonumber \; .
\eea
}
Similarly a specialization to complete depolarization reveals the depolarizing channel 
\be \mathcal{E}_t\, (\rho)  = p(t) \,\rho + (1-p(t))\,\text{tr}(\rho)\,\frac{1}{2}\mathbf{1}  \ee
with $p(t)=e^{-8 L t}$. 

The fact that the above depolarization procedure for continuous dynamics rediscover the standard forms already obtained in Sec.~\ref{NFoneParty}  for CPM can also be understood by the simple fact, that the evolution $e^{\mathcal{Z}'t}$ is invariant under unitary conjugation, i.e. $\mathcal{U}e^{\mathcal{Z}'t}\mathcal{U}^\dagger = e^{\mathcal{Z}'t}$, iff the generator $\mathcal{Z}'$ is invariant under this conjugation. By construction of the protocol above, this definitely holds for the standard form $\mathcal{Z}'$ w.r.t. any of the corresponding twirling unitaries $U_k$. With $u_k$ being the twirling {\it constant} we thus obtain 
\be \mathcal{E}'_t = e^{\mathcal{Z}'t} = \sum_k\, u_k \,\mathcal{U}_k \,e^{\mathcal{Z}'t} \,\mathcal{U}_k^\dagger \; .\ee
Since this Eq. precisely reflects the corresponding depolarization procedure for CPM of Sec.~\ref{NFoneParty}, the resulting time dependent CPM $\mathcal{E}'_t$ has to be of standard form for CPMs.

The depolarization protocols for decoherence processes (described in this subsection) can be readily generalized to the multi-party setting.  The Lamb shift $H_\text{l}'$  in the standard form can be neglected again whereas the Liouvillian is of a standard form  that corresponds to the multi-party Pauli channel Eq.~(\ref{MultiPauli}) or the multi-party depolarizing channel Eq.~(\ref{isotropic1}). Note, however, that each additional party causes a finer splitting of the time interval $\Delta t$, yielding an exponential increase of the number of intermediate control operations. More precisely for $N$ parties the system dynamics in each time interval $\Delta$ has to be interrupted by $4^N$ [$12^N$] control operations in order to achieve a dynamics corresponding to a Pauli channel [depolarizing channel]. In this case, the alternative method of applying random local unitary operations rather than the complete sequence of unitaries is certainly privileged.

\subsection{Standard forms for noisy Ising-type interactions}\label{SF_Dynamics_Ising}

Let us now move to the case where the ideal operation is given by some Ising-type interaction 
\be
H^{AB} = g \sigma_y^A \sigma_y^B. 
\ee
For depolarization we can essentially consider the same protocol as in the previous section, except that we now take the twirling unitaries $U_k$, that were used in Sec.~\ref{PhaseGateSF}. More precisely  $u_k=\frac{1}{32}$ and the $U_k$ are given by the $32$ unitaries from the product set $\mathcal{U}_1 \cdot \mathcal{U}_2 \cdot\mathcal{U}_3 $, where
\bea 
\mathcal{U}_1 & = & \{\mathbf{1}_A\mathbf{1}_B,\,  e^{-i\frac{\pi}{4}\sigma_y^A}\mathbf{1}_B,\, \mathbf{1}_A e^{-i\frac{\pi}{4}\sigma_y^B},\, e^{-i\frac{\pi}{4}\sigma_y^A}e^{-i\frac{\pi}{4}\sigma_y^B}\} \nonumber \\
\mathcal{U}_2 & = & \{\mathbf{1}_A\mathbf{1}_B,\, \sigma_x^A \sigma_x^B \} \\
\mathcal{U}_3 & = & \{\mathbf{1}_A\mathbf{1}_B,\, \sigma_y^A\mathbf{1}_B,\, \mathbf{1}_A \sigma_y^B, \,\sigma_y^A \sigma_y^B\}\; , \nonumber
\eea
 e.g. $U_k=  e^{-i\frac{\pi}{4}\sigma_y}\sigma_x \otimes \sigma_x \sigma_y$. Recall that a twirling with these unitaries brings the phase gate $U(\alpha)=e^{-i H \alpha} $ with an arbitrary angle $\alpha$ into a standard form Eq.~(\ref{NoisyPhaseGateSF}) described by only $17$ independent parameters.

Apart from the different choice of twirling unitaries the depolarization reads exactly as in Sec.~\ref{SF_Dynamics_Dec}:
The overall interaction time $t$ is divided into sufficiently small time intervals $\Delta t$, in which the system dynamics is interrupted for short local unitary control operations $U_k$. The resulting dynamics approximates an evolution $\mathcal{E}'_t=e^{\mathcal{Z}'t}$ with new Markovian generator 
\be \mathcal{Z}'= \sum_{k=1}^{32}\, u_k\,\mathcal{U}_k \mathcal{Z} \mathcal{U}_k^\dagger = -i \mathcal{H} -i \mathcal{H}'_\text{l} + \mathcal{L}' \; .\ee
Note that due to the choice of the twirling unitaries the ideal Ising-type interaction Hamiltonian $H'=H$ is in fact not changed by this protocol. The GKS matrix $\mathbf{L}'$ is of standard form Eq.~(\ref{NoisyPhaseGateSF}), except that in Eq.~(\ref{NoisyPhaseGateSF}) the first row and column are disregarded. Since no normalization constraints are involved, the smaller $15\times 15$-matrix is still specified by $17$ real parameters.
For the new Hamiltonian of the lamb shift one can compute 
that the twirling yields $H_\text{l}' = H^\text{l}_{00} \sigma^A_0\sigma^B_0 + H^\text{l}_{22} \sigma^A_2\sigma^B_2 $. Omitting the term $H^\text{l}_{00} \, \mathbf{1}_{AB}$, which would contribute only as global phase $e^{-iH^\text{l}_{00}t}$ to the system dynamics, the Lamb shift in the standard form thus can be specified by a single real parameter $H^\text{l}_{22}$. Moreover we will include this term into the ideal interaction term rewriting $H'= g' \sigma_y^A\sigma_y^B$ with $g'= g+ H^\text{l}_{22}$ and will disregard any Lamb shift in the following. The change in coupling strength demands a reinterpretation of the interaction time. In order to simulate the actual dynamics $e^{\mathcal{Z}t}$ running for some time $t$ by a dynamics of the standard form $e^{\mathcal{Z}'t}$ the simulation actually has to run for the time $ t_s = c t$, where $c=\frac{g}{g'}$ represents the {\it time cost} of the simulation.  

A similar argument as in the previous subsection shows that the obtained standard form $e^{\mathcal{Z}'t }$ seen as a CPM is actually in the standard form considered in Sec.~\ref{PhaseGateSF}.

\subsection{Standard forms for arbitrary noisy evolutions by means of sacrificing}\label{SF_Dynamics_Gen}

Let us now consider standard forms for arbitrary ideal unitary evolutions $\mathcal{H}$. We make use of the fundamental fact \cite{HamSim} that by a stroboscopic application of a sequence of local unitaries any (entangling) two--qubit Hamiltonian $H$ can simulate the Hamiltonian $H_{y}=\sigma_y\otimes \sigma_y$  of the phase gate operation $U(\alpha)=e^{-i H_y \alpha}$ in Sec.~\ref{PhaseGateSF} to arbitrary good approximation (and vice versa).  
Before going into detail we shortly sketch the procedure of deriving a standard form for any noisy unitary evolution. Along each single stroboscopic step $\Delta t =\frac{t}{M}$ ($M$ number of steps) of the simulation protocols we propose to 
\begin{itemize}
\item[(1)] first apply the sequence of unitary operations in order to obtain the evolution according to $-i\mathcal{H}_y -i\mathcal{H}_{\text{l}y}+ \mathcal{L}_y$,
\item[(2)] depolarize the CPM $\mathcal{F}_t$ described by  $-i\mathcal{H}_y -i\mathcal{H}_{\text{l}y}+ \mathcal{L}_y$ according to the procedure derived in Sec.~\ref{SF_Dynamics_Ising} yielding a standard form $\mathcal{F}'_t$ given by  $-i\mathcal{H}_y -i\mathcal{H}'_{\text{l}y}+ \mathcal{L}'_y$ and finally
\item[(3)]  transform it back to the original Hamiltonian $H$ accompanied by some decoherence process of desired standard form $\mathcal{H}'_\text{l}$ and $\mathcal{L}'$.\\
\end{itemize}
\[ \begin{CD}
  \mathcal{H} @<\text{approximates}<< \mathcal{Z} @>\text{simulate}>> \mathcal{Z}_y  \\
 @.                              @V\text{standard form}VV  @VV\text{standard form}V\\
  \mathcal{H} @<<\text{approximates}< \mathcal{Z'} @<<\text{simulate}< \mathcal{Z}'_y 
\end{CD} \]

We remark that steps (1) and (3) require in general a 'time cost', i.e. the simulation of the action of a desired Hamiltonian for some time $c t$ requires a time $t$. This time cost translates into a smaller pre--factor for the interaction Hamiltonian in the corresponding generator $\mathcal{Z'}$, ultimately leading to an increased noise. That is, the ratio of the strength of desired interaction (described by $\mathcal{H}$) to strength of noise (described by $\mathcal{L}$) decreases, leading to a reduction of the fidelity. We note that for two--qubit systems the time cost is at most 3.

In order to simulate the Hamiltonian $H_y$ by the (entangling) two--qubit Hamiltonian $H$ and fast local unitary transformations (see \cite{HamSim}) one considers the decomposition of $H_y$ in terms of $H$ (term isolation):  
\be\label{forthDecomp} H^{AB}_y = c_v \, \sum_{i=1}^{R_v} v_i V_i \, H^{AB}\,  V_i^\dagger  + Q_1^{A} + Q_2^{B} \; ,\ee
where $V_i$ are the local unitaries with probabilities $v_i>0$ ($\sum_i v_i=1$), $Q_1$ and $Q_2$ are some local Hamiltonian on qubit $A$ and $B$ respectively and $c_v>0$ is some factor to adjust the coupling 'strength' of the Hamiltonians $H$ and $H_y$. If the unitary evolution $e^{-i H_y t}$ is supposed to be simulated for the time $t$, the simulation has to be carried out for the time $t_s=c_vt$. Since the local unitary control operation can be performed on negligible time scales, the factor $c_v$ thus determines the time cost for the following simulation: One chooses a split of the time $t_s$ into $M$ time intervals $\Delta t$, such that the sequence $\left( \prod_i e^{-i v_i \mathcal{V}_i \mathcal{H} \mathcal{V}_i^\dagger \Delta t} \, e^{-i \mathcal{Q}_1  \Delta t} \, e^{-i \mathcal{Q}_2 \Delta t}\right)^M $ is a sufficient approximation for $e^{-i \mathcal{H}_y t_s }$ as discussed in (ii). Note that in each time step $\Delta t$ the original dynamics according to $H$ is simply interrupted after the time $v_i\, \Delta t$ in order to apply the local unitaries $V^\dagger_{i+1}V_{i}$  ($V_0=\mathbf{1}$). This corresponds to the system evolving according to a sequence of Hamiltonians $V_i H V_i^\dagger$ for the time intervals $v_i \Delta t$. At the end of each simulation step $\Delta t$ the local unitary $ e^{-i \mathcal{Q}_1 \Delta t} \otimes e^{-i\mathcal{Q}_2 \Delta t}$ has to be applied in order to cope with the single qubit dynamics in the Hamiltonian simulation.

Similarly, one can consider a Hamiltonian simulation for step (3) according to the decomposition 
\be\label{backDecomp} H^{AB} = c_w \, \sum_{j=1}^{R_w} w_j W_j \, H^{AB}_y \,  W_j^\dagger  + \tilde{Q}_1^{A} + \tilde{Q}_2^{B} \; \ee
with local unitaries $W_j$,  single qubit Hamiltonians $ \tilde{Q}_1$, $ \tilde{Q}_2$ and time cost $c_w$ for the 'backward' simulation. For step (2) we use the twirling protocol derived in Sec.~\ref{SF_Dynamics_Ising} providing a standard form Eq.~(\ref{NoisyPhaseGateSF}) described by $17$ independent parameters. With $c_y$ we denote the corresponding time cost of this depolarization procedure.

With these notations at hand we can now specify the protocol to achieve the standard form $\mathcal{E}'_t=e^{\mathcal{Z}' t}$ for an arbitrary noisy two--qubit evolution $\mathcal{E}_t=e^{\mathcal{Z} t}$. Let $\mathcal{E}_t=e^{\mathcal{Z}t}$ be a Markovian evolution with the generator $\mathcal{Z}=-i \mathcal{H} -i \mathcal{H}_\text{l} + \mathcal{L}$, where $\mathcal{H}\,\rho=-i[H,\rho]$ corresponds to the ideal evolution with Hamiltonian $H$, $\mathcal{H}_\text{l}\,\rho=-i[H_\text{l},\rho]$ represents the Lamb shift with Hamiltonian $H_\text{l}$ and 
\be  \mathcal{L}\, \rho = \sum_{\mathbf{k},\mathbf{l} \neq \mathbf{0}} L_\mathbf{kl}\left( [\sigma_\mathbf{k} \rho,  \sigma_\mathbf{l}] + [\sigma_\mathbf{k} ,\rho\sigma_\mathbf{l}]\right)
\ee 
corresponds to the Liouvillian with GKS-matrix $\mathbf{L}$. 
For notational simplicity we will in the following restrict to the case where in both steps (1) and (3) no single qubit dynamics has to be corrected, i.e. the terms $Q_1$, $Q_2$, $\tilde{Q}_1$ and $\tilde{Q}_2$ in the decompositions Eq.~(\ref{forthDecomp}) and  Eq.~(\ref{backDecomp}) vanish. If the system evolves for some time $t$, the following protocol requires the time $t_s:=c_v c_y c_w t$ and thus has time cost $c_v c_y c_w$. 
The 'simulation' $t_s$ again has to be divided into sufficiently small time steps $\Delta t=\frac{t_s}{M}$. In these time intervals we consider the following sequence of $R= 32 \,R_v \,R_w $ operations: \be \prod_{i,j,k=1}^{R} \, \mathcal{W}_j \mathcal{U}_k \mathcal{V}_i  \,e^{\mathcal{Z}\,  w_j u_k v_i \,\Delta t}\, \mathcal{V}_i^\dagger \mathcal{U}_k^\dagger \mathcal{W}_j^\dagger  \; .\ee
This sequence of operations corresponds to a splitting of the time interval $\Delta t$ into smaller intervals of length $ w_j u_k v_i \,\Delta t$, in which at the beginning the (fast) local unitary $W^\dagger_j U^\dagger_k V^\dagger_i$ is performed, the system then evolves according to the given dynamics $\mathcal{Z}$ and finally the inverse unitary 'pulse' $W_jU_kV_i $ is applied at the end of the interval $ w_j u_k v_i \,\Delta t$. In the limit $\Delta t \rightarrow 0$ we obtain the Markovian dynamics $\mathcal{E}'_t=e^{\mathcal{Z}' t}$ with  
\be \mathcal{Z}' =  \sum_{i,j,k=1}^{R} \,  w_j u_k v_i \, \mathcal{W}_j \mathcal{U}_k \mathcal{V}_i  \,\mathcal{Z}\, \mathcal{V}_i^\dagger \mathcal{U}_k^\dagger \mathcal{W}_j^\dagger \; . \ee
It is straightforward to show that the ideal operation $\mathcal{H}$ in the generator $\mathcal{Z}'$ remains the same, since the twirling over $U_k$ leaves the Hamiltonian $H_y$ invariant. Moreover $\mathcal{Z}'$ again has a decomposition of the form $\mathcal{Z}'= -i\mathcal{H} -i \mathcal{H}'_\text{l} +\mathcal{L}'$ described by the new lamb shift Hamiltonian $H'_\text{l}$ and the GKS matrix $\mathbf{L}'$, that are obtained as follows:
\begin{widetext}
\be
\begin{CD}
\mathcal{Z} @>{v_i, V_i}>(1)> \mathcal{Z}_y @>{u_k, U_k}>(2)> \mathcal{Z}'_y @>{w_j, W_j}>(3)> \mathcal{Z}' \\
H_\text{l}  @>>>  H_\text{yl}= \sum_i v_i V_i H_\text{l} V_i^\dagger @>>>  H'_\text{yl}= \sum_k u_k U_k H_\text{yl} U_k^\dagger @>>>  H'_\text{l}= \sum_j w_j W_j H'_\text{yl} WU_j^\dagger \\
\mathbf{L}  @>>> \mathbf{L}_y = \sum_i v_i O_{V_i} \mathbf{L} O_{V_i}^T @>>> \mathbf{L}'_y = \sum_k u_k O_{U_k} \mathbf{L}_y O_{U_k}^T @>>> \mathbf{L}' = \sum_j w_j O_{W_j} \mathbf{L}'_y O_{W_j}^T 
\end{CD}
\ee
\end{widetext}
As discussed in Sec.~\ref{SF_Dynamics_Ising} the effect of step (2) on  $\mathbf{L}_y$ is to bring the matrix into the standard form Eq.~(\ref{NoisyPhaseGateSF}). The final standard form $\mathbf{L}'$ of the GKS matrix is obtained from $\mathbf{L}'_y$ by mixing according to $(w_j,W_j)$ and is thus specified by $17$ independent parameters only, although $\mathbf{L}'$ in general is not of the form Eq.~(\ref{NoisyPhaseGateSF}). As seen in Sec.~\ref{SF_Dynamics_Ising}  we can neglect the lamb shift by introducing some time cost $c_y$. 

To summarize, we have shown how to achieve a standard form for arbitrary two--qubit interactions, where the noise process (described by the GKS matrix) is specified by 17 parameters. The above protocol can be affected by different {\it sources of errors}. For this, one can again compare the noise level of the standard form dynamics $\mathcal{Z}'$ with the noise level of the original dynamics $\mathcal{Z}$ in terms of the distance  $d(\mathcal{E}_t,\mathcal{I}_t)$ and $d(\mathcal{E}'_t,\mathcal{I}_t)$ to the ideal unitary evolution $\mathcal{I}_t=e^{i\mathcal{H}t}$ for different times $t$, where $d(\mathcal{E},\mathcal{I})$ is a suitable distance measure (see Sec.~\ref{distance}). Although we have yet not performed a detailed error analysis in this sense, a non--unit time cost (at most a factor of $c_v c_w \leq 3$ from simulating corresponding Hamiltonians - steps (1) and (3), plus the time cost $c_y$ from 'Lamb' shift), in general, corresponds to an increase of the noise level for the evolution. 

As it holds for the depolarization of CPMs in previous chapters and as opposed to the assumptions made in this paper, in practise, the depolarization protocol has to face imperfections in the local control operations, whose extent depends on the physical realization. Additionally, for the depolarization of master equations by means of stroboscopic control operations one also encounters errors of order $O(\Delta t^2)$ due to the finite approximation (see fact {\bf (ii)}) \cite{error}. Note that, in practise, there will be a trade-of between errors in approximation and errors due to imperfect local control operations.

\subsection{Simplified standard forms for arbitrary noisy evolutions}

A further reduction of the number of relevant noise parameters and thus a simpler standard form may be achieved following the ideas developed in Sec.~\ref{Sacrificing_SF} for CPMs. There, by increasing the noise level and hence reducing the fidelity of the operation, we have shown that one can in fact achieve that the noise part of the evolution is described by only a single parameter (white noise). The procedure outlined in Sec.~\ref{tnpg} is based on probabilistically mixing the (already depolarized) noisy CPM ${\cal E}$ with a certain separable map ${\cal D}$, i.e. a map which can be obtained without interactions between particles. That is, one chooses randomly whether one wishes to apply the map ${\cal E}$ corresponding to the noisy operation, or the separable map ${\cal D}$. For a proper choice of ${\cal D}$ the resulting map ${\cal E}_S$ is of the form Eq.~(\ref{simplestandardform}). 

In the case of master equations, one may adopt this procedure in such a way that for each time interval $\Delta t$, one applies the (already depolarized) noisy evolution described by the $\mathcal{Z'}$, together with an appropriate separable evolution (that may e.g. be generated with the help of available local unitary control operations and additional measurements) with corresponding generator $\mathcal{Z}_D$. Both evolutions now have to be applied either sequentially or chosen randomly. This implies that either one has the ability to switch off the evolution $\mathcal{Z'}$, or one can produce a separable evolution of arbitrary strength $\mathcal{Z}_D$. Note that in this case, fast local unitary operations are in general not sufficient, but arbitrary local control operations (including measurements) are required to generate the desired separable operations. As in the case of CPMs this depolarization procedure requires moreover the knowledge of the exact form for $\mathcal{Z'}$ in order to choose an appropriate, separable $\mathcal{Z}_D$. The total evolution is finally described by a Liouvillian with GKS matrix proportional to the identity, which corresponds to global white noise at the level of the respective CPM.

\section{Summary}\label{summary}

In this article, we have introduced the concept of depolarization of noisy evolutions. We have shown how to reduce the relevant number of parameters describing an arbitrary, unknown noise process described by a CPM in such a way that the ideal (unitary) part of the evolution is not altered. For decoherence processes  we have explicitly calculated the corresponding standard forms for multipartite systems of arbitrary number $N$ of parties and arbitrary dimension $d$. We find a reduction of an arbitrary noise process described by $O(d^{4N})$ to local and global white noise processes described by only $2^N$ parameters. For specific two--qubit unitary operations (e.g. Phase gate with arbitrary phase), we obtain a standard form described by at most 17 parameters. For other gates, the standard forms can be further simplified. In particular we find standard forms described by 8 parameters for the CNOT--gate and 3 parameters for the SWAP gate. The depolarization procedures used to obtain these standard forms are universal in the sense that the exact form of the noise process need not be known. 
With knowledge of the exact form of the noise process, and by allowing for a (small) reduction of the fidelity, one can further simplify the standard forms. In fact, we have derived a depolarization protocol that yields a reduction to {\em global white noise}, which is described by only a single noise parameter and where, in the worst case, the noise level is increased by about an order of magnitude.  

We have generalized our results to evolutions described by a master equation of Lindblad form. Standard forms for decoherence processes and interaction Hamiltonians proportional to the Ising Hamiltonian can be derived using similar methods as for CPMs, leading to standard forms with same number of parameters. We have also obtained a standard form described by 17 parameters for {\em arbitrary} two--qubit interaction Hamiltonians, which, in general, goes along with an increase of the noise level. As the basic tool we have used the possibility to simulate the Ising Hamiltonian by an arbitrary Hamiltonian (and vice versa), together with depolarization of the Ising type interaction. Again, a further simplification to a single parameter leading to a GKS matrix proportional to identity is possible under certain circumstances. 

We are confident that such simplified standard forms for noise processes will provide a useful tool to investigate various problems in quantum information processing involving noisy apparatus and interactions with environment. Straightforward applications include the possibility to calculate lower bounds on the channel capacity of arbitrary noise channels (by investigating the corresponding depolarized channels), and a simplified process tomography where only a reduced number of parameters of the noise process needs to be determined. Further conceivable applications include the determination of lower bounds to the lifetime of entangled states, and strict error thresholds for quantum computation that are valid for arbitrary noise processes and are not restricted to certain noise models.

\section*{Acknowledgements}
{We thank F. Verstraete for interesting discussions. This work was supported by the Austrian Science Foundation (FWF), the European Union (IST-2001-38877,-39227,OLAQUI,SCALA), the \"Osterreichische Akademie der Wissenschaften through project APART (W.D.) and the Deutsche Forschungsgemeinschaft (DFG).}


\section*{Appendix A: Spectral decomposition versus Kraus representation}\label{SpectralVsKraus}
For sake of completeness we will shortly discuss the relation of the Jamio\l kowski state $E$ with two common representations of the corresponding CPM $\mathcal{E}$ in terms of its Kraus representation and its purification. In Appendix A we consider a CPM $\mathcal{E}$ described by its Kraus representation. More precisely any CPM $\mathcal{E}$  acting on $\mathbf{H}^A$ allows for a decomposition of the form
\be\label{KrausRepr} \mathcal{E}(M) = \sum_{i=1}^r \, K_i M K_i^\dagger\ee
with $r\leq d_A\times d_{A'}$ ($r=$ Choi rank) {\it Kraus operators} $K_i \in \mathcal{M}(\mathbf{H}^A,\mathbf{H}^{A'})$, where $K_i$ can be chosen to be orthogonal, i.e. $\text{tr} \left[K_i^\dagger K_j \right] =\delta_{ij}$.
$\mathcal{E}$ corresponds to SLOCC operation iff
\be\label{TracePresKraus}  \sum_{\alpha=1}^r \,  K_i^\dagger K_i \leq \mathbf{1}\; ,\ee
where equality holds iff $\mathcal{E}$ is {\it trace preserving}.
Eq.~(\ref{KrausRepr}) is an immediate consequence of the corresponding fact, that any {\it positive} operator $E\geq 0$ (see No.~{\bf 3} in Sec.~\ref{IsoPhysics}) allows a {\it spectral decomposition}
\be\label{spectral} E = \sum_{i=1}^r \, |v_i \rangle \langle v_i | \;  ,\ee
where $|v_i \rangle$ are some unnormalized vectors in  $ \mathbf{H}^{A'}\otimes \mathbf{H}^A$, $ r = \text{rank}(E) \leq d_A\times d_{A'} $, which can be chosen to be orthogonal. Using the decomposition 
\be |v_i \rangle = \sum_{\genfrac{}{}{0pt}{}{\alpha \in \mathbb{N}_{d_{A'}}}{ \beta \in \mathbb{N}_{d_{A}}}} \, v^{\alpha\beta}_i |\alpha\rangle^{A'}|\beta\rangle^A  \ee this correspondence is simply given by
\be\label{Decomp_Rel} K_i = \sqrt{d_A} \;\sum_{\alpha \beta} \; v^{\alpha\beta}_i\; |\alpha\rangle^{A'}\,{}^A\hspace{-0.03cm}\langle \beta| \ee
or $K^{\alpha|\beta}_i=  \sqrt{d_A}\, v^{\alpha\beta}_i$.\\
Tracing out either system $A'$ or $A$ yields:
\bea \label{Tr_sigma}
\text{tr}_{A'} E^{A'A} & = & \sum_{i=1}^r \left[K_i^\dagger K_i\right]^T \\
\text{tr}_{A} E^{A'A} & = & \sum_{i=1}^r K_i K_i^\dagger 
\eea 
Note that No.~{\bf 4} in Sec.~\ref{IsoPhysics} now follows directly from Eq.~(\ref{TracePresKraus}) using Eq.~(\ref{Tr_sigma}). The {\it unitary freedom in the choice of decompositions} for $E$ translates to the corresponding CPM as follows. Two decomposition of $\mathcal{E}$ [$E$] with Kraus operators $K_i$ and $L_j$ [vectors $|v_i\rangle$ and $|w_j\rangle$] correspond to the same CPM [positive operator] iff there exists a unitary matrix $U_{ij}$ such that $K_i = \sum _{i j} U_{ij} L_j$ [$|v_i\rangle = \sum _{ij} U_{ij} |w_j\rangle$].
From these decompositions one can easily obtain a few more results about the relation between positive operators and CPM under the Jamio\l kowski isomorphism:
\begin{itemize}
\item[\bf{10.}] $\mathcal{E}$ is {\it factorizable}, i.e. $\mathcal{E}(M)=K M K^\dagger$, iff $E$ is {\it pure}, i.e. $E=|v \rangle\langle v |$.
\item[\bf{11.}] $\mathcal{E}$ is an {\it isometry}, i.e. $\mathcal{E}(M)=V M V^\dagger$ with $V^\dagger V= \mathbf{1}_A$, iff $d_A\leq d_{A'}$ and $E$ is {\it maximally entangled}, i.e. pure and $\text{tr}_{A'} E^{A'A} = \mathbf{1} _A$. (For a unitary we have $d_A=d_{A'}$.)
\item[\bf{12.}] $\mathcal{E}$ is a {\it projection}, i.e. $\mathcal{E}(M)= d_A\, \text{tr}\left[\tau_2^t M \right]\; \tau_1 $, iff $E$ is a {\it product state}, i.e. $E=\tau_1\otimes\tau_2$. 
\item[\bf{13.}] $\mathcal{E}$ can be decomposed into a {\it sum of projections}, iff $E$ is {\it separable}. In this case $\mathcal{E}$ is {\it entanglement breaking} \cite{Shor}.
\end{itemize}

\section*{Appendix B: Purification for quantum states and operations}\label{purification}
Frequently a CPM $\mathcal{E}$ is also regarded as a description of the (non-unitary) evolution of the system alone (i.e. by tracing the environmental degrees of freedom), where the system $A$ together with its environment $C$ is believed to evolve according to some unitary operation $U_{\mathcal{E}}$. Any such system-environment model $U_{\mathcal{E}}$  then is said to be a {\it purification} of the CPM $\mathcal{E}$, if it yields $\mathcal{E}$ as the evolution of the system alone after tracing out the environment:
\be \mathcal{E}(\rho)=\text{tr}_C\left[\,U_\mathcal{E}^{AC}\,\rho\otimes\rho_C\, (U_\mathcal{E}^{AC})^\dagger\, \right] \; .\ee 
Here we assume that the system $A$ and its environment $A'$ are initially decoupled $\rho\otimes\rho_C$, where $\rho_C=\frac{1}{d_C} \mathbf{1}_{C}$ is the maximally mixed state of the environment.
Like the operator sum decomposition, a purification of a quantum operation $\mathcal{E}$ in terms of unitary evolution $U_{\mathcal{E}}$ can simply be derived from the corresponding purification of the quantum state $E$ in terms of a pure state on the joint system $A'A$ and $C$. For this let $|\psi\rangle^{A'AC}$ be the pure state, such that $E^{A'A}=\text{tr}_{C}\left[\, |\psi\rangle^{A'AC}\langle\psi |\, \right]$, then the corresponding purification $U_{\mathcal{E}}$  is given similarly to Eq.~(\ref{Decomp_Rel}) by
\be \label{Purif_Rel} U_\mathcal{E}^{AC} = \sqrt{d_A\times d_C} \;\sum_{\genfrac{}{}{0pt}{}{i \in \mathbb{N}_{d_{A'}}}{\genfrac{}{}{0pt}{}{ j \in \mathbb{N}_{d_A}}{ k\in \mathbb{N}_{d_C}}}} \; \psi_{i j k}\; |i\rangle^{A'}\,{}^{A}\hspace{-0.03cm}\langle j|  \,{}^{C}\hspace{-0.03cm}\langle k|  \; ,\ee
using the decomposition $|\psi\rangle =\sum_{i,j,k} \psi_{ijk} |i\rangle^{A'}|j\rangle^{A}|k\rangle^{C}$. 
Note that in both cases purifications can be chosen such that $d_C=dim_\mathbb{C} (\mathbf{H}^C) \leq d_A \times d_{A'}$.

\section*{Appendix C: Details for the extension to $d$--level systems}

In this appendix we present some details about the generalization of the standard forms for decoherence to $d$--level systems as it was introduced in Sec.~\ref{NFoneParty}. We proof that 
\begin{itemize}
\item[(i)] the twirling over the Pauli group as in Eq.~(\ref{TwirlPauliGen}) depolarizes any CPM to the standard form of a generalized Pauli channel Eq.~(\ref{PauliChan}),
\item[(ii)] the standard form of a generalized depolarizing channel Eq.~(\ref{DepolChan}) can be achieved by twirling over a finite set of generalized Clifford operations.
\end{itemize}

But let us briefly consider the measurements in the generalized Bell basis Eq.~(\ref{GenBellBasis}) of the isomorphism protocol. It is straight forward to compute, that the other Bell measurement results $|\psi_{kl} \rangle^{A\bar{A}}$  yield $\mathcal{E} \left(U^*_{kl} \rho U_{kl}^T \right)$ instead of $\mathcal{E}(\rho)$ for the outcome of the protocol on system $A'$ but with the same probability $\frac{1}{d^2}$.

Before coming to the proofs, we summarize some useful facts (see also \cite{Wo03}) about the generalized Pauli operators $U_{kl}$ (see Eq.~(\ref{GenPauli})), that are straightforward to proof:
\bea
U_{k'l'} U_{kl} & = & e^{i \frac{2\pi}{d} k' \cdot l} U_{(k+k')(l+l')} \\ 
U_{kl}^\dagger & = & e^{i \frac{2\pi}{d} k \cdot l} U_{(-k)(-l)} \\
{U}^*_{kl} & = &  U_{(-k)(l)} \\
U_{kl}^T & = &   e^{- i \frac{2\pi}{d} k \cdot l} U_{(k)(-l)}\\
U_{k'l'}^A |\psi_{kl}\rangle & = & e^{i \frac{2\pi}{d} k' \cdot l} |\psi_{(k+k')(l+l')}\rangle \\
U_{k'l'}^{ A'} |\psi_{kl}\rangle & = & e^{- i \frac{2\pi}{d} (k+k') \cdot l'} |\psi_{(k+k')(l-l')}\rangle 
 \eea

With these relations at hand it is easy to verify (i). \\
First note that the set of generalized local Pauli operators
\be S\,=\,\{\, g_{kl} \; |\; g_{kl} = U_{kl}^A\otimes U_{(-k)(l)}^{ A'}\, ; \, k,l \in \mathbb{N}_{d} \, \}\ee 
is a commutative subgroup (of the generalized local Pauli group) that stabilizes $|\Omega \rangle$ and is generated by the two elements $g_{10}$ and $g_{01}$.
Moreover a simple calculation shows that 
\be g_{k'l'} |\psi_{kl}\rangle = e^{i \frac{2\pi}{d} \left(k' \cdot l - k\cdot l' \right)} |\psi_{kl}\rangle \; . \ee
For a general state  $E = \sum_{\alpha\beta,\alpha'\beta'} E_{\alpha\beta,\alpha'\beta'}\, |\psi_{\alpha\beta}\rangle \langle\psi_{\alpha' \beta'} | $ we therefore find, that it can be {\it diagonalized} by a probabilistic application of the local unitaries $g_{kl}$ with uniform probability $\frac{1}{d^2}$:
\bea
\mathcal{D} (E) & :=  &\frac{1}{d^2} \; \sum_{k,l =0}^{d-1} \; g_{kl}\,E\, g_{kl}^\dagger \nonumber \\ 
 & = & \frac{1}{d^2}\;  \sum_{\alpha\beta,\alpha'\beta'}\, E_{\alpha\beta,\alpha'\beta'}\, |\psi_{\alpha\beta}\rangle \langle\psi_{\alpha' \beta'} | \, \times \nonumber \\
 &  & \, \times \,\left(\sum_{k=0}^{d-1} e^{i \frac{2\pi}{d} k \cdot (\beta -\beta')} \right) \times \left(\sum_{l=0}^{d-1} e^{i \frac{2\pi}{d} l \cdot (\alpha -\alpha')} \right) \nonumber \\
 & = & \sum_{\alpha\beta}\, E_{\alpha\beta,\alpha\beta}\, |\psi_{\alpha\beta}\rangle \langle\psi_{\alpha \beta} |
\eea
This mixing operation $\mathcal{D} = \mathcal{D}_1 \circ \mathcal{D}_2$ can also be decomposed into a mixing of shift operation $\mathcal{D}_1(E)= \frac{1}{d}  \sum_{l =0}^{d-1}  g_{0l}E g_{0l}^\dagger$ and a mixing of phase multiplication operation $\mathcal{D}_2(E)= \frac{1}{d}  \sum_{k =0}^{d-1}  g_{k0}E g_{k0}^\dagger$.

Let us now consider statement (ii).\\
For this we assume that $\mathcal{E}$ is already brought into the form of a (generalized) Pauli channel of Eq.~(\ref{PauliChan}) by a random application of one of the Pauli operators $U_{kl}$. Now the generalized Clifford operation is a unitary operation $Q$ that maps the generalized Pauli group $\mathcal{P}=\left\{e^{i\frac{2\pi}{d} \delta} U_{kl}\, |\, k,l,\delta \in \mathbb{N}_{d} \right\}$ to itself under conjugation, i.e. $Q\mathcal{P}Q^\dagger=\mathcal{P}$. It is totally specified by its action on the two generators $U_{10}$ and $U_{01}$ of the Pauli group, since 
\bea QU_{kl}Q^\dagger & = &\left(QU_{01} Q^\dagger \right)^l \left(QU_{10} Q^\dagger \right)^k \\
& = &  U_{(ka+lb)(kc+ld)} \, .\eea
Here we have chosen $QU_{10}Q^\dagger= U_{ac}$ and $QU_{01}Q^\dagger= U_{bd}$. In addition we disregard appropriate phase factors $e^{i\frac{\pi}{d} \delta}$ with $\delta \in \mathbb{N}_{2d}$, since they will be irrelevant for our purposes. Up to these phase factors the Clifford unitary $Q$ permutes a Pauli operator $U_{kl}$ to a new element $U_{k'l'}$, which in modular arithmetic (modulo $d$) is related to $U_{kl}$ by the linear transformation 
\be \begin{pmatrix} k' \\ l' \end{pmatrix} = \mathbf{C}_Q\, \begin{pmatrix} k \\ l \end{pmatrix}:=
 \begin{pmatrix} a & b \\ c & d \end{pmatrix} \begin{pmatrix} k \\ l \end{pmatrix} \; .\ee   
This linear transformation needs to be symplectic \cite{symplectic} in order to truly correspond to a Clifford operation \cite{Ho04}. Symplecticity of $\mathbf{C}_Q$ in our single-party case simply reduces to the condition $\text{det} \mathbf{C}_Q = 1 $ (modulo $d$). Applying one of these Clifford unitaries $Q^A\otimes {Q^*}^{A'}$ to the state $E^{AA'}$ gives the state 
\bea & & Q^A\otimes {Q^*}^{A'} E (Q^A\otimes {Q^*}^{A'})^\dagger \nonumber \\
 &= & \sum_{kl}\, E_{kl}\, Q^A U_{kl}^{A} (Q^A)^\dagger\, Q^A\otimes {Q^*}^{A'} \;\times \nonumber \\
& \times & \; |\Phi \rangle^{AA'}\langle \Phi | \left( Q^A\otimes {Q^*}^{A'} \right)^\dagger \, \left(Q^A U_{kl}^{A} (Q^A)^\dagger \right)^\dagger  \nonumber \\
& = & \sum_{k'l'}\, E_{kl}\,  |\psi_{k'l'} \rangle^{AA'} \langle \psi_{k'l'} |  \; ,
\eea 
where as claimed above any phase factor would cancel out. Note that the component $E_{00}$ will remain 'untouched', since any symplectic matrix $\mathbf{C}_Q$ (even over $\mathbb{F}_d$ with $d$ non-prime) is invertible. In the following the Clifford unitary $Q$ will be chosen uniformly at random from a set of all Clifford unitaries, where each $Q$ corresponds only to a single $\mathbf{C}_Q$ (i.e.: fix a choice of phase factor for each $\mathbf{C}_Q$). 

By elementary results from group theory it follows, that by application of the different $\mathbf{C}$ the set of all vectors $(k,l)^T$  with $k\neq 0$ or $l\neq 0$ will be mapped onto itself in such a way, that all vectors will occur equally often. For this let $G$ denote the group of symplectic matrices over $\mathbb{F}_d^2$, that act on the set $X=\mathbb{F}_d^2$. Furthermore for $x\in X$ let $Gx:=\{gx\,|\, g\in G\}$ denote the orbit of $x$ under the group action $G$ and let $G_x:=\{g\in G\,|\, gx = x\}$ denote the stabilizer of $x$. From the stabilizer orbit and Lagrange theorem it follows that for any finite group $G$ acting on a set $X$ we have $|G|=|G_x|\,|Gx|$. This result can be used to show that each non--trivial element $y\in X\setminus \{0\}$ ($0:=(00)^T\in \mathbb{F}_d^2$) is obtained $|G|$ times, if the complete group $G$ is applied to all elements in $X\setminus \{0\}$. Since $G$ consists of invertible matrices, it maps the set $X\setminus \{0\}$ onto itself. Moreover any $y\in  X\setminus \{0\}$ is only obtained from elements of its orbit $Gy$. For a fixed element $x\in Gy$ the set $G_{xy}:=\{g\in G\,|\, gx=y\}$ can be rewritten in terms of only one of its elements $g'$ (i.e. $g'x=y$) and the stabilizer $G_x$ as $G_{xy}=g'G_x$. Thus $y$ is obtained form each of the $|Gy|$ elements (of its orbit) by $|G_x|$ different matrices. Since for two elements in the same orbit we have $Gx=Gy$, any element $y$ is obtained $|G_x|\,|Gx|=|G|$ times. Note that in the case of prime dimension $d$, the set $X$ is (not only a module but also) a vector space over the field $\mathbb{F}_d$ and one can easily show that for all $x\neq0$ the orbits are the same $Gx=X\setminus \{0\}$. This is due to the fact, that for each non--trivial vector $x$ one can find a symplectic (i.e. $\mathbb{F}_d$--invertible) matrix $g$ with the first column being $x$ (and the second the orthonormal vector $x^\perp$).

A random application of the corresponding Clifford operations therefore provides a mixing of all components $E_{kl}$ with $k\neq 0$ or $l\neq 0$. Thus starting with a CPM $\mathcal{E}$ in the form of a Pauli channel ( Eq.~(\ref{PauliChan})) we can achieve the standard form in Eq.~(\ref{DepolChan}) by uniformly choosing a unitary $Q$ form the set of Clifford operations and applying $Q^\dagger$ before and $Q$ after the application of the CPM $\mathcal{E}$. In fact the actual set, which the Clifford operations have to be chosen from in order to achieve a complete mixing of all the components $E_{kl}$ with $k\neq 0$ or $l\neq 0$, might even be decreased, as it is illustrated for the qubit case in Eq.~(\ref{finiteDepolqubit}). 
Note that the Clifford operation $Q_1$, $Q_2$ and $Q_3$ in Eq.~(\ref{finiteDepolqubit}) correspond to the three symplectic matrices $C_1=\begin{pmatrix} 1 & 1 \\ 0& 1\end{pmatrix}$, $C_2=\begin{pmatrix} 0 & 1 \\ 1 & 0\end{pmatrix}$ and $C_2=\begin{pmatrix} 1 & 0 \\ 1& 1\end{pmatrix}$.  



\begin{thebibliography}{99}

\bibitem{Lidar01}
For a short discussion of the relation between completely positive maps and quantum Markovian master equation as well as for further references see e.g. 
D.A. Lidar, Z. Bihary and K.B. Whaley, Chem. Phys. {\bf 268}, 35-53 (2001);
S. Daffer, K. Wodkiewicz, J.D. Cresser and J.K. McIver, E-print quant-ph/0309081.

\bibitem{We89}
R.F. Werner, Phys. Rev. A {\bf 40}, 4277 (1989).

\bibitem{Jam72}
A. Jamio\l kowski, Rep. Mod. Phys. {\bf 3}, 275-278 (1972).

\bibitem{Ci00}
J. I. Cirac, W. D\"ur, B. Kraus and M. Lewenstein, 
Phys. Rev. Lett. {\bf 86}, 544 (2001)

\bibitem{GLN}
A. Gilchrist, N. K. Langford and M. A. Nielsen,  Phys. Rev. A {\bf 71}, 062310 (2005).

\bibitem{Ar03}
P. Arrighi and C. Patricot,
Annals Phys. {\bf 311},  26-52 (2004).

\bibitem{local}
with respect to (w.r.t.) the partitioning $(A',A)$

\bibitem{Notation}
Some notation:  A map $\mathcal{E}$ is called (i) {\it Hermiticity preserving} iff it maps Hermitian matrices to Hermitian matrices, (ii) {\it positivity preserving} iff it  maps positive (semi-definite) matrices $\rho \geq 0$ to positive (semi-definite) matrices and is called {\it completely positive} iff for any state $\rho \in \mathcal{D} \left(\mathbf{H}^{A} \otimes \mathbf{H}^{C}\right)$  over $A$ and an arbitrary ancillary system $C$ $\mathcal{E}\otimes \text{Id}^C \left( \rho \right)\geq 0$. Equivalently complete positivity is given iff the map $\mathcal{E}$ has a Kraus representation $\mathcal{E}(\rho)=\sum_i K_i \rho K_i^\dagger$ with $\sum_i K_i^\dagger K_i \leq \mathbf{1}$. A complete positive map (CPM) is called {\it trace-preserving} iff it maps density matrices to density matrices, i.e. the Kraus representation fulfills $\sum_i K_i^\dagger K_i = \mathbf{1}$


\bibitem{Be98}
C. H. Bennett, D. P. DiVincenzo, C. A. Fuchs, T. Mor, E. Rains, P. W. Shor, J. A. Smolin, W. K. Wootters, eprint quant-ph/9804053.

\bibitem{ancEquiv}
This equivalence can be shown as follows: The protocol representing the isomorphism (P1) can easily be seen to be of the type (P2): 
(i)  performing operations on the input state plus ancillary system (ii) application of $\mathcal{E}$ at the input system (iii) performing operations on the output system plus some ancillary system that contains the ancillary system used before (iv) tracing out (forgetting) all information but the state at the output system. Concerning the possibility to simulate P1 by P2 one simply needs the maximally entangled state $P_\Phi$ at the ancillary system, whereas the operations before and after the application of $\mathcal{E}$ consist in swapping the state at the input/output system with one side of $P_\Phi$ and performing a Bell measurement. Conversely, since we are only interested in the comparison of the outcomes of the two protocols after tracing out all ancillary systems, one has to simulate the CPM $\mathcal{D}$ according to some protocol of type P2 by some protocol of type P1, which has the state $E$ available. But since the protocol P1 allows to perform arbitrary operations on the state $E$ before the Bell measurement is carried out, we certainly can change the state $E$ into $D$ representing the protocol P2. The same argumentation extends to the multi-party setting (Sec.~\ref{MultiPartyIso}), in which the operations performed before and after the application of $\mathcal{E}$ as well as the ancillary systems have to be local w.r.t. some given partitioning. The converse direction can in this case be regarded as an application of statement {\bf{7.}}, since the outcome of both protocols P1 and P2 can in general only be achieved probabilistically.



\bibitem{Oz00}
M. Ozawa, Phys. Lett. A {\bf 268}, 158 (2000).

\bibitem{Ki02}
A.Yu. Kitaev, A.H. Shen and M.N. Vyalyi, {\it Classical and Quantum Computation}, Graduate Studies in Mathematics Vol. {\bf 47}, American Mathematical Society (2002);

\bibitem{Ni00}
M.A. Nielsen and I.L. Chuang, {\it Quantum Computation and Quantum Information}, Cambridge University Press (2000).

\bibitem{metric}
A metric $d (x,y)$ is (i) positive $d (x,y)\geq 0$ with $d (x,y)=0$ iff $x=y$, (ii) symmetric  $d (x,y)=d (y,x)$ and (iii) obeys the triangle inequality $d (x,z) \leq d (x,y) + d (y,z)$.

\bibitem{chainingCond}
In fact, the chaining property $\Delta (\mathcal{E}_1\circ\mathcal{E}_2,\mathcal{F}_1\circ\mathcal{F}_2) \leq \Delta(\mathcal{E}_1,\mathcal{F}_1) + \Delta(\mathcal{E}_2,\mathcal{F}_2)$ has only been shown \cite{GLN} for $\Delta_1$ and $\Delta_2$, provided $\mathcal{F}_1$ is not only trace-preserving but also bi-stochastic, i.e. $\mathcal{F}_1(\mathbf{1})=\mathbf{1}$. Note that every unitary process is automatically bi-stochastic. Since we will in the following always compare noisy operations with ideal unitary processes, we will omit this extra assumption.

\bibitem{HoNi}
P. Horodecki, M. Horodecki and  R. Horodecki, Phys. Rev. A {\bf 60},1888 (1999);
M.A. Nielsen, Phys. Lett. A {\bf 303} (4), 249-252 (2002).


\bibitem{Du0102}
W. D\"ur and J. I. Cirac, Quantum Information and Computation, Vol. {\bf 2}, No. 3, 240-254 (2002);
W. D\"ur, G. Vidal and J. I. Cirac, Phys. Rev. Lett. {\bf 89}, 057901 (2002). 

\bibitem{PPT_preserving}
A state $\rho$ has non-negative partial transpose (PPT) w.r.t. some party $A_k$, iff $\rho^{T_{A_k}}\geq 0$. A CPM is PPT preserving, if it maps states, that are PPT w.r.t. party $A_k$, to states, that are PPT w.r.t. party $A'_k$;

\bibitem{slocc}
SLOCC $=$ stochastic local operation and classical communication;

\bibitem{Du00}
W. D\"ur and J. I. Cirac, Phys. Rev. A {\bf 64}, 012317 (2001). 

\bibitem{Du03}
W. D\"ur and H.-J. Briegel, Phys. Rev. Lett. {\bf 90}, 067901 (2003).


\bibitem{Conversion}
A bipartite state $|\psi_1\rangle$ can be transformed  into the state $|\psi_2\rangle$ by means of (i) SLOCC (ii) LOCC or (iii) LU operations iff for the corresponding Schmidt decomposition of $|\psi_1\rangle^{AB}=\sum_{i=1}^{R_1} \lambda^1_i |i\rangle^A |i\rangle^B$ and 
$|\psi_1\rangle^{AB}=\sum_{j=1}^{R_2} \lambda^2_j |j'\rangle^A |j'\rangle^B$ (i) $R_1\geq R_2$, (ii) the coefficient list $(\lambda^1_i)_{i=1}^{R_1}$ is a majorization of $(\lambda^2_j)_{j=1}^{R_2}$ or (iii) the coefficient list $(\lambda^1_i)_{i=1}^{R_1}$ coincides with $(\lambda^2_j)_{j=1}^{R_2}$ up to permutations.

\bibitem{Kr00}
B. Kraus and J. I. Cirac, Phys. Rev. A {\bf 63}, 062309 (2001). 


\bibitem{IBM}
C.H. Bennett, G. Brassard, S. Popescu, B. Schumacher, J.A. Smolin and W.K. Wootters,
Phys. Rev. Lett. {\bf 76}, 722–725 (1996).


\bibitem{parametercount}
Note that for Pauli channels $\mathcal{E}(\rho)=\sum_{kl} E_{kl} U_{kl}\rho U_{kl}^\dagger$ the trace-preserving-CPM condition (No.~4 in Sec.~\ref{IsoPhysics}) is simply that (i) $E_{kl}\geq 0$ and (ii) $\sum_{kl} E_{kl}=1$, i.e. that $E$ is a density matrix. This is because also for $d$-level systems $|\psi_{kl}\rangle$ are maximally entangled states and therefore $\text{tr}_{A'} \left( |\psi_{kl}\rangle^{AA'}\langle\psi_{kl} |\right)=\frac{1}{d} \mathbf{1}_{A}$ holds.

\bibitem{Hor99}
M. Horodecki and P. Horodecki, Phys. Rev. A {\bf 59}, 4206 (1999). 

\bibitem{Wo03}
M. Wolf, Ph.D. thesis, TU Braunschweig, Germany (2003).


\bibitem{IsoOf1}
Note that the CPM corresponding to $E=\frac{1}{d_A d_{A'}} \mathbf{1}_{A A'}$ is $\mathcal{E}(\rho)=\text{tr}(\rho)\, \frac{1}{d_{A'}}\mathbf{1}_{A'}$.

\bibitem{dir/indir}
As it was already mentioned above, if we follow an direct application of the isomorphism as stated in {\bf (A)}, we essentially allow an additional ancillary system, that is accessible before and after the application of $\mathcal{E}$ and that is separable w.r.t. the given partitioning.


\bibitem{MultiIndex}
Note that we again use the multi-index notation $\mathbf{k}=(k_1,\ldots,k_N)$ . 



\bibitem{noIncrease}
For any (probabilistic) operation $E\mapsto K E K^\dagger$ we have $K^\dagger K\leq \mathbf{1}$ and thus $\|K\|\leq 1$, where $\|X\|:=\max \{ \sqrt{\lambda} | \lambda \, \text{is eigenvalue of}\, X^\dagger X\}$ is the spectral norm. Thus $f'=\langle \Phi | K E K^\dagger \otimes \mathbf{1} | \Phi\rangle \leq \| K E K^\dagger\otimes \mathbf{1} \| \leq \|K\|^2 \|E\| \leq f$, if $f=\langle \Phi | E | \Phi\rangle$ corresponds to the largest eigenvalue of $E=E^\dagger$. If $|\Phi\rangle$ is the only eigenvector of $E$ to the largest eigenvalue $f$ and $K| \Phi\rangle \neq | \Phi\rangle$ it is easy to see that even $f'<f$ holds.

\bibitem{AllInvTrafo}
From $\mathbf{1}\otimes B | \Phi\rangle = B^T \otimes \mathbf{1} | \Phi\rangle$ it follows that any $C_1^A\otimes C_2^{\bar{A}}$ with $C_1 \otimes C_2|\Phi \rangle =|\Phi\rangle$ fulfills $C_1 C_2^T \otimes \mathbf{1} |\Phi \rangle =|\Phi\rangle$ and thus $C_1 C_2^T  = \mathbf{1}$, i.e. $C_1^{-1}= C_2^T$.




\bibitem{Du01}
W. D\"ur, G. Vidal and J. I. Cirac, Phys. Rev. Lett. {\bf 89}, 057901 (2002).

\bibitem{Pe96}
A. Peres, Phys. Rev. Lett. {\bf 77}, 1413 (1996).

\bibitem{Ho96} M. Horodecki, P. Horodecki and
R. Horodecki, Phys. Lett. A{\bf 223}, 8 (1996).

\bibitem{Markovian}
More precisely the family $\mathcal{E}_t$ of quantum operation, parametrized by real $t\geq 0$, is a continuous one--parameter semigroup, i.e. (i) $\mathcal{E}_0=\text{Id}$, (ii) $\mathcal{E}_s\circ \mathcal{E}_t =\mathcal{E}_{s+t}$ and (iii) the map $(t,\rho)\mapsto \mathcal{E}_t(\rho)$ from $[0,\infty)\times\mathcal{M}_A$ to $\mathcal{M}_{A'}$ is jointly continuous. The family $\mathcal{E}_t$ can equivalently be represented by its generator $\mathcal{Z}(\rho)= \lim_{t\downarrow 0} \frac{\mathcal{E}_t(\rho)-\rho}{t}$ in the exponential form $\mathcal{E}_t=e^{\mathcal{Z}t}$ or by the differential equation $\dot \rho= \mathcal{Z}\rho $.

\bibitem{GKS}
V. Gorini, A. Kossakowski and E.C.G. Sudarshan, L. Math. Phys. {\bf 17}, 821 (1976);
G. Lindblad, Comm. Math. Phys. {\bf 48}, 119–130 (1976).

\bibitem{HamSim}
J.L. Dodd, M.A. Nielsen, M.J. Bremner and R.T. Thew,  Phys. Rev. A {\bf 65}, 040301 (2002).
P. Wocjan, D. Janzing and Th. Beth,  Quant. Inf. Comp. {\bf 2}, 117 (2002). 
W. D\"ur , G. Vidal, J. I. Cirac, N. Linden and S. Popescu, Phys. Rev. Lett. {\bf 87}, 137901 (2001);
C. H. Bennett, J. I. Cirac, M. S. Leifer, D. W. Leung, N. Linden, S. Popescu, and G. Vidal, Phys. Rev. A {\bf 66}, 012305 (2002);
E. Jan\' e, G. Vidal, W. D\"ur, P. Zoller, and J.I. Cirac, Quant. Inf. Comp., Vol. {\bf 3}, No. 1, 15-37 (2003).
 

\bibitem{MarkovSim}
D. Bacon, A.M. Childs, I.L. Chuang, J. Kempe, D.W. Leung and X. Zhou, Phys. Rev. A {\bf 64}, 062302 (2001). 

\bibitem{error}
More precisely for $M$ time steps $\Delta t =\frac{t}{M}$ one obtains \bea & & \|e^{\sum_i p_i \mathcal{Z}_i \Delta t } - \prod_i e^{p_i \mathcal{Z}_i \Delta t}\| \\
 &\leq  &   \sum_{j,k=1}^R \|[\mathcal{Z}_j,\mathcal{Z}_k] \|\, \frac{\Delta t^2}{2} + O(\Delta t^3) \\
& \leq &  \sum_{j,k=1}^R \|\mathcal{Z}_j - \mathcal{Z}_k \|\, \Delta t^2 + O(\Delta t^3) \\
& \leq & C \, R^2 \, \Delta t^2  + O(\Delta t^3)\; ,
\eea
where $\|\mathcal{Z}\|:= \text{max}_{M: |M|_\text{tr}=1} |\mathcal{Z}(M)|$ ($|M|_\text{tr} =\text{tr}\left(\sqrt{M^\dagger M}\right)$) is the maximum norm w.r.t. the trace norm (for details about this worst case distance see \cite{GLN}) and $C:= 2\, \text{max}_k \|\mathcal{Z}_k \|$ is a constant depending on the generators $\mathcal{Z}_i$.
Since $e^{\mathcal{Z}'t}= \left(e^{\mathcal{Z}'\Delta t}\right)^M$ and because $\|\mathcal{Z}\|$ obeys the chaining property \cite{GLN}, we can estimate the total error
\be \|e^{\mathcal{Z}' t } - \left(\prod_i e^{p_i \mathcal{Z}_i \Delta t}\right)^M \| \leq M \, C \, R^2 \, \Delta t^2  + O(\Delta t^3) \; . \ee 

\bibitem{note_randomops}
This can be seen by considering a taylor series expansion of the expression $\left(\sum_{i=1}^R p_i e^{\mathcal{Z}_i  \frac{t}{M}}\right)^M$, which can be re--written as 
\be
\sum_{k=0}^{\infty} \frac{t^k}{k!}\left(\sum_{i=1}^R p_i e^{\mathcal{Z}_i  \frac{t}{M}}\right)^k \frac{M!}{(M-k)! M^k} + O(\frac{1}{M}).
\ee
Taking the limit $M \to \infty$, one finds that $\frac{M!}{(M-k)! M^k} \to 1$, and the $O(\frac{1}{M})$ terms vanish, although the overall convergence is rather slow. The resulting series exactly corresponds to $e^{\sum^R_{i=1} p_i \mathcal{Z}_i t}$.




\bibitem{BriegelEnglert}
H.J. Briegel and B.G. Englert, 
Phys. Rev. A {\bf 47}, 3311 (1993).

\bibitem{Shor} P.W.Shor, J. Math. Phys. Vol. {\bf 43}, 4334-4340 (2002). 


\bibitem{symplectic}
A matrix $C$ is symplectic iff it leaves the symplectic form $P=\begin{pmatrix} 0 & -1 \\ 1 & 0 \end{pmatrix}$ invariant, i.e. $P=C^TPC$. Note that its inverse is given by $C^{-1}=-PC^TP$.

\bibitem{Ho04}
see e.g. E. Hostens, J. Dehaene and B. De Moor, eprint quant-ph/0408190 (2004).



\end{thebibliography}
\end{document}